\documentclass[manuscript,screen, anonymous=False]{acmart}

\PassOptionsToPackage{dvipsnames}{xcolor}
\PassOptionsToPackage{colorlinks=true,linkcolor=RubineRed,breaklinks=True,citecolor=RubineRed}{xcolor}

\usepackage{enumitem}
\setlist[itemize]{noitemsep, topsep=0pt}

\usepackage[most]{tcolorbox}

\usepackage{adjustbox}
\usepackage[linesnumbered,algo2e,ruled]{algorithm2e}
\usepackage{multirow}
\usepackage{booktabs,subcaption,dcolumn}
\usepackage{tikz}
\usepackage{enumitem}
\usepackage{tabularx}
\usepackage[font=small]{caption}

\usepackage{pifont}
\usepackage{svg}
\usepackage{soul}
\usepackage{subcaption} 
\usepackage[utf8x]{inputenc}

\usepackage{amsmath}
\usepackage{amsfonts} 

\usepackage{colortbl}

\newcommand{\cmark}{\color{ACMGreen}{\ding{51}}}
\newcommand{\xmark}{\color{ACMRed}{\ding{55}}}

\newcolumntype{?}{!{\vrule width 1.5pt}}

\newcommand{\takeaway}[1]{
    \vspace{1em}
    \noindent\fbox{%
        \parbox{\columnwidth}{%
            \textbf{Takeaway}: {#1}
        }%
    }
}

\newtcolorbox{cooltextbox}[1][]{%
    colback=black!5,
    colframe=black!5,
    notitle,
    sharp corners,
    borderline west={1pt}{0pt}{red!80!black},
    enhanced,
    breakable,
    }

\newcommand{\overbar}[1]{\mkern 1.5mu\overline{\mkern-1.5mu#1\mkern-1.5mu}\mkern 1.5mu}

\newcommand\smamath[1]{{\small $#1$}}
\newcommand\smacal[1]{{\small $\mathcal{#1}$}}

\newcommand\ftcal[1]{{\footnotesize $\mathcal{#1}$}}
\newcommand\ftmath[1]{{\footnotesize $#1$}}

\newcommand\scmath[1]{{\scriptsize $#1$}}

\newcommand\revisionblue[1]{%
  \bgroup
  \hskip0pt\color{blue!80!black}%
  #1%
  \egroup
}

\newcommand\revisiongreen[1]{%
  \bgroup
  \hskip0pt\color{green!80!black}%
  #1%
  \egroup
}

\newcommand\dataset[1]{{\fontfamily{pcr}\selectfont {\footnotesize #1}}}

\newcommand\atk[1]{{\fontfamily{qhv}\selectfont {\small #1}}}
\newcommand\aatk[0]{$\widehat{\text{\atk{WA}}}$}
\newcommand\atks[2]{${\text{\atk{#1}}^{#2}}$}
\newcommand\aatks[1]{$\widehat{\text{\atk{WA}}^{#1}}$}

\newcommand\ftatk[1]{{\fontfamily{qhv}\selectfont {\scriptsize #1}}}

\newcommand\ftaatk[0]{$\widehat{\text{\ftatk{WA}}}$}

\newcommand\ftatks[2]{{\scriptsize ${\text{{\fontfamily{qhv}\selectfont #1}}^{#2}}$}}

\newcommand\res[2]{\small{$#1$}\tiny{\text{$\pm #2$}}}
\newcommand\bestres[2]{\small{$\mathbf{#1}$}\tiny{\text{$\mathbf{\pm #2}$}}}

\AtBeginDocument{%
  }

\setcopyright{acmcopyright}
\copyrightyear{2018}
\acmYear{2018}
\acmDOI{XXXXXXX.XXXXXXX}

\acmConference[Digital Threats: Research and Practice]{}{}{}
\acmPrice{15.00}
\acmISBN{978-1-4503-XXXX-X/18/06}

\begin{document}


\title[Multi-SpacePhish]{Multi-SpacePhish: Extending the Evasion-space of Adversarial Attacks against Phishing Website Detectors using Machine Learning}


\author{Ying Yuan}
\email{ying.yuan@math.unipd.it}
\orcid{0000-0001-9530-4725}
\affiliation{%
    \department{Department of Mathematics}
  \institution{† University of Padua}
  \country{Italy}
}

\author{Giovanni Apruzzese}
\orcid{0000-0002-6890-9611}
\email{giovanni.apruzzese@uni.li}
\affiliation{%
  \department{School of Business Informatics}
  \institution{University of Liechtenstein}
  \country{Liechtenstein}
}

\author{Mauro Conti}\authornotemark[2]
\email{conti@unipd.it}
\orcid{0000-0002-3612-1934}
\affiliation{%
  \department{Department of Computer Science}
  \institution{Delft University of Technology}
  \country{Netherlands}
}

\begin{abstract}

Existing literature on adversarial Machine Learning (ML) focuses either on showing attacks that break every ML model, or defenses that withstand most attacks. Unfortunately, little consideration is given to the actual \textit{cost} of the attack or the defense.
Moreover, adversarial samples are often crafted in the ``feature-space'', making the corresponding evaluations of questionable value. 
Simply put, the current situation does not allow to estimate the actual threat posed by adversarial attacks, leading to a lack of secure ML systems.

We aim to clarify such confusion in this paper. By considering the application of ML for Phishing Website Detection (PWD), we formalize the ``evasion-space'' in which an adversarial perturbation can be introduced to fool a ML-PWD---demonstrating that even perturbations in the ``feature-space'' are useful. Then, we propose a realistic threat model describing evasion attacks against ML-PWD that are cheap to stage, and hence intrinsically more attractive for real phishers. After that, we perform the first statistically validated assessment of state-of-the-art ML-PWD against 12 evasion attacks. Our evaluation shows (i)~the true efficacy of evasion attempts that are more likely to occur; and (ii)~the impact of perturbations crafted in different evasion-spaces; 
Our realistic evasion attempts induce a statistically significant degradation (3--10\% at $p\!<$0.05), and their cheap cost makes them a subtle threat. Notably, however, some ML-PWD are immune to our most realistic attacks ($p$=0.22). 

Finally, as an additional contribution of this journal publication, we are the first to propose and empirically evaluate the intriguing case wherein an attacker introduces perturbations in multiple evasion-spaces \textit{at the same time}. These new results show that simultaneously applying perturbations in the problem- and feature-space can cause a drop in the detection rate from 0.95 to 0.

Our contribution paves the way for a much needed re-assessment of adversarial attacks against ML systems for cybersecurity. 
\end{abstract}

\maketitle

\section{Introduction}
\label{sec:introduction}
After more than a decade of research~\cite{Biggio:Wild} and thousands of papers~\cite{carlini:papers}, it is well-known that Machine Learning (ML) methods are vulnerable to adversarial attacks. Specifically, by introducing imperceptible perturbations (down to a single pixel or byte~\cite{Su:One, Apruzzese:Evading}) in the input data, it is possible to compromise the predictions made by a ML model. Such vulnerability, however, is more dangerous in settings that implicitly assume the presence of adversaries. A cat will not try to fool a ML model. An attacker, in contrast, will actively try to evade a ML detector---the focus of this paper.

On the surface, the situation portrayed in research is vexing. The confirmed successes of ML~\cite{Jordan:Machine} are leading to large-scale deployment of ML in production settings (e.g.,~\cite{Darktrace:CyberML, tang2021survey, rathore2018malware}). At the same time, however, dozens of papers showcase adversarial attacks that can crack `any' ML-based detector (e.g.,~\cite{liang2016cracking, Apruzzese:Addressing}). Although some papers propose countermeasures (e.g.,~\cite{papernot2016distillation}), they are quickly defeated (e.g.,~\cite{carlini2016defensive}), and typically decrease the baseline performance (e.g.~\cite{demontis2017yes, Apruzzese:Addressing}). 
As a result, recent reports~\cite{fischer2021stakeholder, kumar2020adversarial} focusing on the integration of ML \textit{in practice} reveal that: ``I Never Thought About Securing My Machine Learning Systems''~\cite{boenisch2021never}. This is not surprising: if ML can be so easily broken, then why invest resources in increasing its security through --unreliable-- defenses?

Sovereign entities (e.g.,~\cite{Europe:AI, DHS2021AI}) are endorsing the development of ``trustworthy'' ML systems; yet, any enhancement should be economically justified. No system is foolproof (ML-based or not~\cite{carlini2021poisoning}), and guaranteeing protection against omnipotent attackers is an enticing but unattainable objective. In our case, a security system should increase the \textit{cost} incurred by an attacker to achieve their goal~\cite{moore2010economics}. Real attackers have a cost/benefit mindset~\cite{wilson2014some}: they may try to evade a detector, but only if doing so yields positive returns. In reality, worst-case scenarios are an exception---not the norm.

Our paper is inspired by several recent works that pointed out some `inconsistencies' in the adversarial attacks carried out by prior studies. Pierazzi et al.~\cite{Pierazzi:Intriguing} observe that real attackers operate in the ``problem-space'', i.e., the perturbations they can introduce are subject to physical constraints. If such constraints are not met, and hence the perturbation is introduced in the ``feature-space'' (e.g.,~\cite{nasr2021defeating}), then there is a risk of generating an adversarial example that is not physically realizable~\cite{tong2019improving}. Apruzzese et al.~\cite{apruzzese2021modeling}, however, highlight that even `impossible' perturbations can be applied, but \textit{only if} the attacker has internal access to the data-processing pipeline of the target system. Nonetheless, Biggio and Roli suggest that ML security should focus on ``anticipating the most likely threats''~\cite{Biggio:Wild}. Only \textit{after} proactively assessing the impact of such threats a suitable countermeasure can be developed---if required.

We aim to promote the development of secure ML systems. However, meeting Biggio and Roli's recommendation presents two tough challenges for research papers. First, it is necessary to devise a \textit{realistic threat model} which portrays adversarial attacks that are not only physically realizable, but also economically viable. Devising such a threat model, however, requires a detailed security analysis of the \textit{specific cyberthreat} addressed by the detector---while factoring the resources that attackers are willing to invest. Second, it is necessary to \textit{evaluate the impact} of the attack by crafting the corresponding perturbations. Doing so is difficult if the threat model assumes an attacker operating in the problem-space, because such \textit{perturbations must be applied on raw-data}, i.e., before any preprocessing occurs---which is hard to find.

In this paper, we tackle both of these challenges. In particular, we focus on ML-systems for Phishing Website Detection (PWD). 
Countering phishing -- still a major threat today~\cite{proofpoint2022phish, Kettani:Threats} -- is an endless struggle. Blocklists can be easily evaded~\cite{tian2018needle}, and to cope against adaptive attackers some detectors are equipped with ML (e.g.~\cite{tang2021survey}). Yet, as shown by Liang et al.~\cite{liang2016cracking}, even such ML-PWD can be ``cracked'' by oblivious attackers---if they invest enough effort to reverse engineer the entire ML-PWD.
Indeed, we address ML-PWD because prior work (e.g.,~\cite{bahnsen2018deepphish, shirazi2019adversarial, gressel2021feature, lee2020building}) assumed threat models that hardly resemble a real scenario. Phishing, by nature, is meant to be cheap~\cite{kim2021security} and most attempts end up in failure~\cite{oest2020sunrise}. It is unlikely\footnote{It is unlikely, but \textit{not impossible}. Hence, as recommended by Arp et al~\cite{arp2022dos}, it is positive that such cases have also been studied by prior work.} that a phisher invests many resources \textit{just to evade} ML-PWD: even if a website is not detected, the user may be `hooked', but is not `phished' yet. As a result, the state-of-the-art on adversarial ML for PWD is immature---from a pragmatic perspective.

\textbf{Contribution and Organization.}
Let us explain how we aim to spearhead the security enhancements to ML-PWD. 
We begin by introducing the fundamentals concepts (PWD, ML, and adversarial ML) at the base of our paper in §\ref{sec:background}, which also serves as a motivation. Then, we make the following four contributions.
\begin{itemize}
    \item We \textit{formalize the evasion-space} of adversarial attacks against ML-PWD (§\ref{sec:evasion}), rooted in exhaustive analyses of a generic ML-PWD. Such evasion-space explains `where' a perturbation can be introduced to fool a ML-PWD. Our formalization highlights that even adversarial samples created by direct feature manipulation can be realistic, \textit{validating all the attacks performed by past work}.

    \item By using our formalization as a stepping stone, we propose a \textit{realistic threat model} for evasion attacks against ML-PWD~(§\ref{sec:threat}). Our threat model is grounded on detailed security considerations from the viewpoint of a typical phisher, who is confined in the `website-space'. Nevertheless, our model can be relaxed by assuming attackers with greater capabilities (which require a higher cost). 
    
    \item We combine and practically demonstrate the two previous contributions. We perform an extensive, reproducible, and statistically validated \textit{evaluation of adversarial attacks} against state-of-the-art ML-PWD. By using diverse datasets, ML algorithms and features, we develop 18 ML-PWD (§\ref{sec:evaluation}), each of which is assessed against 12 different evasion attacks built upon our threat model (§\ref{sec:attacks}). 
    
    \item By analyzing the results (§\ref{sec:results}) of our evaluation: (i)~we \textit{show the impact of attacks that are very likely to occur} against both baseline and adversarially robust ML-PWD; and (ii)~we are the first to \textit{fairly compare the effectiveness} of evasion attacks in the problem-space with those in the feature-space.
    
    \item As an an additional contribution of this journal paper, we propose and empirically assess 37 new perturbations that envision an attacker who can \textit{operate in multiple spaces} (§\ref{sec:new}).
    
\end{itemize}
\textbf{Our results highlight that more realistic attacks are not as disruptive as claimed by past works (§\ref{sec:related})} but their low-cost makes them a threat that induces statistically significant degradations. Intriguingly, however, some ``cheap'' perturbations can lead to devastating impact.
Finally, our evaluation serves as a `benchmark' for future studies: we provide the complete results in the Appendix, whereas the source-code is publicly available in a dedicated website: \url{https://spacephish.github.io}.

\section{Background and Motivation}
\label{sec:background}
Our paper lies at the intersection of Phishing Website Detection (PWD) and Machine Learning (ML) security. To set-up the stage for our contribution and motivate its necessity, we first summarize PWD (§\ref{ssec:pwd}), and then explain the role of ML in PWD (§\ref{ssec:mlpwd}). Finally, we provide an overview of the adversarial ML domain (§\ref{ssec:aml}).

\subsection{Phishing Website Detection} 
\label{ssec:pwd}
\begin{figure}[!htbp]
    \centering
    \includegraphics[width=0.8\columnwidth]{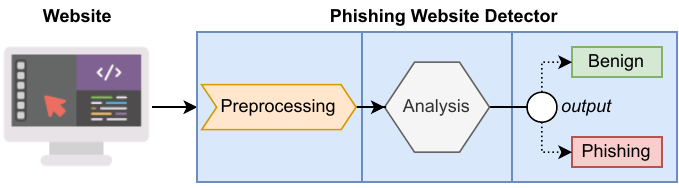}
    \caption{Exemplary PWD. After preliminary preprocessing, a website is analyzed by a detector to determine its legitimacy.}
    \label{fig:pwd}
\end{figure}

Although having been studied for nearly two decades~\cite{kirda2005protecting}, phishing attacks are still a rampant menace~\cite{Kettani:Threats}: according to the FBI~\cite{fbi2020icr}, the number of reported phishing attempts has increased by 900\% from 2018 to 2020 (26k up to 240k). Aside from the well-known risks to single users (e.g., fraud, credential theft~\cite{guo2022safer}), phishing is still one of the most common vectors to penetrate an organization's perimeter. Intuitively, the best countermeasure to phishing is its prevention through proper \textit{education}~\cite{xiong2019embedding}. Despite recent positive trends, however, such education is far from comprehensive: the latest ``State of the Phish'' report~\cite{proofpoint2022phish} states that more than 33\% of companies do not have any training program for their employees, and more than 50\% only evaluate such education through simulations. As a result, there is still a need of IT solutions that mitigate the phishing threat by its early \textit{detection}. In our case, this entails identifying a phishing website before a user lands on its webpage, therefore defusing the risk of falling victim to a phishing attack. We provide in Fig.~\ref{fig:pwd} an exemplary architecture of a Phishing Website Detector (PWD).

Despite extensive efforts, PWD remains an open issue. This is due to the intrinsic limitations of the most common detection approaches reliant on \textit{blocklisting} (e.g.,~\cite{prakash2010phishnet, oest2020phishtime}). Such techniques have been improved and nowadays they even involve automatic updates with recent feeds (e.g., PhishTank~\cite{phishtank}). However, blocklists are a double-edged sword: on the good side, they are very precise and are hence favored due to the low rate of false alarms; on the bad side, they are only effective against known phishing websites~\cite{acharya2021phishprint}. The latter is a problem: expert attackers are aware of blocklists and hence move their phishing `hooks' from site to site, bypassing most PWD. As shown by Tian et al.~\cite{tian2018needle}, such strategies can elude over 90\% of popular blocklists for more than one month. To counter such \textit{adaptive} attackers, much attention has been given to data-driven detection schemes---including those within the Machine Learning (ML) paradigm~\cite{tang2021survey}. Indeed, ML allows to greatly enhance the detection capabilities of PWD. Let us explain why.

\subsection{Machine Learning for PWD}
\label{ssec:mlpwd}
\begin{figure}[!htbp]
    \centering
    \includegraphics[width=0.5\columnwidth]{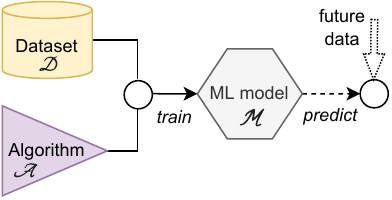}
    \caption{Machine Learning workflow. By training~\ftcal{A} on \ftcal{D}, a ML model \ftcal{M} is developed. Such \ftcal{M} can be used to predict future data.}
    \label{fig:ml}
\end{figure}

The cornerstone of ML is having ``machines that automatically learn from experience''~\cite{Jordan:Machine}, and such experience comes in the form of \textit{data}. By applying a given ML \textit{algorithm} \smacal{A}, e.g. Random Forest (RF), to analyze a given \textit{dataset} \smacal{D}, it is possible to \textit{train} a ML \textit{model} \smacal{M} that is able to `predict' previously unseen data. We provide a schematic of such workflow in Fig.~\ref{fig:ml}.
In the case of PWD, a ML model \smacal{M} can be deployed in a detector (e.g., in the hexagon in Fig.~\ref{fig:pwd}) to \textit{infer} whether a given webpage is benign or phishing. 

The main advantage of ML models is their intrinsic ability of noticing weak patterns in the data that are overlooked by a human, and then leveraging such patterns to devise `flexible' detectors that can counter even adaptive attackers. As a matter of fact, Tian et al.~\cite{tian2018needle} show that a ML model based on RF is effective even against ``squatting'' phishing websites---while retaining a low-rate of false alarms (only 3\%). Moreover, acquiring suitable data (i.e., recent and labelled) for ML-PWD is not difficult---compared to other cyber-detection problems for which ML has been proposed~\cite{apruzzese2022sok}.

Such advantages have been successfully leveraged by many research efforts (e.g.,~\cite{niakanlahiji2018phishmon, tan2016phishwho}). Existing ML-empowered PWD can leverage different types of \textit{information} (i.e., \textit{features}) to perform their detection. Such information can pertain either to a website's \textit{URL}~\cite{verma2015character} or to its \textit{representation}, e.g., by analyzing the actual image of a webpage as rendered by the browser~\cite{haruta2019novel}, or by inspecting the HTML~\cite{jain2019machine}. For example, Mohammad et al.~\cite{mohammad2014intelligent} observed that phishing websites usually have long URLs; and often contain many `external' links (pointing to, e.g., the legitimate `branded' website, or the server for storing the phished data), which can be inferred from the underlying HTML. Although some works use only URL-related features (e.g.,~\cite{butnaru2021towards}) -- which can also be integrated in phishing \textit{email} filters (e.g.,~\cite{gupta2021novel}) -- more recent proposals use combinations of features (e.g.,~\cite{Corona:Deltaphish, van2021combining}); potentially, such features can be derived by querying third-party services (e.g., DNS servers~\cite{jain2018towards}).

The cost-effectiveness of ML-PWD increased their adoption: even commercial browsers (e.g., Google Chrome~\cite{liang2016cracking}) integrate ML models in their phishing filters (which can be further enhanced via customized add-ons~\cite{tang2021survey}); moreover, ML-PWD can also be deployed in corporate SIEM~\cite{howell2015building}. However, it is well-known that no security solution is foolproof: in our case, ML models can be thwarted by exploiting the so-called adversarial attacks~\cite{Apruzzese:Addressing}.

\subsection{Adversarial Attacks against ML}
\label{ssec:aml}
The increasing diffusion of ML led to question its security in adversarial environments, giving birth to ``adversarial machine learning'' research~\cite{Biggio:Wild, chen2017adversarial}. Attacks against ML exploit \textit{adversarial samples}, which leverage perturbations to the input data of a ML model that induce predictions favorable to the attacker. Even imperceptible perturbations can mislead proficient ML models: for instance, Su et al.~\cite{Su:One} modify a single pixel of an image to fool an object detector; whereas Apruzzese et al.~\cite{Apruzzese:Evading} evade botnet detectors by extending the network communications with few junk bytes. 

An adversarial attack is described with a \textit{threat model}, which explains the relationship of a given \textit{attacker} with the \textit{defender's system}. The attacker has a \textit{goal} and, by leveraging their \textit{knowledge} and \textit{capabilities}, they will adopt a specific \textit{strategy}~\cite{Biggio:Wild}. Common terms associated with the attacker's knowledge are \textit{white-box} and \textit{black-box}: the former denotes attackers who know everything about the defender; whereas the latter denotes attackers who know nothing~\cite{Papernot:Practical, zheng2021black}. The capabilities describe how the attacker can interact with the target system, e.g., they: can influence only the \textit{inference} or also the \textit{training} stage of the ML model; can use the ML model as an ``oracle'' by inspecting the output to a given input; and can be subject to constraints on the creation of the adversarial perturbation (e.g., a limited amount of queries).

Despite thousands of papers focusing on this topic, a universal and pragmatic solution has not been found yet. Promising defenses are invalidated within the timespan of a few months (e.g. distillation was proposed in~\cite{papernot2016distillation} and broken in~\cite{carlini2016defensive}). Even ``certified'' defenses~\cite{jia2020almost} can only work by assuming that the perturbation is bounded within some magnitude---which is not a constraint to which real attackers must abide (as pointed out by Carlini et al.~\cite{carlini2019evaluating}). 
From a pragmatic perspective, \textit{any defense has a cost}: first, because it must be developed; second, because it can induce additional overhead. The latter is particularly relevant in cybersecurity, because it may decrease the performance of the ML model when no adversarial attack occurs. For instance, a well-known defense is \textit{feature removal}~\cite{smutz2012malicious}, which entails developing ML models that do not analyze the features expected to be targeted by a perturbation. Doing this, however, leads to less information provided to the ML model, hence inducing performance degradation (e.g.,~\cite{Apruzzese:Addressing}). Even when countermeasures have a small impact (e.g.,~\cite{demontis2017yes}), this is not negligible in cyber-detection: attacks are a ``needle in a haystack''~\cite{tian2018needle}, and even a 1\% increase in false positives is detrimental~\cite{veeramachaneni2016ai}.
Therefore, ML engineers will not devise any protection mechanism unless the corresponding threat is shown to be dangerous in reality~\cite{kumar2020adversarial}. 

\textbf{The Problem.} Unfortunately, research papers intrinsically impair the development of secure ML systems, because the aim is often to ``outperform the state-of-the-art''. In adversarial ML, this leads to papers that either showcase devastating attacks stemming from extremely powerful adversaries (i.e., white-box~\cite{Su:One}); or viceversa, i.e., show that even oblivious attackers can thwart ML systems~\cite{Papernot:Practical}. However, real `adaptive' attackers (i.e., those that ML methods should be protected against) do not conform to these two extremes. Indeed, having complete knowledge of the target system requires a huge resource investment (especially if such system is devoted to cybersecurity), which may be better spent elsewhere; conversely, it is unlikely that opponents will launch attacks while knowing nothing of the defender. 
Hence, to provide \textit{valuable} research, efforts on adversarial ML should start focusing on the gray area within these two extremes---which implicitly are more likely to occur~\cite{apruzzese2021modeling}. In the context of ML-PWD, our paper is a first step in this direction: as we will show, evasion attempts evaluated in literature (§\ref{sec:related}), despite being devastating, are costly to launch---even in black-box settings.

\section{The Evasion-space of Adversarial Attacks against ML-PWD}
\label{sec:evasion}
We aim to spearhead valuable research in adversarial attacks against ML-PWD. To this purpose, we first elucidate the internal functionalities of a ML-PWD (§\ref{ssec:analysis}). Then, we propose our original formalization of the \textit{evasion-space} of adversarial perturbations (§\ref{ssec:evasion}). Finally, we explain why our contribution validates \textit{all} prior work (§\ref{ssec:validation}).

\subsection{Analysis of a ML-PWD}
\label{ssec:analysis}
\begin{figure}[!htbp]
    \centering
    \vspace{-1em}
    \includegraphics[width=\columnwidth]{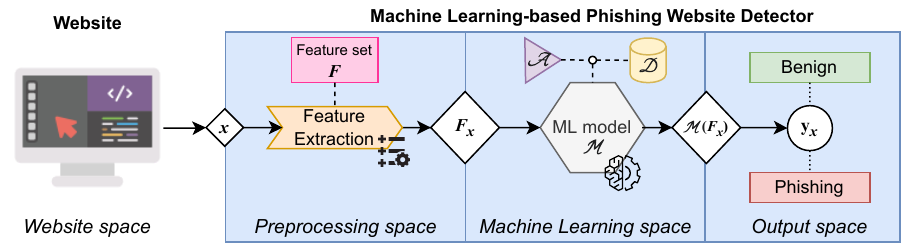}
    \caption{Architecture of a ML-PWD. A website, $x$, is preprocessed into $F_x$. A ML model \ftcal{M} analyzes such feature representation and predicts its ground truth as $\mathcal{M}(F_{x})=y_x$.}
    \label{fig:scenario}
\end{figure}

We connect the previously introduced concepts (§\ref{ssec:pwd} and §\ref{ssec:mlpwd}) and provide an overview of a generic ML-PWD in Fig.~\ref{fig:scenario}.

A sample (i.e., a website), $x$, `enters' the ML-PWD and is subject to some preprocessing aimed at transforming any input into a format accepted by the ML model---according to a given feature set, $F$. (We assume that $x$ is not blocklisted.) The result of such preprocessing is the feature representation of the website $x$, i.e. $F_x$, which can now be analyzed by the ML model \smacal{M}. 
We consider a ML model focused on \textit{binary classification}. Hence, training \smacal{M} requires: a dataset, \smacal{D}, whose samples are labelled as \textit{benign} or \textit{phishing}; and any ML algorithm, \smacal{A}, supporting classification tasks (e.g., RF).  

The ML model \smacal{M} predicts the ground truth of $F_x$ as $y_x$, i.e., $\mathcal{M}(F_x)=y_x$. 
Hence, we can summarize the workflow of our ML-PWD through the following Expression:
\begin{align}
    x \rightarrow F_x \rightarrow \mathcal{M}(F_x) = y_x.
    \label{eq:normal}
\end{align}
If $x$ is a phishing (benign) webpage and $y_x$ is also phishing (benign), then we have a true positive (true negative); otherwise, we have an incorrect classification (either a false positive or a false negative). We assume that \smacal{M} has been properly trained, so that its deployment performance yields a high true positive rate (\smamath{tpr}) while maintaining a low false positive rate (\smamath{fpr})---under the assumption that no adversarial attack occurs.

\subsection{Evasion Attacks against ML-PWD}
\label{ssec:evasion}
Adversarial attacks exploit a perturbation, $\varepsilon$, that induces a ML model \smacal{M} to provide an output favoring the attacker (§\ref{ssec:aml}). 
In our case, \smacal{M} is a (binary) classifier that analyzes $F_x$, hence we can express an adversarial attack as follows:
\begin{align}
    \text{find}~\varepsilon~\text{s.t.}~\mathcal{M}(F_x)=y^{\varepsilon}_{x} \neq y_x.
    \label{eq:aa}
\end{align}
In other words, the objective is finding a perturbation $\varepsilon$ that induces a ML model \smacal{M} (that is assumed to work well) to misclassify a given sample $x$ (i.e., $y^{\varepsilon}_x \neq y_x$).
Because our focus is on \textit{evasion} attacks, such misclassification entails having a positive (i.e., phishing) classified as a negative (i.e., benign). It is implicitly assumed that such $\varepsilon$ must: (i)~preserve the \textit{ground truth}\footnote{E.g., changing a URL from ``go0gle.com'' to ``google.com'' is not a valid $\varepsilon$.} (i.e., $y^{\varepsilon}_x$ should be the same as $y_x$); and (ii)~preserve the \textit{phishing logic} of a webpage~\cite{panum2020towards}.
Such $\varepsilon$, however, can lead to different effects on $y^{\varepsilon}_{x}$ depending on `where' it is applied during the workflow described by Exp.~\ref{eq:normal}. We describe such occurrence by formalizing the \textit{evasion-space} of an attacker.

\textsc{\textbf{Evasion-Space.}} Let us observe Fig.~\ref{fig:scenario}. We can see that the figure is divided into four `spaces', each allowing the introduction of a perturbation $\varepsilon$ that can affect the output of the ML-PWD. Of course, a perturbation in the last space, i.e., the \textit{output-space}, cannot be considered as an `adversarial ML attack', because it will have no relationship with the ML model \smacal{M}. Hence, the evasion-space of an attacker that wants to induce a misclassification by \smacal{M} is confined to the first three spaces. 
Let us analyze each of these. 

\begin{enumerate}
\item \textit{Website-space Perturbations (WsP).}
The entire detection workflow begins in the `website-space', in which the website (i.e., $x$) is generated. Such space is accessible by any attacker, because they are in control of the generation process of their (phishing) website. As an example, the attacker can freely modify the URL or the representation of a website (subject to physical constraints\footnote{Which depend on the semantics of websites, e.g., URLs cannot be 1 character long.}).
Introducing a perturbation $\varepsilon$ in this space (i.e., a WsP) yields an adversarial sample $\overbar{x}\!=\!x\!+\!\varepsilon$, and the effects of such $\varepsilon$ \textit{can} affect all the operations performed by the ML-PWD (cf. Exp~\ref{eq:normal}). We emphasize the word ``can'': this is because what happens \textit{after} $\overbar{x}$ enters the ML-PWD strictly depends on the implementation of such ML-PWD---which may, or may not, `notice' the corresponding $\varepsilon$ (e.g., \smacal{M} can analyze an $F$ that is not influenced by $\varepsilon$).

\item \textit{Preprocessing-space Perturbations (PsP).}
After $x$ is acquired by the ML-PWD, it is first transformed into $F_x$. \textit{An attacker with write access} to the `preprocessing-space' can introduce a PsP $\varepsilon$ that affects the \textit{process} that yields the feature representation of a website, leading to $\overbar{F_x}\!=\!F_x\!+\!\varepsilon$. For instance, a website $x$ with an URL of 40 characters can be turned into a $\overbar{F_x}$ that has the \textit{URL\_length} feature=$20$. Intuitively, attackers able to introduce PsP are powerful, but are still subject to constraints: before any $F_x$ is sent to the ML model \smacal{M}, such $F_x$ is checked to ensure that it is not corrupted~\cite{apruzzese2021modeling}. Indeed, $\overbar{F_x}$ must not violate any inter-feature dependencies or physical constraints. With respect to WsP, PsP are guaranteed to be `noticed' by the ML-PWD; however, they do not necessarily influence the predictions of \smacal{M}: making a URL shorter may not be enough to fool the detection process.

\item \textit{ML-space Perturbations (MsP).}
After the preprocessing, the feature representation of a website $F_x$ enters the Machine Learning-space in order to be analyzed by \smacal{M}. \textit{If an attacker has write access} to this space, they can introduce an MsP, i.e., a perturbation $\varepsilon$ that affects $F_x$ \textit{immediately before} it reaches \smacal{M}. An MsP is the `strongest' type of perturbation because it affects the $F_x$ after all integrity checks\footnote{Indeed, a ML model \ftcal{M} is agnostic to the generation process of a given input.} have been performed---potentially leading to corrupted values, or which have no relationship to any real $x$. We hence denote MsP as $\overbar{F}_x = F_x + \varepsilon$. As an example, a MsP can yield a $\overbar{F}_x$ having an \textit{URL\_length}=$0$. As such, MsP are very likely to induce uncanny responses by \smacal{M} (but do not guarantee evasion).

\end{enumerate}

\textbf{Summary and Cost.} 
From Exp.~\ref{eq:aa}, we observe that any perturbation $\varepsilon$ should ultimately affect the feature representation $F_x$ of a given sample $x$. Hence, the crux is determining `where' such perturbation is introduced---which can happen in three spaces. We formally define adversarial attacks by means of introducing a perturbation in each of these spaces (i.e., WsP, PsP and MsP) through the following Expression (which extends Exp.~\ref{eq:normal}):
\begin{equation}
\resizebox{0.65\columnwidth}{!}{
  {{find $\varepsilon$ s.t. }}$
    \begin{cases}
      \overbar{x} = x+\varepsilon~\Rightarrow~
     x \rightarrow \overbar{x} \rightarrow F_{\overbar{x}} \rightarrow \mathcal{M}(F_{\overbar{x}}) = y^{\varepsilon}_x \neq y_x, & \text{\small{WsP}};\\
     \overbar{F_x} = F_{x}+\varepsilon~\Rightarrow~
     x \rightarrow \overbar{F_x} \rightarrow \mathcal{M}(\overbar{F_x}) = y^{\varepsilon}_x \neq y_x, & \text{\small{PsP}};\\
     \overbar{F}_x = F_{x}+\varepsilon~\Rightarrow~
     x \rightarrow F_{x} \rightarrow \overbar{F}_x \rightarrow \mathcal{M}(\overbar{F}_x) = y^{\varepsilon}_x \neq y_x, & \text{\small{MsP}}.\\
    \end{cases}$
    }
    \label{eq:models}
\end{equation}
We remark that the effects of WsP can match those of PsP---which can also match those of MsP. For instance, a MsP can yield a sample with an \textit{URL\_length} of 20 which -- as long as it does not violate any inter-feature dependency -- can represent a valid website (hence MsP=PsP)\footnote{Of course MsP=PsP if there is no `integrity check'.}; to obtain an equivalent WsP, the attacker would have to modify the actual URL and make it of exactly 20 characters (which is doable). Hence, in some cases, $F_{\overbar{x}}$=$\overbar{F_x}$=$\overbar{F}_x$. As such, although some MsP cannot be crafted in the website-space, it is also unfair to consider all MsP (or PsP) as being not physically realizable. 
Finally, from a \textit{cost} viewpoint, WsP$\ll$PsP$<$MsP, because realizing MsP requires the attacker to have more control\footnote{Our formalization is orthogonal to the one by Šrndic and Laskov.~\cite{Laskov:Practical}: while~\cite{Laskov:Practical} focus on the attacker's \textit{knowledge} (``what does the attacker know about the ML system?''), we focus on the \textit{capabilities} (i.e., ``where can the attacker introduce a perturbation affecting the ML system?''). Moreover, our PsP are semantically different than the ``adversarial preprocessing'' by Quiring et al.~\cite{quiring2020adversarial}: while ~\cite{quiring2020adversarial} affect the preprocessing phase \textit{from outside} the ML system, our PsP affect such phase \textit{from the inside}.} on the ML-PWD (i.e., they must obtain write-access to deeper segments of the ML-PWD).

\subsection{Validation of Previous Work}
\label{ssec:validation}

An important contribution of our evasion-space is that it \textit{validates all past research} that consider perturbations in the ``feature-space'' (i.e., PsP or MsP). Let us explain why.

\textbf{Context.}
By using Pierazzi et al.~\cite{Pierazzi:Intriguing} notation, our WsP can be seen as perturbations in the ``problem-space''; whereas PsP and MsP are perturbations in the ``feature-space''. The main thesis of Pierazzi et al.~\cite{Pierazzi:Intriguing} is that evaluations carried out in the feature space are unreliable due to the ``inverse mapping problem'': some changes in the feature representation of a sample (i.e., $F_x$) may not be physically realizable when manipulating the original sample (i.e., $x$)---therefore exposing the ``weakness of previous evasion approaches.''

\textbf{Intuition.} Our original formalization elucidates that the ``weaknesses'' of past work are not, in fact, weaknesses---therefore overturning some of the claims of Pierazzi et al.~\cite{Pierazzi:Intriguing}. Our thesis is rooted in the following observation: the ``inverse mapping problem'' is irrelevant \textit{if the attacker has write access to the ML-PWD}. 

\textbf{Explanation.}
Any attacker is able to craft WsP by manipulating their own phishing webpages (to some degree). In contrast, \textit{reliably} realizing PsP and MsP can only be done by assuming an attacker that can manipulate the corresponding space (i.e., either the preprocessing- or the ML-space). Achieving this in practice presents a high barrier of entry---but \textit{it is not impossible}. For instance, consider the case of an attacker who has compromised a given device integrating a client-side ML-PWD: such attacker can interfere with any of the ML-PWD operations---especially if it is open-source (e.g.,~\cite{hr2020development}). Of course, realizing PsP or MsP if the ML-PWD is deployed in an organization-wide intrusion detection system is harder, but not unfeasible (as pointed out by~\cite{apruzzese2021modeling}).

\vspace{-0.5em}
\takeaway{Our formalization validates all evasion attacks previously evaluated through perturbations in any internal `space' of the ML-PWD. This requires to \textit{revise the attacker's capabilities}, implicitly increasing the cost of the attack.}

\textbf{Consequences.} Simply put, we \textit{restore} the value (partially `lost' after the publication of~\cite{Pierazzi:Intriguing}) of the evaluations performed by prior work (§\ref{sec:related}). By assuming that the considered attacker can access a given space of the ML-PWD (either for PsP or MsP), then there is no risk of falling into the ``inverse mapping problem''---because it is a constraint that such attacker is not subject to. Such different assumptions, however, implicitly raise the cost of the corresponding attack. 
For example. Corona et al.~\cite{Corona:Deltaphish} craft perturbations in the ML-space: according to~\cite{Pierazzi:Intriguing}, the resulting perturbations are, hence, unreliable. However, by assuming that the attacker \textit{can manipulate the ML-space}, then such adversarial examples (deemed unreliable by~\cite{Pierazzi:Intriguing}) would become realistic (thanks to our contribution).

\section{Proposed Realistic Threat Model}
\label{sec:threat}
We use our evasion-space formalization to devise our proposed adversarial ML threat model---describing attractive strategies for real phishers. We first provide its definition (§\ref{ssec:definition}), and then support its realisticness via security analyses (§\ref{ssec:security}). Next, we provide some considerations (§\ref{ssec:considerations}) that set-up the stage for the additional contribution of this paper (§\ref{subsec:extensions}). Finally, we show how to apply WsP on \textit{real} phishing webpages (§\ref{ssec:pragmatic}).

\subsection{Formal Definition}
\label{ssec:definition}
We define our threat model according to the following four criteria (well-known in adversarial ML~\cite{Biggio:Wild}).

\begin{itemize}

\item  \textit{Goal.} The adversary wants to \textit{evade} an ML-PWD that uses \smacal{M} (i.e., the attacker wants to satisfy Exp.~\ref{eq:aa}).

\item  \textit{Knowledge.} The adversary has \textit{limited knowledge} of the target system, the ML-PWD. They know nothing about: the ML model \smacal{M}, its training data \smacal{D}, and its underlying ML algorithm \smacal{A} (except that it supports binary classification). However, the adversary knows a subset of the feature set $F$ analyzed by \smacal{M}. Let $K \subseteq F$ be such a subset. The adversary is also aware that the ML-PWD will likely detect phishing websites if no evasion attempt is made (otherwise, there would be no reason to do so). Finally, the adversary implicitly knows that no blocklist includes their phishing webpages (otherwise, the attacker would be \textit{forced} to manipulate the URL).

\item \textit{Capability.} The adversary has \textit{no access} to the ML-PWD. They cannot use the ML-PWD as an ``oracle'' (i.e., inspect the output to a given input); and they are hence limited to perturbations in the website-space (i.e., WsP).

\item \textit{Strategy.} The adversary uses their knowledge of $K$ to craft WsP that may lead to evasion (at inference stage). 

\end{itemize}
We observe that our threat model is \textit{general} because no specific set of features ($F$) or ML model \smacal{M} (and hence \smacal{D} and \smacal{A}) is provided. Therefore, our threat model can cover any ML-PWD that resembles the one in Fig.~\ref{fig:scenario}. Potentially, it can even be a ML-PWD used by email filters if the corresponding \smacal{M} analyzes URL-related information (e.g.,~\cite{duman2016emailprofiler, gupta2021novel}).

\subsection{Security Analysis}
\label{ssec:security}
Let us analyze our threat model and explain why it portrays a \textit{realistic} attacker---especially if compared to typical `white-/black-box' adversarial scenarios (cf. §\ref{ssec:aml}). We intend to justify that our threat model describes attacks that are \textit{interesting} to investigate, and hence \textit{valuable} for the security of ML-PWD.

\textbf{Phishing in a nutshell.}
We start by focusing the attention on the intrinsic nature of phishing. Indeed, phishing attempts -- and especially those involving phishing websites -- are `cheap' in nature~\cite{kim2021security}. Considering that real attackers operate with a cost-benefit mindset, it is unlikely that such attackers will invest extensive resources just to have their webpages evade a ML-PWD. Firstly, because such evasion will be temporary (as soon as the webpage is reported in a blocklist, any adversarial attack will be useless); secondly, because, even if a website evades a ML-PWD, the phishing attempt is not guaranteed to succeed (a user still has to input its sensitive data). Indeed, despite the exponential proliferation of phishing~\cite{proofpoint2022phish}, most phishing attempts are prone to failure~\cite{oest2020sunrise}---and the attackers are well aware of this fact. Of course, attackers can opt for more expensive spear-phishing campaigns~\cite{caputo2013going} (which still have a success rate of barely 10\%~\cite{ho2019detecting}), but in this case they will likely design entirely new phishing webpages---and not rely on cheap perturbations on pre-existing samples.

\textbf{Limited Knowledge.} Our attacker knows \textit{something} (i.e., $K$) about the ML-PWD, but they are not omniscient---hence, our threat model can be considered as a gray-box scenario. Such `box', however, is the entire ML-PWD, i.e., the blue rectangle in Fig.~\ref{fig:scenario}. Our scenario is \textit{more interesting} to investigate than white-box scenarios. The reason is simple: ours is \textit{more likely} to occur, because `phishers' with complete knowledge of the entire ML-PWD are extremely unlikely. Furthermore, extensive adversarial ML literature~\cite{Biggio:Wild} has ably demonstrated that white-box attacks can break most systems---including ML-PWD (e.g.,~\cite{song2021advanced, gressel2021feature, abdelnabi2020visualphishnet, lin2021phishpedia}). 

\textbf{Realistic Capabilities.} Our `standard' attacker has no access to the ML-PWD, which is a realistic assumption. For instance, the attacker can share a phishing website via social media, but without knowing which device (and, hence, ML-PWD) is being used by potential victims to open such website. Therefore, the attacker cannot reliably use \smacal{M} as an oracle. They could opt for querying a surrogate ML-PWD to reverse-engineer its functionalities and then leverage the transferability of adversarial attacks~\cite{Demontis:Adversarial}. However, such `black-box' scenario is both (i)~unlikely to occur; and (ii)~ultimately not interesting to consider for a research paper. \textit{Unlikely}, because it would \textit{defeat the purpose} of phishing attacks: reverse-engineering operations require a huge resource investment---which can be invalidated via a simple re-training of \smacal{M} (a common cybersecurity practice~\cite{apruzzese2022role}). \textit{Not interesting}, because such attacks \textit{have been investigated before}~\cite{sabir2020evasion, al2021generating}. For instance, Liang et al.~\cite{liang2016cracking} demonstrated that attackers with access to client-side detectors can crack and evade the corresponding ML-PWD; doing this, however, required \textit{more than 24 hours} of queries~\cite{liang2016cracking}.

\takeaway{Phishing attempts have an intrinsic low rate of success. Attackers that aim to evade a ML-PWD will favor `cheap' tactics---which can be represented by our proposed threat model.}

\subsection{Technical Considerations} 
\label{ssec:considerations}

Let us enhance our threat model with four considerations.

\begin{enumerate}
    \item The attacker can easily acquire a rough idea of the feature set $F$ analyzed by the ML-PWD. For instance, the descriptions of many state-of-the-art solutions are openly accessible. However, it is unlikely that the attacker knows the \textit{exact} feature set $F$: the actual implementation of a ML-PWD (including the feature extractor) can -- or, rather, should! -- differ from the publicly available information. This is why we consider an attacker that only knows $K \subseteq F$.

    \item We note that it is also possible that $K=\varnothing$. In this case, the attacker expects the ML-PWD to analyze some features that are \textit{not} actually analyzed by \smacal{M} (for instance, the attacker can modify the URL, but nothing about the URL is analyzed by \smacal{M}). This can happen, e.g., against an `adversarially robust' ML-PWD that leverages the well-known \textit{feature removal} strategy (cf. §\ref{ssec:aml}). As a result, WsP targeting such $K$ will likely result in a negligible impact. Furthermore, it is also possible that some features in $K$ simply \textit{cannot} be influenced by an attacker operating in the website-space (e.g., features that depend on third-party sources, such as DNS logs). 

    \item Since our attacker cannot access the ML-PWD, they cannot observe the output-space and, thus, cannot optimize their perturbations to find the best WsP that guarantees evasion; and cannot even verify whether their WsP evade the ML-PWD or not. The attacker is, however, not subject to strict boundaries on WsP (§\ref{ssec:evasion}). 

    \item Our threat model considers attacks at \textit{inference}-time (i.e., after \smacal{M} has been deployed in the PWD). This is because the dataset used to devise ML-based security systems is typically well-protected~\cite{apruzzese2021modeling}. Compromising such dataset would significantly raise the cost of the offensive campaign (as also highlighted in~\cite{liao2018server}). Therefore, phishers are unlikely to launch attacks at training-time.
\end{enumerate}

The last two are significant: lack of access (and, hence, knowledge) on the training set prevents from achieving the no-box attacks of~\cite{li2020practical}; furthermore, the impossibility of witnessing the output of \smacal{M} prevents enacting typical black-box strategies (e.g.,~\cite{mu2021hard}). Finally, as pointed out also by~\cite{zuo2019exploiting, carlini2019evaluating, apruzzese2023real}, achieving `minimal' perturbations may be an unrealistic objective.

\subsection{Extensions}
\label{subsec:extensions}
Our threat model can be \textit{extended} by relaxing some of its assumptions. Indeed, in its current formulation, our threat model envisions an attacker that is ``weak'' (and, hence, very likely to appear in reality). However, some adversaries may be willing to invest more resources to ensure that their attacks come to fruition (i.e., increasing the chances that their phishing webpages are misclassified by the ML-PWD, and hence displayed to the end-user). Abundant prior work in the adversarial ML domain considers attacks having different levels of \textit{knowledge} (i.e., the so-called ``black-box'' and ``white-box''~\cite{apruzzese2023real}). However, given that our original formalization focuses on the attacker's \textit{capabilities} (§\ref{sec:evasion}), we identify two types of extensions that portray a stronger attacker. Namely:
\begin{itemize}
    \item \textit{Deeper spaces.} An attacker who manages to obtain write-access to the ML-PWD (or part of its elements) can tamper with its internal functionalities, thereby realizing either PsP or MsP.
    \item \textit{Mixed spaces.} If the attacker can obtain some control on either the Preprocessing- or Machine Learning-space, then -- alongside being able to apply PsP or MsP -- they are also able to apply WsP. Indeed, the attacker will \textit{always} be able to manipulate the phishing webpage, since it is (by definition) under their complete control. Hence, an attacker who can inject PsP can also inject a WsP; furthermore, an attacker who can inject a MsP can also inject a PsP (since they can overlap), and can, of course, also inject a WsP.
\end{itemize}
We will empirically evaluate all the abovementioned cases in our evaluation (§\ref{sec:results}) in which we compare the effects of attacks using WsP against those entailing PsP and MsP (by assuming the same knowledge, i.e., limited to $K$). Moreover, we will also assess attacks entailing perturbations in different spaces (§\ref{sec:new}).

\subsection{Pragmatic Use-case}
\label{ssec:pragmatic}
Let us showcase \textit{how} an attacker can physically realize WsP leading to adversarial samples. We intend to demonstrate that WsP ``can be done'', and hence represent a (likely) threat that must be considered in a proactive development lifecycle of ML-PWD.

\textbf{Target System.}
We consider the ML-PWD proposed in~\cite{jain2018towards}, whose architecture aligns with the one in Fig.~\ref{fig:scenario}. The corresponding \smacal{M} is a \smamath{RF} classifier trained on a dataset created ad-hoc through public feeds. 
The complete feature set $F$ analyzed by \smacal{M} is reported in Table~\ref{tab:features}, which includes features related to both the URL and the representation of the website (based on the HTML). The ML-PWD extracts such features by inspecting the raw webpage according to the thresholds proposed in~\cite{mohammad2014intelligent} (and also used in~\cite{jain2018towards}). We observe that such methodology (and, hence, $F$) is also adopted by very recent works (e.g.,~\cite{hannousse2021towards, sharma2020feature}). We provide more details in the next section (§\ref{sssec:features}).

\begin{table}[htpb!]
\centering
\caption{Features $F$ of the considered ML-PWD.}
\resizebox{0.7\columnwidth}{!}{%
        \begin{tabular}{|c|c||c|c||c|c|}
            \toprule
            \# & \textbf{Feature Name} & \# & \textbf{Feature Name} & \# & \textbf{Feature Name} \\
            \midrule
            1 & URL\_length & 20 & URL\_shrtWordPath & 39 & HTML\_commPage\\ 
            2 & URL\_hasIPaddr & 21 & URL\_lngWordURL & 40 & HTML\_commPageFoot \\
            3 & URL\_redirect & 22 & URL\_DNS & 41 & HTML\_SFH\\
            4 & URL\_short & 23 & URL\_domAge & 42 & HTML\_popUp\\
            5 & URL\_subdomains & 24 & URL\_abnormal & 43 & HTML\_rightClick \\
            6 & URL\_atSymbol & 25 & URL\_ports & 44 & HTML\_domCopyright\\
            7 & URL\_fakeHTTPS & 26 & URL\_SSL & 45 & HTML\_nullLnkWeb \\
            8 & URL\_dash & 27 & URL\_statisticRe & 46 & HTML\_nullLnkFooter \\
            9 & URL\_dataURI & 28 & URL\_pageRank & 47 & HTML\_brokenLnk \\
            10 & URL\_commonTerms & 29 & URL\_regLen & 48 & HTML\_loginForm \\
            11 & URL\_numerical & 30 & URL\_checkGI & 49 & HTML\_hiddenDiv \\
            12 & URL\_pathExtend & 31 & URL\_avgWordPath & 50 & HTML\_hiddenButton \\
            13 & URL\_punyCode & 32 & URL\_avgWordHost & 51 & HTML\_hiddenInput \\
            14 & URL\_sensitiveWrd & 33 & URL\_avgWordURL & 52 & HTML\_URLBrand \\
            15 & URL\_TLDinPath & 34 & URL\_lngWordPath & 53 & HTML\_iframe \\
            16 & URL\_TLDinSub & 35 & URL\_lngWordHost & 54 & HTML\_favicon \\
            17 & URL\_totalWords & 36 & HTML\_freqDom & 55 & HTML\_statBar \\
            18 & URL\_shrtWordURL & 37 & HTML\_objectRatio & 56 & HTML\_css \\
            19 & URL\_shrtWordHost & 38 & HTML\_metaScripts & 57 & HTML\_anchors \\
            \bottomrule
        \end{tabular}
}
\label{tab:features}
\end{table}

\begin{figure}[!htbp]
    \centering
    \frame{\includegraphics[width=0.8\columnwidth]{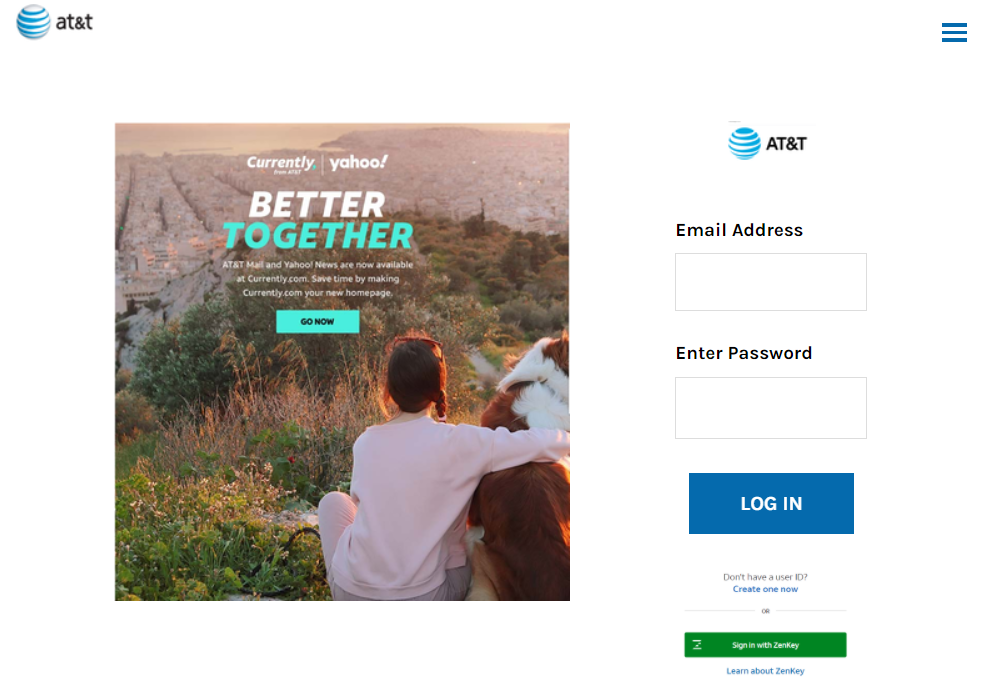}}
    \caption{An exemplary (and true) Phishing website, whose URL is {\small{https://www.63y3hfh-fj39f30-f30if0f-f392.weebly.com/}}.}
    \label{fig:example}
\end{figure}

\textbf{Attacker.}
The attacker expects the usage of a ML-PWD, but they are agnostic of anything about the ML model \smacal{M}, i.e., they are oblivious of the ML algorithm (i.e., \smamath{RF}) and its training data. The attacker, however, follows the state-of-the-art and hence knows the most popular feature sets used by ML-PWD (e.g.,~\cite{sharma2020feature}). In particular, the attacker correctly guesses that the ML-PWD analyzes features related to both the URL and the representation of the webpage, and specifically the URL length and the objects embedded in the HTML. Formally: $K$=(\textit{URL\_length, HTML\_objectRatio}). The attacker, however, does not know the \textit{exact} functionality of the feature extractor, the complete feature set $F$, and which features are more important for the final classification (the latter requires knowledge of \smacal{M}).
To provide a concrete example, we assume that the attacker owns the phishing\footnote{PhishTank reports such webpage to be a true and verified phishing (March 2022).} webpage shown in Fig.~\ref{fig:example}, whose URL is ``https://www.63y3hfh-fj39f30-f30if0f-f392.weebly.com/''.

\begin{figure}[!htbp]
    \centering
    \frame{\includegraphics[width=\columnwidth]{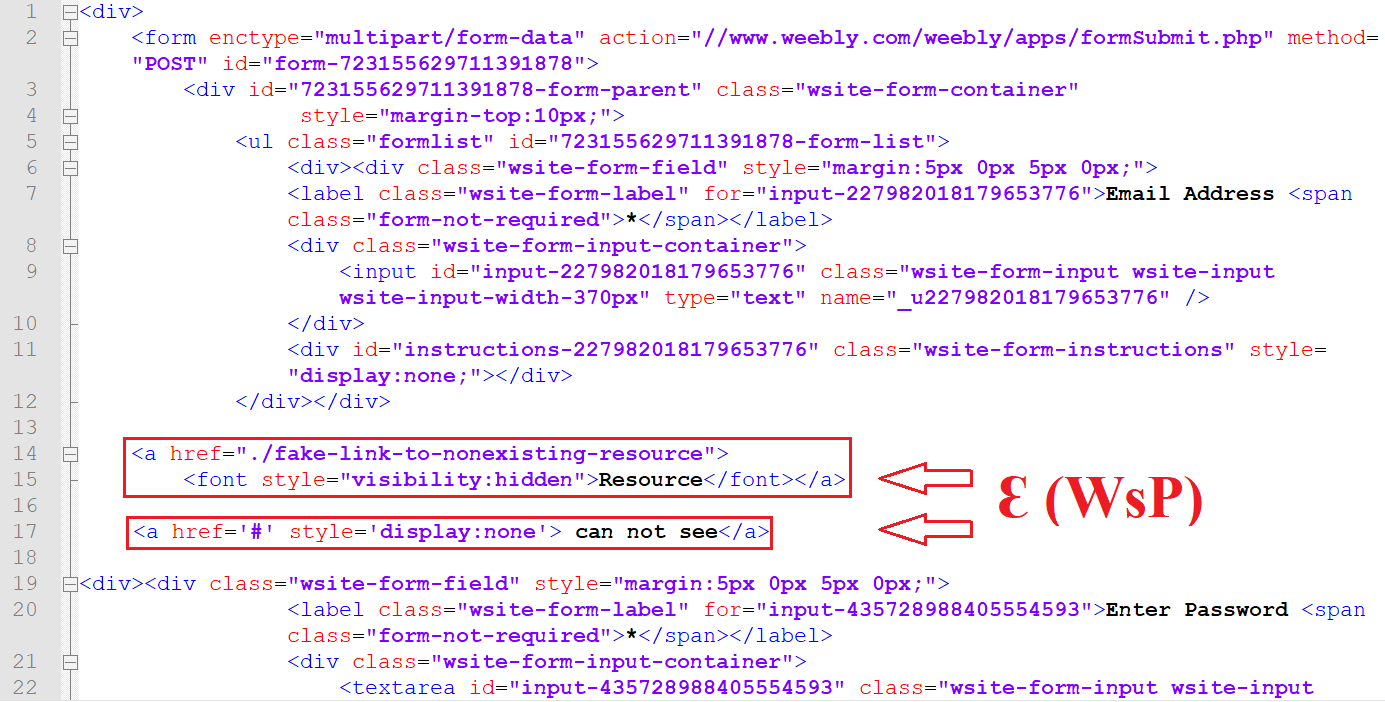}}
    \caption{A perturbation $\varepsilon$ in the website-space (WsP). The original HTML (related to the website in Fig.~\ref{fig:example}) is modified by introducing \textit{hidden link}(s). Such WsP will not be noticed by a user.}
    \label{fig:html}
    \vspace{-1em}
\end{figure}

\textbf{Real Perturbations.}
To craft perturbations in the website-space (i.e., WsP) that affect $K \subset F$, the attacker can:

\begin{itemize}

\item \textit{Modify the HTML.} The attacker knows that phishing websites have many links that point to external domains\footnote{E.g., phishing associated with AT\&T will have many links pointing to the real AT\&T.} with respect to internal resources (which would require to invest more into webhosting). Hence, the attacker can introduce (in the HTML) a high number of `fake links' that point to non-existent internal resources, which will affect the ratio of internal-to-external objects (making it more even). Such fake links, however, are can be made invisible (by exploiting some CSS properties) to users, who will not notice any difference\footnote{N.b.: complete `invisibility' is not a strict requirement. Some WsP can be `spotted' by a detailed analysis, but users may not notice them while still being phished. E.g., a link can be deleted; or a WsP can wrap: \textit{<a href='link'>} into  \textit{<a onclick="this.href='link'">}.}. We provide a visual representation of such WsP in Fig.~\ref{fig:html}, showing a snippet of the HTML of the original phishing webpage (cf. Fig.~\ref{fig:example}); the red rectangles denote two exemplary `perturbations', i.e., the introduction of (hidden) links pointing to an internal resource (which may not exist). Note that such WsP does not break the website's functionality, and can be cheaply introduced anywhere (and many times) in the source HTML.
Similar WsP are feasible and will\footnote{In theory, similar WsP could be detected by analyzing whether a given link is valid or not. Doing so, however, would pose an extremely high overhead: it requires checking every single link for every webpage that is analyzed by the ML-PWD.} influence the \textit{HTML\_objectRatio} (included in $K$). 

\item \textit{Modify the URL}. The attacker knows that long URLs are suspicious. So the attacker can, e.g., use a URL-shortening service (e.g., bit.ly) to alter the length of the phishing URL. In our case, the original URL (of 52 characters) can be shrunk to ``https://bit.ly/3MZHjt7'' (of 14 characters), thereby resulting in a completely different URL. Such a WsP will affect many features analyzed by \smacal{M} (cf. Table~\ref{tab:features}). Such features are not included in $K$, and hence their modifications are beyond the attacker's knowledge. The shrunk URL can then be shared in the wild\footnote{The ML-PWD will be fooled if it does not visit all the redirections of the shortening service. Nevertheless, there are many ways to reduce the URL\_length.}.

\item \textit{Both of the above.} The attacker can perturb both the URL and HTML to induce perturbations of higher impact.
\end{itemize}
We observe that none of these WsP are guaranteed to evade the ML-PWD. Indeed, a short URL is not necessarily benign, and having a non-suspicious ratio of internal-to-external objects is also not a strict requirement for being a benign webpage. The WsP could even be useless in the first place, e.g., the original URL could be already `short'. Indeed, our attacker is not aware of what happens inside the ML-PWD. The problem, however, is that such uncertainty is shared by both the attacker (who cannot observe the ML-PWD) and the defender (who cannot exactly pinpoint what the attacker does). To reveal\footnote{\textbf{Remark.} Attacking ML-PWD through (potentially unreliable) WsP is not the only way to `realistically' evade ML-PWD. This is clearly evidenced by prior work---whose validity is restored thanks to our evasion-space formalization. However, our proposed `cheap' attacks (through WsP) have never been investigated before in adversarial ML literature on PWD (§\ref{sec:related}). We hence set out to proactively assess the impact of feasible WsP on state-of-the-art ML-PWD; and comparing such impact to `less realistic' (hence, less likely to occur) attacks performed through PsP and MsP. Therefore, our evaluation will also consider such worst-case scenarios. We stress, however, that our threat model shall not envision attackers who: (i)~can observe or manipulate \smacal{D} (for poisoning attacks); (ii)~can observe the output-space (for black-box attacks); (iii)~have full knowledge of the ML-PWD (for white-box attacks): all of these scenarios have already been investigated by past work (§\ref{sec:related}), and are hence outside our scope. } the uncanny effects of such WsP, we assess them in §\ref{sec:results}.

\section{Evaluation: experimental setup and technical implementation}
\label{sec:evaluation}
As a constructive step forward, we assess the robustness of 18 ML-PWD against 12 evasion attacks---all based on our threat model, but performed in different evasion spaces. We have three goals:
\begin{itemize}
    \item assess state-of-the-art ML-PWD against \textit{feasible attacks};
    \item compare perturbations introduced in \textit{distinct evasion-spaces};
    \item provide a \textit{statistically validated benchmark} for future studies.
\end{itemize}
Achieving all such goals is challenging in research. Indeed, \textit{crafting} perturbations in the three distinct spaces (i.e., WsP, PsP, MsP) requires: (i)~datasets containing raw-data (for WsP), which are difficult to find; (ii)~devising custom feature extractors (for developing the ML-PWD); as well as (iii)~foreseeing the effects of WsP on such extractor (for PsP). Furthermore, to derive statistically sound conclusions, we must repeat our experiments multiple times~\cite{apruzzese2022sok}.

We present our experimental testbed (§\ref{ssec:testbed}), and then describe our technical implementation (§\ref{ssec:implementation}). Finally, we measure the baseline performance of our ML-PWD (§\ref{ssec:baseline}).

\subsection{Testbed}
\label{ssec:testbed}
We consider 18 ML-PWD, which vary depending on the \textit{source dataset} (2), the \textit{ML algorithm} (3), and the \textit{feature set} (3) used to develop the corresponding ML model. Such a wide array allows one to draw more generalizable conclusions.

\subsubsection{\textbf{Source Datasets}}
\label{sssec:datasets}
We rely on two datasets for ML-PWD: \dataset{$\delta$phish} and \dataset{Zenodo}~\cite{van2021combining,Corona:Deltaphish}. The reason is threefold. 
\begin{itemize}
    \item Both datasets include \textit{raw information} of each sample (specifically, its URL and its HTML). This is necessary because most of our attacks leverage WsP, for which we must modify the raw webpage, i.e., before its features are extracted. 
    \item Both datasets have been \textit{used by the state-of-the-art}. Prior research~\cite{van2021combining,Corona:Deltaphish} has demonstrated the utility of both datasets for ML-PWD, allowing for fair and significant comparisons.
    \item They enable experimental \textit{reproducibility}. Indeed, collecting ad-hoc data through public feeds (e.g., AlexaTop/PhishTank) prevents fair future comparisons: phishing webpages are taken down quickly, and it is not possible to retrieve the full information of webpages `blocklisted' years before. 
\end{itemize}
We provide an overview of our datasets in Table~\ref{tab:dataset}, which shows the number of samples (benign and phish) and the performance (\smamath{tpr} and \smamath{fpr}) achieved by their creators (in the absence of evasion).

\begin{table}[!ht]
    \centering
    \caption{Statistics and state-of-the-art of our datasets.}
    \resizebox{0.5\columnwidth}{!}{
        \begin{tabular}{c||c|c?c|c}\toprule
            \textbf{Dataset} & \textbf{\#Benign} & \textbf{\#Phish} & \textbf{\textit{fpr}}& \textbf{\textit{tpr}}\\ \midrule

            \dataset{$\delta$phish}~\cite{Corona:Deltaphish} & 5511 & 1012 & $0.01$ & $0.98$ \\ 
            \dataset{Zenodo}~\cite{van2021combining} & 2000 & 2000 & $0.08$ & $0.99$ \\ 
            
            \bottomrule
            
        \end{tabular}
    }
    
    \label{tab:dataset}
\end{table}
We mention that the original \dataset{Zenodo} contains 100k phishing, and almost 4M benign webpages. To make our evaluation ``humanly feasible,'' we randomly sample 4000 webpages from \dataset{Zenodo}, equally split between benign and phishing. In such a way, we can analyze the response of ML-PWD having diverse \textit{balancing}: while \dataset{Zenodo} is perfectly balanced, \dataset{$\delta$Phish} has significantly more benign samples.

\subsubsection{\textbf{ML Algorithms}}
\label{sssec:algorithms}
We consider ML-PWD based on shallow and deep learning algorithms~\cite{Apruzzese:Addressing}. Our selection aims to provide a meaningful assessment of ML-PWD based on exemplary ML methods. 
We consider:
\begin{itemize}
    \item Logistic Regression (\smamath{LR}). One of the simplest ML algorithms, we consider \smamath{LR} because it was (assumed to be) used by the ML-PWD embedded in Google Chrome~\cite{liang2016cracking}.
    \item Random Forests (\smamath{RF}). An ensemble technique, \smamath{RF} often outperforms other contenders for ML-PWD~\cite{tian2018needle}.
    \item Convolutional neural Network (\smamath{CN}). We consider this well-known deep learning technique~\cite{Lecun:Deep} due to its demonstrated proficiency also in ML-PWD (e.g.,~\cite{wei2020accurate}).
\end{itemize}
All of these algorithms support binary classification, making them appropriate for our ML-PWD.

\subsubsection{\textbf{Feature Sets}}
\label{sssec:features}
We consider ML-PWD that use three feature sets ($F$), all resembling the one described in our use-case (§\ref{ssec:pragmatic}). Specifically, our ML-PWD analyze one of the following:
\begin{itemize}
    \item URL-only ($F^u$), i.e., the first 35 features in Table~\ref{tab:features}.
    \item Representation-only ($F^r$), i.e., the last 22 features in Table~\ref{tab:features}.
    \item Combined ($F^c$), corresponding to all features in Table~\ref{tab:features}.
\end{itemize}

\textit{Rationale.} Analyzing more information (i.e., larger feature sets, such as $F^c$) leads to superior detection performance---as shown, e.g., in~\cite{Corona:Deltaphish}. However, in some cases this may not be possible: for instance, phishing \textit{email} filters may make their decisions only by analyzing the URL (cf. §\ref{ssec:mlpwd}). Nevertheless, modifying the URL is one of the easiest ways to trick a ML-PWD~\cite{o2021generative}: hence, a defender may develop an `adversarially robust' detector that analyzes only the representation of a webpage. Such detector will have a lower performance (w.r.t. $F^c$) in non-adversarial scenarios, but will counter evasion attacks that manipulate the URL (cf. §\ref{ssec:aml}). 

\textit{Observation.} Our feature sets are not only popular in research (e.g.,~\cite{jain2018towards, mohammad2014intelligent, hannousse2021towards, sharma2020feature}), but also used in \textit{practice}. Indeed, several leading security companies yearly organize MLSEC, an ML evasion competition~\cite{mlsec}. In 2021 and 2022, MLSEC also involved evading ML-PWD \textit{which specifically analyzed the HTML}~\cite{gao2023evading} representation of a webpage---i.e., our $F^r$. We will also refer to MLSEC in our evaluation.

\subsection{Technical Implementation}
\label{ssec:implementation}

\begin{figure}[!htbp]
    \centering
    \includegraphics[width=0.94\columnwidth]{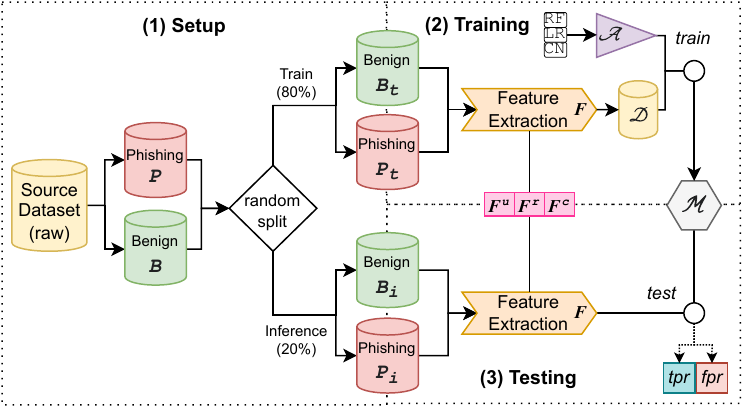}
    \caption{Experimental workflow. Each source dataset (containing benign, $B$, and phishing, $P$, samples) is randomly split into the training ($B_t$ and $P_t$) and inference ($B_i$ and $P_i$) partitions, used to train and test each ML-PWD. We use $P_i$ as basis for our adversarial samples.}
    \label{fig:workflow}
\end{figure}

Let us describe how we combined all the elements described insofar to devise our ``baseline'' ML-PWD. We provide a schematic of our workflow in Fig.~\ref{fig:workflow}.
Each source dataset (\dataset{Zenodo} and \dataset{$\delta$phish}) represents a different setting---which we use to extract the corresponding training and inference partitions for our ML-PWD.
Such ML-PWD are based on one among three ML algorithms, encompassing either shallow (\smamath{LR} and \smamath{RF}) or deep learning (\smamath{CN}) classifiers. Each of these classifiers presents three variants, depending on the analyzed features ($F^u$, $F^r$, or $F^c$), yielding a total of 9 `baseline' ML-PWD per source dataset. Finally, we ensure that such 9 ML-PWD maximize their performance (high \smamath{tpr} and low \smamath{fpr}, at least for $F^c$). 

We now describe the implementation of our feature extractor (§\ref{sssec:extractor} and the development of our ML-PWD (§\ref{sssec:mlpwd}).

\subsubsection{\textbf{Feature Extractor}}
\label{sssec:extractor}

An important part of our evaluation is represented by the feature extractor, for which we rely on the established guidelines provided in~\cite{mohammad2014intelligent, mohammad2014predicting} and still widely employed in recent literature (e.g.,~\cite{jain2018towards}). The underlying principle of such guidelines is to analyze several elements of a webpage (e.g., the length of its URL), and then use threshold-based mechanisms to determine whether such element is `benign' or `phishing' (e.g., a short URL is likely benign, whereas a long one is likely phishing). Any feature can have a value within [-1, 1], where -1 is `benign' and 1 is `phishing'.
Our extractor generates all the features reported in Table~\ref{tab:features}. We explain some of them. 
\begin{itemize}
    \item (\#1) \textit{URL\_length}. We compute the amount of character composing the entire URL. Strings shorter than 53 characters correspond to -1 (likely `benign'), whereas longer strings correspond to +1 (likely `phishing').
    \item (\#4) \textit{URL\_short}. If the URL starts\footnote{Our feature extractor is `stateless'. Once it receives a sample, the only queries performed are those to some third-party services (e.g., PageRank API, DNS servers), which can be cached to save time. Our extractor, however, does not `update' a sample: if, e.g., a URL uses a shortening service, then the extractor uses such `shortened' URL as basis, and if the HTML changes (due to some automatic script) then such change will not be captured. Such choice makes sense because ML-PWD must be fast: a user does not want to wait seconds before visiting each website just because a phishing check is made. Moreover, our decision makes our extractor suitable also to to ML-PWD that analyze \textit{only} the URL, because the webpage will not be opened in the first place (which is common for phishing email filters) due to the high overhead.} with keywords related to popular shortening services (bit.ly, goo.gl, tinyurl, ad.fly) then this feature is set to +1, and to -1 otherwise.
    \item (\#28) \textit{URL\_pageRank}. We use OpenPageRank API to query the URL domain. The response shows the page ranks from 0 to 10: the corresponding feature is normalized between -1 (if the rank is 10) and +1 (if the rank is 0).
    \item (\#37) \textit{HTML\_objectRatio}. We capture all the objects embedded in the webpage, and compute the ratio of internal-to-external objects. An internal object either has its link starting with \textit{../} or with the same `root' as the website's URL. If the ratio is less than 0.15, then this feature is -1 (likely benign), and +1 otherwise (likely phishing).
    \item (\#38) \textit{HTML\_metaScripts}. Same as \#37, but for scripts, links and metas. If the ratio is more than 0.61, the feature value is +1 (likely phishing); if the ratio is less than 0.52, the feature value is -1 (likely benign); and 0 otherwise.
    \item (\#45) \textit{HTML\_nullLnkWeb}. We check how many links are useless, i.e., they point to the exact same page (e.g., \textit{href=\#}). The count can be normalized between +1 (high number of useless links) and -1 (no useless links).
    \item (\#51) \textit{HTML\_hiddenInput}. We check if there are any hidden \textit{input} tags in the webpage. If there are, the feature value is +1 (likely phishing), and -1 otherwise (likely benign). 
    \item (\#52) \textit{HTML\_URLBrand}. We check (in the HTML) if the webpage \textit{title} includes the brand name in the URL. If included, the feature value is -1 (benign); otherwise, is +1 (phishing).
\end{itemize}
(Our repository includes the source-code of our feature extractor.)

\noindent
We use similar thresholds as those by Mohammad et al.~\cite{mohammad2014intelligent, mohammad2014predicting}, and are the same used to create the popular \dataset{UCI} dataset~\cite{UCIphishing}. 
To validate our choice of using the same thresholds (which play a crucial role in our evaluation), we find instructive to report the length of URLs contained in our chosen datasets, i.e., \dataset{Zenodo} and \dataset{$\delta$phish}. The results are as follows: for \dataset{Zenodo}, there are 1500 URLs (out of 4000) which are \textit{longer} than 54 characters; for \dataset{$\delta$phish}, there are 1909 URLs (out of 6523) which are \textit{longer} than 54 characters. Hence, such a threshold is still sensible for more recent datasets.

\subsubsection{\textbf{Development of the ML-PWD}}
\label{sssec:mlpwd}

We follow three phases (i.e., the three dotted squares in Fig.~\ref{fig:workflow}). Namely:

\begin{enumerate}
    \item \textit{Setup.} The first phase is choosing a given source dataset (i.e., \dataset{Zenodo} or \dataset{$\delta$phish}) and partition its samples into \textit{benign} and \textit{phishing} ($B$ and $P$, respectively). Then, we perform a \textit{random split} (to avoid bias) on each of these partitions by using a 80:20 ratio (common in related literature~\cite{bac2021pwdgan, al2021generating}). In other words, we randomly select 80\% of the samples in both $B$ and $P$ (i.e., $B_t$ and $P_t$ respectively), which will be used to train the ML model. The leftout samples, $B_i$ and $P_i$ (corresponding to 20\% of $B$ and $P$, respectively), are used to assess the inference performance of the resulting ML model. We will also use $P_i$ as basis to craft our adversarial samples.
    
    \item \textit{Training.} To train \smacal{M}, we recall that the source data is in raw format. Hence, before obtaining the training dataset \smacal{D}, the corresponding \textit{training} partitions $B_t$ and $P_t$ must be transformed into their feature representation. Hence, we develop a feature extractor (described in §\ref{sssec:features}) that is based on a given feature set $F$ (either $F^u$, $F^r$, or $F^c$). Then, we preprocess both $B_t$ and $P_t$ to obtain the actual training data \smacal{D}. At this point, we apply a given ML algorithm \smacal{A} (either \smamath{RF}, \smamath{LR} or \smamath{CN}) to such \smacal{D}; the resulting ML model \smacal{M} (a binary classifier, which we fine tune via grid-search) will be the detection component of the considered ML-PWD.
    
    \item \textit{Testing.} The last phase is measuring the performance of \smacal{M}. In our case, a ML-PWD must exhibit both a high detection rate and a low false positive rate: indeed, no one is interested in detectors that block legitimate websites due to excessive false alarms. Hence, we preprocess the \textit{inference} partitions $B_i$ and $P_i$ (by considering the proper $F$) and measure the \smamath{fpr} and \smamath{tpr}---in the absence of adversarial attacks.
\end{enumerate}
The topmost priority is ensuring that \smacal{M} analyzing $F^c$ achieve optimal performance: indeed, models using either $F^u$ or $F^r$ are expected to exhibit a lower performance as they are provided with less information; however, using $F^u$ or $F^r$ is expected to yield a superior robustness in the presence of evasion attacks. (Our repository includes the best parameter configurations of each ML algorithm.)

\textbf{Statistical Validation.}
To provide results that are devoid of experimental bias and also to serve as a reliable benchmark for future researches, \textit{we repeat all the abovementioned operations 50 times}. This means that each source dataset is randomly sampled 50 times, each resulting in a different training partition \smacal{D} and, hence, a different \smacal{M}. Such \smacal{M} is, in turn, assessed on different data (i.e., different inference partitions), yielding different \smamath{fpr} and \smamath{tpr}. Furthermore, all\footnote{Overall, for our experiments we develop 900 \ftcal{M} (given by: 2 source datasets * 50 random draws * 3 $F$ * 3 \ftcal{A}).} such \smacal{M} are also assessed against all our considered attacks (which we will discuss in the next section). 
Such a large evaluation allows one to perform \textit{statistically validated comparisons} by leveraging well-known techniques~\cite{apruzzese2022sok}. We will do this to infer whether some attacks induce a performance degradation that is statistically significant. To the best of our knowledge, we are the first to use statistical tests to validate the impact of adversarial attacks against ML-PWD.

\subsection{Baseline Performance}
\label{ssec:baseline}

We report the performance of our ML-PWD (\textit{in the absence} of adversarial attacks) in Table~\ref{tab:normal}. This table shows that the best ML-PWD on both datasets use \smamath{RF}. We appreciate that the `true' baseline ML-PWD (using $F^c$) exhibit similar results as the state-of-the-art (cf. Table~\ref{tab:dataset}). In contrast, the `robust' baselines (using either $F^r$ or $F^u$) are slightly inferior\footnote{Focusing on the ML-PWD using $F^r$ (which are similar to the real ML-PWD in MLSEC~\cite{mlsec}), we appreciate that \scmath{RF} achieves a remarkable 0.935 \scmath{tpr} and 0.01 \scmath{fpr} (averaged on both datasets), making such ML-PWD a valid baseline.}. For instance, on \dataset{Zenodo}, the \smamath{RF} using $F^u$ has almost the same performance as $F^c$, but the one using $F^r$ has 5\% less \smamath{tpr} and 2\% more \smamath{fpr}; whereas on \dataset{$\delta$phish}, the \smamath{RF} using $F^u$ has 50\% less \smamath{tpr} (but similar \smamath{fpr}), while the one using $F^r$ has 0.5\% more \smamath{fpr}, but only 3\% less \smamath{tpr}. Such degradation is the \textbf{cost} of using defenses based on \textit{feature removal} on the considered ML-PWD. The expected benefit, however, is a superior resilience to evasion attempts.

\begin{table}[!htbp]
    \centering
    \caption{Performance in non-adversarial settings, reported as the average (and std. dev.) \smamath{tpr} and \smamath{fpr} over the 50 trials.}
    \label{tab:normal}
    \resizebox{0.6\columnwidth}{!}{
        \begin{tabular}{c|c?cc||cc}
             \toprule
             \multirow{2}{*}{\smacal{A}}& \multirow{2}{*}{$F$} & \multicolumn{2}{c||}{\dataset{Zenodo}} & \multicolumn{2}{c}{\dataset{$\delta$phish}} \\ \cline{3-6}
              & & $tpr$ & $fpr$ & $tpr$ & $fpr$ \\
             \toprule

            \multirow{3}{*}{{\smamath{CN}}}
             & $F^u$ & \res{0.96}{0.008} & \res{0.021}{0.0077} & \res{0.55}{0.030} & \res{0.037}{0.0076}\\
             & $F^r$ & \res{0.88}{0.018} & \res{0.155}{0.0165} & \res{0.81}{0.019} & \res{0.008}{0.0020}\\
             & $F^c$ & \res{0.97}{0.006} & \res{0.018}{0.0088} & \res{0.93}{0.013} & \res{0.005}{0.0025}\\

            \midrule 
             
            \multirow{3}{*}{{\smamath{RF}}}
             & $F^u$ & \res{0.98}{0.004} & \res{0.007}{0.0055} & \res{0.45}{0.022} & \res{0.003}{0.0014} \\
             & $F^r$ & \res{0.93}{0.013} & \res{0.025}{0.0118} & \res{0.94}{0.016} & \res{0.006}{0.0025} \\
             & $F^c$ & \bestres{0.98}{0.006} & \bestres{0.007}{0.0046} & \bestres{0.97}{0.007} & \bestres{0.001}{0.0011} \\
             
            \midrule
             
            \multirow{3}{*}{{\smamath{LR}}}
             & $F^u$ & \res{0.95}{0.009} & \res{0.037}{0.0100} & \res{0.24}{0.017} & \res{0.011}{0.0026} \\
             & $F^r$ & \res{0.82}{0.017} & \res{0.144}{0.0171} & \res{0.74}{0.025} & \res{0.018}{0.0036} \\
             & $F^c$ & \res{0.96}{0.007} & \res{0.025}{0.0077} & \res{0.81}{0.020} & \res{0.013}{0.0037} \\
             
            \bottomrule
             
        \end{tabular}
    }
\end{table}

Finally, by comparing Table~\ref{tab:normal} with Table~\ref{tab:dataset}, we appreciate that our ML-PWD using $F^c$ achieve comparable performance as prior work (even after our subsampling on \dataset{Zenodo}), confirming their relevance as baseline. Our repository includes the 4K pages we used for \dataset{Zenodo}.
\section{Evaluation: Attacks (Rationale and Implementation)}
\label{sec:attacks}

We now focus on our considered attacks. We begin by providing an extensive overview (§\ref{ssec:attacks}), and then summarize the workflow for their empirical evaluation (§\ref{ssec:workflow}). Finally, we describe their technical implementation (§\ref{ssec:attacks_implementation})

\subsection{\textbf{Considered Attacks}}
\label{ssec:attacks}
In our paper, we consider a total of 12 evasion attacks, divided in four families. One of these families is an \textit{exact replica} of our `standard' threat model. The remaining three families, however, are \textit{extensions} of our threat model, which assume more `advanced' adversaries who have superior knowledge and/or capabilities. 

Two of our families involve WsP (\atk{WA} and \aatk{}), but assume attackers with different knowledge; whereas the remaining two families involve either PsP or MsP (\atk{PA} and \atk{MA}). Each family has three variants depending on the features `targeted' by the attacker, i.e., either those related to the URL, the HTML, or a combination of both (\smamath{u}, \smamath{r}, or \smamath{c}). For WsP, the underlying `attacked' features are always the same for all variants, which are assumed to be known by the attacker: \smamath{u} is always the \textit{URL\_length}; for \smamath{r} is the \textit{HTML\_objectRatio}; and for \smamath{c} they are both of these. (Do note that our WsP will affect also features beyond the attacker's knowledge.)
\begin{itemize}
    \item \textit{Cheap Website Attacks} (\atk{WA}) perfectly align with our threat model (and resemble the use-cases in §\ref{ssec:pragmatic}). The perturbations are created in the website-space (WsP), realizing either \atks{WA}{u}, \atks{WA}{r}, or \atks{WA}{c}. Specifically for \smamath{r} (and \smamath{c}), we consider two semantically equivalent WsP: ``add fake link'' for \dataset{$\delta$Phish}, and ``link wrapping'' for \dataset{Zenodo}. Such WsP attempt to balance the object ratio: the former by adding (invisible) links to (fake) internal objects, whereas the latter by eluding the preprocessing mechanism---thereby having a link not being counted among the total links shown in a webpage.

    \item \textit{Advanced Website Attacks} (\aatk{}), which envision a more knowledgeable attacker than \atk{WA}. The attacker knows how the feature extractor within the ML-PWD operates (i.e., they know the specific thresholds used to compute some features). The attacker -- who is still confined in the website-space -- will hence craft more sophisticated WsP because they know how to generate an adversarial sample that is more likely to influence the ML-PWD. Thus, the attacker will modify either the URL, the HTML, or both (i.e., \aatks{u}, \aatks{r}, \aatks{c}), but in more elaborate ways---e.g., by ensuring that the \textit{HTML\_objectRatio} exactly resembles the one of a `benign' sample; or by making an URL to be `long enough' to be considered short.

    \item \textit{Preprocessing Attacks} (\atk{PA}), which are an extension of our threat model, and assume an even stronger attacker that is able to access the preprocessing stage of the ML-PWD, and hence introduce PsP. Such an attacker is capable of direct feature manipulation---subject to integrity checks (i.e., the result must reflect a ``physically realizable'' webpage). Since the attacker does not know anything about the actual \smacal{M}, the attacker must still guess their PsP. 
    Such PsP will target features based on either \smamath{u}, \smamath{r}, \smamath{c} (i.e., \atks{PA}{u}, \atks{PA}{r}, \atks{PA}{c}) by accounting for inter-dependencies between other features.

    \item \textit{ML-space attacks} (\atk{MA}), representing a worst-case scenario. The attacker can access the ML-space of the ML-PWD, and can hence freely manipulate the entire feature representation of their webpage through MsP. However, the attacker is still oblivious of \smacal{M}, and must hence still guess their WsP. Thus, the MsP applied by the attacker completely `flip' many features related to \smamath{u}, \smamath{r}, \smamath{c} (i.e., \atks{MA}{u}, \atks{MA}{r}, \atks{MA}{c}). 
\end{itemize}

\textbf{Motivation.} We consider these 12 attacks for three reasons. First, to assess the effects of diverse \textit{evasion attacks at increasing `cost'}. For instance, the simplicity of \atk{WA} makes them the most likely to occur; whereas \atk{MA} can be disruptive, but are very expensive (from the attacker's viewpoint). Second, to study the response of ML-PWD to WsP targeting the same features (\atks{WA}{r}), but in different ways (one per dataset), leading to alterations of \textit{different features beyond the attacker's knowledge}. Third. to highlight the \textit{effects of potential `pitfalls'} of related researches. Indeed, we observe that all three remaining families (\aatk{}, \atk{PA}, \atk{MA}) envision attackers with similar knowledge which they use to target \textit{similar features}. Such peculiarity allows comparing attacks carried out in different `spaces.' A particular focus is on \atk{PA}, for which we apply PsP by \textit{anticipating} how a WsP can yield a physically realizable~\cite{tong2019improving} PsP. Put differently, our evaluation shows what happens if the perturbations are applied without taking into account \textit{all} preprocessing operations that transform a given $x$ into the $F_x$ analyzed by \smacal{M}.

\textbf{Effectiveness and Affordability.}
In terms of effectiveness, assuming the same targeted features, \atk{WA}$<$\aatk{}$<$\atk{PA}$\ll$\atk{MA} (as confirmed by our results in §\ref{ssec:comparison}). This is justified by the higher investment required by the attacker, who must either perform extensive intelligence gathering campaigns (to understand the exact feature extractor for \aatk{}) or gain write-access to the ML-PWD (for \atk{PA} and \atk{MA}). Let us provide a high-level summary of the requirements to implement all our attacks---all of which are \textit{query-less} and rely on \textit{blind} perturbations.
\begin{itemize}
    \item \atk{WA}: they require as little as a dozen lines of elementary code, and a very rough understanding of how ML-PWD operate (which can be done, e.g., by reading research papers).
    \item \aatk{}: they also require a few lines of code to implement. However, determining the exact thresholds requires a detailed intelligence gathering campaign (or many queries to reverse-engineer the ML-PWD, if it is client-side).
    \item \atk{PA}: they require a compromise of the ML-PWD. For example, introducing a special `backdoor' rule that ``if a given URL is visited, then do not compute its length and return that the URL is \textit{short}''. Doing this is costly, but it is not unfeasible if the feature extractor is open-source (e.g.,~\cite{bagdasaryan2021blind}).
    \item \atk{MA}: they also require a compromise of the ML-PWD. In this case, the `backdoor' is introduced \textit{after} all features have been computed---and irrespective of their relationships. Hence. the cost is very high: the ML model is likely to be tailored for a specific environment, thereby increasing the difficulty of successfully introducing such backdoors in one of the deepest segments of the ML-PWD.
\end{itemize}
Hence, in terms of affordability: \atk{WA}$\gg$\aatk{}$\gg$\atk{PA}$>$\atk{MA} (i.e., the relationship is the reverse of the effectiveness). For this reason, in our evaluation we will put a greater emphasis on \atk{WA}, because `cheaper' attacks are more likely to occur \textit{in the wild}: while \atk{WA} can be associated with ``horizontal phishing'' (the majority), the others are tailored for ``spear phishing'' (the minority).

\subsection{Evaluation Workflow}
\label{ssec:workflow}
The procedure to assess the adversarial attacks involves three steps.
\begin{enumerate}
    \item \textit{Isolate}. Our threat model envisions evasion attacks that occur during inference, hence our adversarial samples are generated from those in $P_i$. Furthermore, we recall that the attacker expects the ML-PWD to be effective against `regular' malicious samples. To meet such condition, we isolate 100 samples from $P_i$ that are detected successfully by the best\footnote{This ensures that all ML-PWD are assessed against the \textit{same} adversarial samples. We provide such sampmles in the source-code.} ML-PWD (typically using $F^c$) during one of our runs. Such samples are then used as basis to craft the adversarial samples corresponding to each of the 12 considered types of evasion attacks.
    
    \item \textit{Perturb}. We apply the perturbations as follows. For \atk{WA} and \aatk{}, we craft the corresponding WsP, apply them to each of the 100 samples from $P_i$, and then preprocess such samples by using the feature extractor. For \atk{PA} and \atk{MA}, we first preprocess the 100 samples with the feature extractor, and then apply the corresponding PsP or MsP. Overall, these operations result in 1200 adversarial samples (given by 12 attacks, each using 100 samples).
    
    \item \textit{Evade}. The 1200 adversarial samples are sent to the 9 ML-PWD (for each dataset), and we measure the $tpr$ again. 
\end{enumerate}
The expected result it that the \smamath{tpr} obtained on the adversarial samples (generated as a result of any of the 12 considered attacks) will be lower than the \smamath{tpr} on the original 100 phishing samples.

\subsection{Attacks Implementation}
\label{ssec:attacks_implementation}
Let us discuss how we implement our perturbations, and provide some insight as to which features are influenced as a result of our attacks. We recall that each attack family presents three variants, depending on which features the attacker is `consciously' trying to affect. Namely: \smamath{u}, \smamath{r} and \smamath{c}, i.e., features involving the URL, the representation (HTML) or a combination thereof.
All attacks are created by manipulating (phishing) samples taken from $P_i$. In particular, during our first trial we isolate 100 samples from $P_i$ that are correctly detected by the best ML-PWD: such samples are then used as basis for all their adversarial variants (to ensure consistency). For simplicity, we will denote any of such samples as $p$.

We start by describing \atk{MA} which are the easiest to implement. Then, we describe \atk{WA} and \aatks{}. Finally, we describe \atk{PA}, which are the most complex to implement because they must consider several implications (e.g., inter-feature dependencies). (Our repository includes the exact implementation of \atk{MA} and \atk{PA}, and also all the pre-processed variant of the samples generated via \atk{WA} and \aatks{}.)

\subsubsection{\textbf{ML-space attacks}}
\label{sapp:msp}
The attacks (i.e., \atk{MA}) are the easiest to implement. Indeed, we simply follow the same procedure as done by most prior works (e.g.,~\cite{Corona:Deltaphish, lee2020building}) that directly manipulate the feature representation $F_p$ of a sample $p$ right before it is analyzed by the ML-PWD. We do this without taking into account any inter-dependency between features and/or any physical property that the actual webpage must preserve: this is compliant with our assumption that the attacker has access to the ML-space. Specifically, for each \atk{MA} we apply the following MsP:
\begin{itemize}
    \item \atks{MA}{u}: The attacker targets URL-related features. Hence, we manipulate $F_p$ by setting features based on $F^u$ equal to -1, which denotes a value that is more likely associated with a benign sample. In particular, we set to -1 the features in Table~\ref{tab:features} with the following numbers: (1-17,19-21,27,30-35)
    \item \atks{MA}{r}: Same as above, but the targeted features are within $F^r$. Hence, we set to -1 the features in Table~\ref{tab:features} with the following numbers: (36-40,42-52,54-57)
    \item \atks{MA}{c}: We set to -1 all features involved in \atks{MA}{u} and \atks{MA}{r}.
\end{itemize}
We remark that the attacker is not aware of the feature importance (because it would require knowledge of \smacal{M}). Hence, although some manipulations will likely `move' $F_p$ towards a benign webpage, it is not guaranteed that \smacal{M} will actually classify such $F_p$ as benign: if the manipulated features are not important, then even MsP may have no effect (and such phenomenon \textit{does} happen in our evaluation, e.g., the ML-PWD using \smamath{RF} with $F^c$ on \dataset{Zenodo} against \atks{MA}{r}).

Of course, we could set \textit{all} features to -1 (e.g., all $F^u$ and $F^r$). Doing this, however, would obviously result in a perfect misclassification (and hence not interesting to show). Moreover, it would not be sensible \textit{even for the attacker}. Indeed, \atk{MA} assume no knowledge of \smacal{M} and of \smacal{D}, meaning that an attacker may suspect the existence of a honeypot~\cite{shan2020gotta}. For instance, \smacal{D} may contain some samples with all features set to -1 (i.e., benign) that are labelled as phishing---for the sole purpose of defeating similar attacks in the ML-space. Hence, it is realistic to assume that even an attacker capable of \atk{MA} would not exaggerate with their perturbations.

\subsubsection{\textbf{Website attacks}}
\label{sapp:wsp}
We recall that we perform two families of attacks in the website-space: \atk{WA} and \aatk{}. 
The peculiarity of both of these attacks (both relying on WsP) is that the attacker does not have access to the ML-PWD. Hence, they are not able to manipulate $F_p$, and they are not even able to \textit{observe} $F_p$. 

\begin{itemize}
    \item \atk{WA}
These attacks resemble the pragmatic example (§\ref{ssec:pragmatic}).

\begin{itemize}
    \item \atks{WA}{u}: We set the URL to a random string starting with ``www.bit.ly/'', followed by 7 randomly chosen characters (which what this popular URL shortener does).
    \item \atks{WA}{r}: For \dataset{$\delta$Phish}, we change the HTML by adding 50 invisible internal links (i.e., having the same root domain of the website);\footnote{The exact string we inject is: ``\textit{<a href=`\#' style='display:none'> can not see</a>}'', which is the second string shown in our pragmatic example (§\ref{ssec:pragmatic}).} for \dataset{Zenodo}, we wrap all links within an ``onclick'', i.e., we change: \textit{<a href=`link'>} into  \textit{<a onclick=``this.href=`link'">}.\footnote{This WsP, if applied to textual link, would remove the underline of such a link, therefore being visible to a user; however, it is possible to make it invisible by editing the CSS properties. Our feature extractor is agnostic of such properties, so we do not do this: the results would be equivalent.}
    \item \atks{WA}{c}: We do both of the above for each dataset.
\end{itemize}

\item \aatk{}: These attacks envision an attacker that knows how the feature extractor within the ML-PWD operates (see §\ref{sssec:features}).  Such knowledge can be acquired, e.g., if the attacker has (or is) an insider that provided them with such intelligence. However, the attacker is still confined in the website-space, and hence can only apply WsP (to generate $\overbar{p}$). For a meaningful comparison, we assume an attacker who is aware of how the features targeted in \atk{WA} are ``extracted'' within the ML-PWD.
Hence, we craft each \aatk{} as follows:
\begin{itemize}
    \item \aatks{u}: The attacker, having knowledge of the extractor, knows that by using an URL shortener they will affect all features related to the URL (i.e., $F^u$); furthermore, they know the threshold (53) that makes an URL to be considered as `benign'. Such length is well above that of an URL generated via any shortening service. As such, these attacks are an exact replica as \aatks{u} (the only difference is that the attacker of \aatks{u} is more confident than the one in \atks{WA}{u}).
    \item \aatks{r}: The attacker manipulates the HTML in the same was as in \atks{WA}{c}. However, the attacker also knows the threshold (0.15) of internal-to-external links that yields a benign value of the \textit{HTML\_objectRatio} feature. Hence, the WsP manipulate the HTML of each $p$ by introducing as many links (or wrappings) as necessary to meet such threshold.
    \item \aatks{c}: The attacker does both of the above.
\end{itemize}
We stress that the attacker cannot observe $F_{\overbar{p}}$. Indeed, doing this would require the attacker to completely replicate the feature extractor, which is costly, and may not even be possible (some third-party services may require subscriptions to be used). As such, the attacker is aware of how to craft WsP that are more likely noticed by the ML-PWD, but evasion is not guaranteed.
\end{itemize}

\subsubsection{\textbf{Preprocessing attacks}}
\label{sapp:psp}
These attacks are the hardest to realize \textit{from a research perspective} and \textit{in a fair way}.

\textbf{Challenges.} The underlying principle of PsP (the backbone of \atk{PA}) is affecting the preprocessing space of the ML-PWD. Technically, since we are the developers of our own feature-extractor (i.e., the component of the ML-PWD devoted to data preprocessing), we could simply directly manipulate our own extractor, i.e., by introducing a `backdoor'. However, doing this would prevent a fair generalization of our results: for instance, it is possible to develop another feature extractor, having the same functionality but whose operations are executed in a different order.
Hence, to ensure a more fair evaluation, we apply the perturbations \textit{at the end} of the preprocessing phase, but we do so by anticipating how a perturbation in the website-space (a WsP) could affect the preprocessing-space, thereby turning a WsP into a ``physically realizable'' PsP. To this purpose, we \textit{assume the viewpoint of an attacker}. For instance, we ask ourselves: ``if an attacker wants to affect URL features by using an URL shortener, how would the feature extractor react?''. 

\textbf{Scenario.}
In \atk{PA} the attacker \textit{knows} and \textit{can interfere} (through PsP) with the feature extraction process of the targeted ML-PWD. However, the attacker is \textit{not} aware of what happens next: the ML-space and the output-space are both inaccessible by the attacker (from both a \textit{read} and \textit{write} perspective). Hence, once the PsP has been applied and $\overbar{F_p}$ is generated, the attacker cannot influence $\overbar{F_p}$ any longer.
For each \atk{PA} we do the following:
\begin{itemize}
    \item \atks{PA}{u}: we anticipate an attack that targets URL features, and specifically \textit{URL\_length}, by using an URL shortener. Hence, we can foresee that operations (in the website-space) can lead to alterations of \textit{all} the features involved with the URL (i.e., $F^u$). For instance, doing this would make weird characters (if present) to disappear from the URL. However, doing this would induce to alterations also to $F^r$. For instance, some objects originally considered to be `internal' would become `external'. Hence, we implement \atks{PA}{u} by setting the following features (from Table~\ref{tab:features}) to -1: (1-3,5,6,8,10-16,22,23,25,26,28-30), whereas the following features are set to +1: (4,27,36-38,41,44,48,52,54,56).
    \item \atks{PA}{r}: we anticipate an attack that targets features related to the representation of a website---in our case the HTML, and specifically the \textit{HTML\_objectRatio} feature. We foresee that an attacker can interfere with such feature in many ways, for instance by removing links, adding new ones, or changing those already contained in the webpage. All such changes will affect many features, such as the \textit{HTML\_freqDom}: because populating the HTML with (fake) internal links would change the `frequent domains' included in the HTML. Such changes can also affect the links in the footer of the webpage (\textit{HTML\_nullLnkFooter}); or the anchors (\textit{HTML\_anchors}); but also others. We implement \atks{PA}{r} by setting the following features (from Table~\ref{tab:features}) to -1: (36--38,41,51,54,56,57); whereas we set (39,40) to 1 and 46 to 0.
    \item \atks{PA}{c}: they are a combination of the two above. We expect the attacker to use a URL shortener, and also infterfer with the HTML\_objectRatio. However, we cannot simply set the features to the same values as \atks{PA}{r} and \atks{PA}{u}, because one of the two will prevail. In our case, shortening the URL will be `stronger', because the URL will change (to that of the URL shortener) and hence the internal objects will become `external'. Hence, we implement \atks{PA}{c} by setting the following features (from Table~\ref{tab:features}) to -1: (1-3,5,6,8,10-16,22,23,25,26,28-30), whereas the following features are set to +1: (4,27,36-38,41,44,48,52,54,56).
\end{itemize}
We remark that our PsP may not yield an $\overbar{F_p}$ that is a perfect match with a $F_{\overbar{p}}$ generated via WsP (i.e., those of \aatks{}). Indeed, some inconsistencies may be present---likely due to `inaccurate' anticipations from our (i.e., the attacker's) side.
Such inconsistencies are sensible. An attacker with access to the preprocessing-space can theoretically \textit{replicate} the entire feature extractor, and use it to exactly pinpoint how to generate PsP that are an exact match with WsP (i.e., $\overbar{F_p}$=$F_{\overbar{p}}$). However, doing this would be \textit{very expensive}. Furthermore, it would \textit{defeat the purpose} of using PsP: the attacker does not want that $\overbar{F_p}$=$F_{\overbar{p}}$, rather, they want a PsP that is `stronger'; otherwise, why use PsP in the first place?

\section{Results and Discussion}
\label{sec:results}
We present the results of our evaluation. We aim at answering two questions:
\begin{itemize}
    \item (§\ref{ssec:wa}) how dangerous are the most likely attacks (i.e., \atk{WA})?
    \item (§\ref{ssec:comparison}) what is the effectiveness of attacks carried out in different evasion spaces (i.e., \aatk{}, \atk{PA}, \atk{MA})?
\end{itemize}
We also perform a proof-of-concept experiment on a competition-grade ML-PWD (§\ref{sec:mlsec}). Finally, we discuss our evaluation and potential for future work in §\ref{ssec:limitations}. We report our full `benchmark' results in Appendix~\ref{app:benchmark}.

\subsection{Effectiveness of the most likely attacks (\atk{WA})}
\label{ssec:wa} 

\begin{figure*}[!htbp]
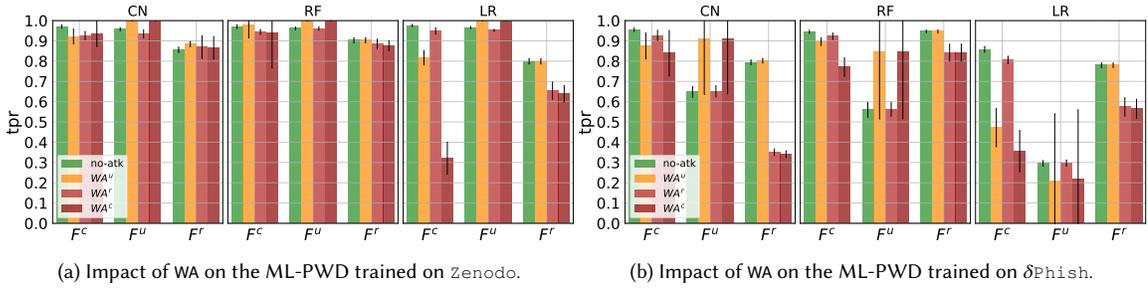

    \centering
    \begin{subfigure}[t]{0.5\textwidth}
        \centering
        \includegraphics[width=1\columnwidth]{figures/results/base_compare_wa_zenodo.pdf}
        \caption{Impact of \ftatk{WA} on the ML-PWD trained on \footnotesize{\dataset{Zenodo}}.}
         \label{sfig:wa_zen}
    \end{subfigure}%
    ~ 
    \begin{subfigure}[t]{0.5\textwidth}
        \centering
        \includegraphics[width=1\columnwidth]{figures/results/base_compare_wa_delta.pdf}
        \caption{Impact of \ftatk{WA} on the ML-PWD trained on \footnotesize{\dataset{$\delta$Phish}}.}
         \label{sfig:wa_delta}
    \end{subfigure}
    \vspace{-3mm}
    \caption{Effectiveness of the most likely attacks (\ftatk{WA}). The three plots in each subfigure represent the algorithm used by a specific ML-PWD. Each plot has bars divided in three groups, each denoting a specific $F$ used by the ML-PWD. The green bars show the \ftmath{tpr} on the original samples, while the others show the \ftmath{tpr} against a specific variant of \ftatk{WA}.}
    \label{fig:wa}
    \vspace{-1em}
\end{figure*}

Let us focus the attention on the most likely attacks. We report in Figs.~\ref{fig:wa} the \smamath{tpr} achieved by all our ML-PWD against all our \atk{WA} attacks (red bars), and compare it with the \smamath{tpr} (\textit{no-atk}, shown in green bars) achieved by the same ML-PWD on the original set of samples used as basis for \atk{WA}. Some intriguing phenomena occur.

\textbf{True Baseline ($F^c$)}. We first consider the ML-PWD using $F^c$ (leftmost group of bars in each plot), since they are the ML-PWD with the best performance in the absence of attacks (cf. Table~\ref{tab:normal}). 
\begin{itemize}
    \item On \dataset{$\delta$phish} (Fig.~\ref{sfig:wa_delta}), all ML-PWD are affected by the `strongest' cheap attack, i.e., \atks{WA}{c}. Specifically, the ML-PWD using \smamath{LR} is completely defeated (from 0.86 \smamath{tpr} down to 0.36); in contrast, those using \smamath{CN} or \smamath{RF} suffer a smaller, but still significant drop (from nearly 0.95 down to $\sim$0.8). Notably, the \smamath{CN} despite being worse than the \smamath{RF} in non-adversarial settings (cf. Table~\ref{tab:normal}), appears to be slightly more robust.
    
    \item The situation is different on \dataset{Zenodo} (Fig.~\ref{sfig:wa_zen}). Here, while the \smamath{LR} is still defeated, the \smamath{CN} and \smamath{RF} appear not to be very affected by \atks{WA}{c}. However, considering that both \smamath{CN} and \smamath{RF} exhibit very high performance in non-adversarial settings (cf. Table~\ref{tab:normal}), it is crucial to determine whether \atks{WA}{c} poses a real threat to such ML-PWD. To this purpose, we carry out a Welch t-test, which we can do thanks to our large amount of trials. We set our null hypothesis as ``\atks{WA}{c} and \textit{no-atk} are equal''. The findings are valuable: against \smamath{RF}, the \smamath{p}\textit{-value} is 0.221; whereas against \smamath{CN}, the \smamath{p}\textit{-value} is 0.002. By using the common statistical significance threshold of 0.05, we can hence provide the following answer: the \smamath{RF} \textit{is not affected} by \atks{WA}{c}, whereas the \smamath{CN} \textit{is affected} by \atks{WA}{c}.
\end{itemize}
The latter finding is intriguing, because it suggests that \textit{shallow learning methods can be more resilient} than deep learning ones for PWD---against our proposed attacks. Finally, we also observe that \atks{WA}{r} clearly defeat \smamath{LR} on both datasets, whereas the impact on \smamath{RF} and \smamath{CN} is significant on \dataset{$\delta$Phish}, but small on \dataset{Zenodo}.

\textbf{Robust Baselines ($F^u$, $F^r$).} 
The robust baselines are, in general, reliable against \atk{WA}. The ML-PWD using $F^u$ counter \atks{WA}{r} (and viceversa), because the \smamath{tpr} is exactly the same as the original one. Notably, however, ML-PWD using $F^r$ (similar to the ML-PWD of\footnote{We also successfully attacked the competition-grade ML-PWD of~\ref{sec:mlsec} with \ftatks{WA}{r}, achieving similar results than the one shown in our custom-built ML-PWD. A demonstrative video (of 140s) can be found at the \href{https://spacephish.github.io}{homepage} of our website.} MLSEC~\cite{mlsec}) are affected by \atks{WA}{r}: the \smamath{LR} is clearly defeated on both datasets, whereas \smamath{RF} suffers a 10\% and 3\% drop on \dataset{$\delta$phish} and \dataset{Zenodo}, respectively.
Nevertheless, we observe a fascinating phenomenon: in some cases, the \smamath{tpr} under attack \textit{is higher} than in \textit{no-atk}; e.g., on \dataset{$\delta$phish} the \smamath{RF} analyzing $F^u$ has its \smamath{tpr} to increase from 0.56 to $\sim$0.84 against both \atks{WA}{u} and \atks{WA}{c}. Such phenomenon occurs because the attacker (in any variant of \atk{WA}) does not know `what to do' to reliably evade the ML-PWD: the attacker guesses some WsP, which can have no impact, or even make the website closer to a `malicious' one (from the viewpoint of \smacal{M}). 

\vspace{-2mm}

\takeaway{The realistic attacks in the website-space (\atks{WA}{c}) can evade five (out of six) ML-PWD. Despite being small, the performance degradation is statistically significant: hence, due to their cheap cost, \atks{WA}{c} represent a threat to state-of-the-art ML-PWD.}

\begin{figure*}[!htbp]
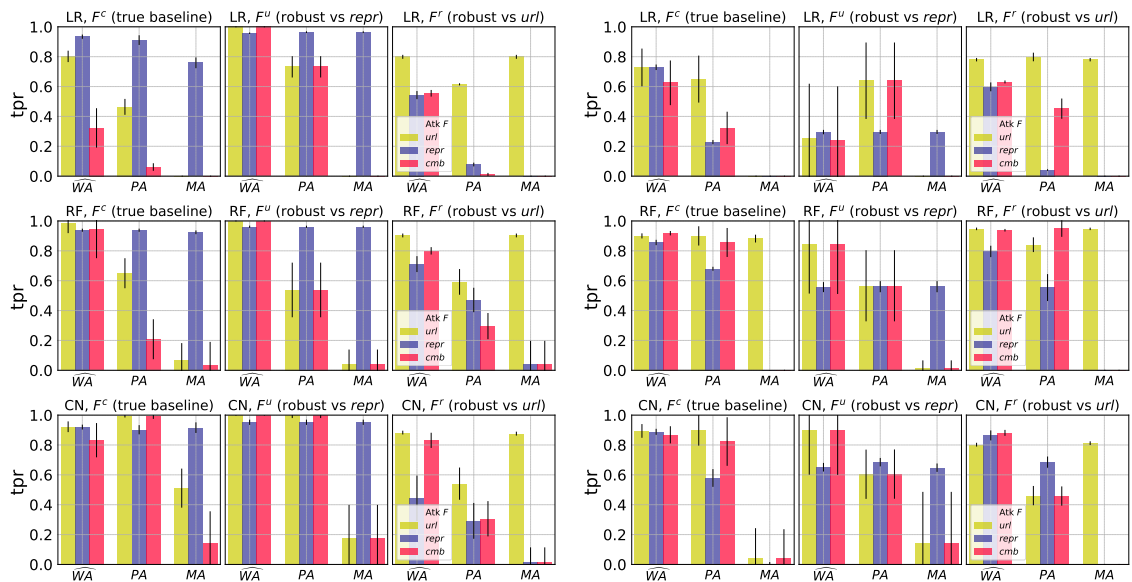

    \centering
    \begin{subfigure}[t]{0.49\textwidth}
        \centering
        \includegraphics[width=0.98\columnwidth]{figures/results/atk_results_zenodo_change_MA.pdf}
        \caption{\dataset{Zenodo}. Each plot reports the \ftmath{tpr} resulting from the 9 advanced attacks (i.e., \ftaatk{}, \ftatk{PA}, \ftatk{MA}) across the 50 trials. Colors denote the targeted features (\ftmath{u}, \ftmath{r}, \ftmath{c}).}
         \label{sfig:comparison_zenodo}
    \end{subfigure}
    \hfill
    \begin{subfigure}[t]{0.49\textwidth}
        \centering
        \includegraphics[width=0.98\columnwidth]{figures/results/atk_results_delta_change_MA.pdf}
        \caption{\dataset{$\delta$phish}. Each plot reports the \ftmath{tpr} resulting from the 9 advanced attacks (i.e., \ftaatk{}, \ftatk{PA}, \ftatk{MA}) across the 50 trials. Colors denote the targeted features (\ftmath{u}, \ftmath{r}, \ftmath{c}).}
         \label{sfig:comparison_deltaphish}
    \end{subfigure}
    \vspace{-1em}
    \caption{Comparison of attacks carried out in different evasion-spaces. Each subfigure refers to a specific dataset, and presents 9 plots. Such plots are organized in three rows and three columns. Rows denote a specific ML algorithm (\ftmath{LR}, \ftmath{RF}, \ftmath{CN}). Columns denote a specific feature set: the `true' baseline (using $F^c$) is on the left; the others are the `robust' baselines (using $F^u$ or $F^r$).}
    \label{fig:comparison}
    \vspace{-1em}
\end{figure*}

\subsection{Comparing the evasion-space (\aatk{}, \atk{PA}, \atk{MA})}
\label{ssec:comparison}

We now focus on comparing the effectiveness of attacks that aim at influencing the same features (i.e., either \smamath{u}, \smamath{r}, \smamath{c}), but whose perturbations are introduced in different spaces (i.e., either WsP, PsP, or MsP).
We visualize such results in Fig.~\ref{fig:comparison}.

The `true' baselines (using $F^c$, i.e., the leftmost plots in Fig.~\ref{fig:comparison}) are defeated by \atk{MA}. However, there are some notable exceptions: on \dataset{Zenodo}, the \smamath{RF} and \smamath{CN} are resilient to \atks{MA}{r} (this is because the HTML features have little importance for $F^c$). In contrast, on \dataset{$\delta$phish}, \smamath{RF} can withstand \atks{MA}{u}. The `robust' baselines counter the corresponding \atk{MA}, but unsurprisingly suffer against the others.

In general, \atk{PA} tend to have a larger impact than \aatk{} against the `true' baselines. However, this is not always true: we find enlightening that the \smamath{CN} on \dataset{Zenodo} is more robust to \atk{PA} than to \aatk{}. What is even more surprising is that such \smamath{CN} significantly outperforms the \smamath{RF} against \atk{PA}, but \textit{also} against \atk{MA}. Such finding could inspire deployment of ML-PWD using deep learning on \dataset{Zenodo}---despite being inferior to \smamath{RF} in the no-atk (Table~\ref{tab:normal}) and against \atks{WA}{c} (§\ref{ssec:wa}). 

We note that \aatks{u} perfectly match \atks{WA}{u}, which makes sense as they involve exactly the same WsP (cf. §\ref{sec:attacks}). We can also see some discrepancies between \aatk{} and \atk{PA}: as a matter of fact, our anticipation of the preprocessing-space (i.e., the PsP of \atk{PA}) did not exactly match what truly happened in the website-space . However, in some cases (e.g., the \smamath{RF} using $F^c$ and $F^r$ on \dataset{$\delta$phish}) we observe that the effectiveness of \aatk{} and \atk{PA} tend to be similar. Such crucial finding demonstrates that perturbations applied directly to $F_x$ (which we use for \atk{PA}) can induce the same effects as those applied to $x$ (which we use for \aatk{}). In other words: if properly crafted, then even perturbations in the ``feature-space'' can resemble adversarial examples that are physically realizable~\cite{tong2019improving}. 

Let us compare our attacks with those considered by \dataset{$\delta$phish} creators. Specifically, the attacks in~\cite{Corona:Deltaphish} manipulate increasingly higher amounts of features (up to 10), and all ultimately evade target ML-PWD (which analyzes the HTML). Such finding is confirmed by our results on the ML-PWD analyzing $F^r$ on \dataset{$\delta$phish} against \atks{MA}{r}, which all misclassify the adversarial samples. However, \textit{if the perturbations are applied in different spaces} (i.e., PsP or WsP), \textit{then the ML-PWD is significantly less affected.}

\subsection{Discussion}
\label{ssec:limitations}

Our evaluation is a proof-of-concept, and we do not claim that \textit{all} ML-PWD will respond in the same way as ours, and neither we claim novelty in the `generic' method used to to evade PWD (attackers have been manipulating the HTML or URL for decades~\cite{Biggio:Wild}).
Indeed, our goal was to validate our primary contribution (whose focus is on machine learning) by performing a fair comparison of attacks (each having a different \textit{cost}) in diverse evasion-spaces.

\textbf{Warning on \atk{$\mathbf{WA}$}.} A legitimate observation is that our cheap attacks, despite affecting most ML-PWD, have a small impact---even if statistically significant (§\ref{ssec:wa}). Such results, however, must not induce conclusions such as ``these attacks are not interesting'' or (worse) ``these attacks can be overlooked in the security lifecycle''. Indeed, \textit{the main threat of} \atk{WA} \textit{is represented by the cheap cost}: thousands of phishing websites are created every day~\cite{proofpoint2022phish}, and in such big numbers even a 1\% difference can be the separation between a compromised and secure system~\cite{apruzzese2022role}. Our goal is not to propose devastating attacks that bypass any ML-PWD; rather, we focus on those attacks that are more likely to occur in reality. As a matter of fact, \atk{WA}s \textit{can be automatized and implemented within seconds and few lines of code}; in contrast, the advanced attacks (including those of past work, e.g.,~\cite{Corona:Deltaphish, liang2016cracking}) require to compromise or reverse-engineer the ML-PWD (§\ref{ssec:analysis}). The \textit{cost} of an attack should also account for the \textit{effort} required for its implementation. Most related literature focuses on measuring `queries' (e.g.,~\cite{Demontis:Adversarial}): our \atk{WA} do not require any query. Nonetheless, we invite future work to explore metrics to estimate the cost of attacks in terms of human effort.

\textbf{Future Work.}
The main purpose of our evaluation is to highlight how state-of-the-art ML-PWD respond to diverse evasion attacks. There are, however, millions of ways to do the above. For instance, the attacks can target different features (and in different ways) than the ones considered in our evaluation (i.e., \smamath{u}, \smamath{r}, \smamath{c}); the ML-PWD can analyze different features, which can be generated via different preprocessing mechanisms (e.g.,~\cite{kondracki2021catching}). Additional defenses can also be considered (e.g., adversarial training~\cite{tramer2018ensemble, pang2020advmind}).
For instance, we did not consider ML-PWD that analyze the visual representation of a webpage (e.g.,~\cite{abdelnabi2020visualphishnet, lin2021phishpedia}): such attacks would resemble those conducted in computer vision, which are well-known to be effective (e.g.,~\cite{Papernot:SoK, tramer2019adversarial}). Nevertheless, our threat model is agnostic of the data-type, so we endorse future work to also consider ML-PWD analyzing images. Finally, our evasion-space formalization can be applied even to settings beyond phishing (e.g., malware), which may entail attackers more likely to use PsP or MsP.

\begin{figure*}[!htbp]
    \centering
    \begin{subfigure}[t]{0.45\textwidth}
        \centering
        \includegraphics[width=1\columnwidth]{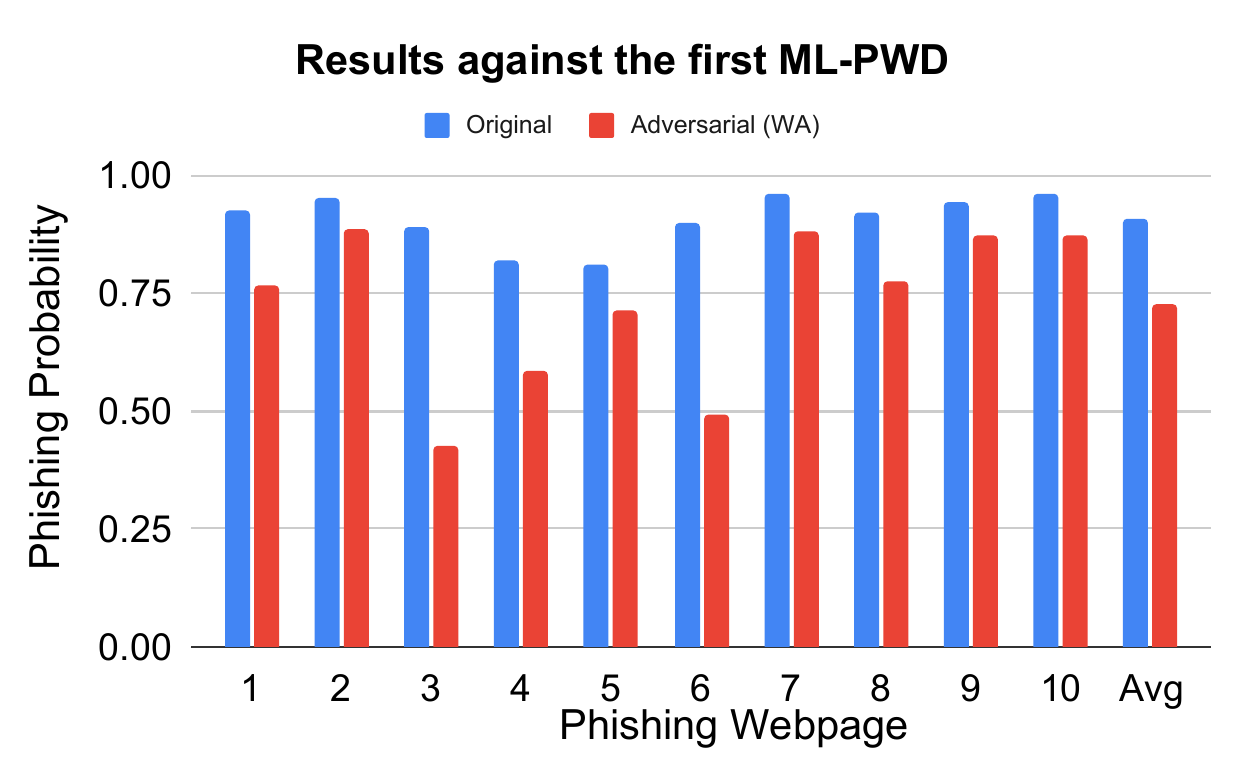} 
        \caption{Impact of \ftatks{WA}{r} on the \textit{first} ML-PWD.}
         \label{sfig:pwd1}
    \end{subfigure}%
    ~ 
    \begin{subfigure}[t]{0.45\textwidth}
        \centering
         \includegraphics[width=1\columnwidth]{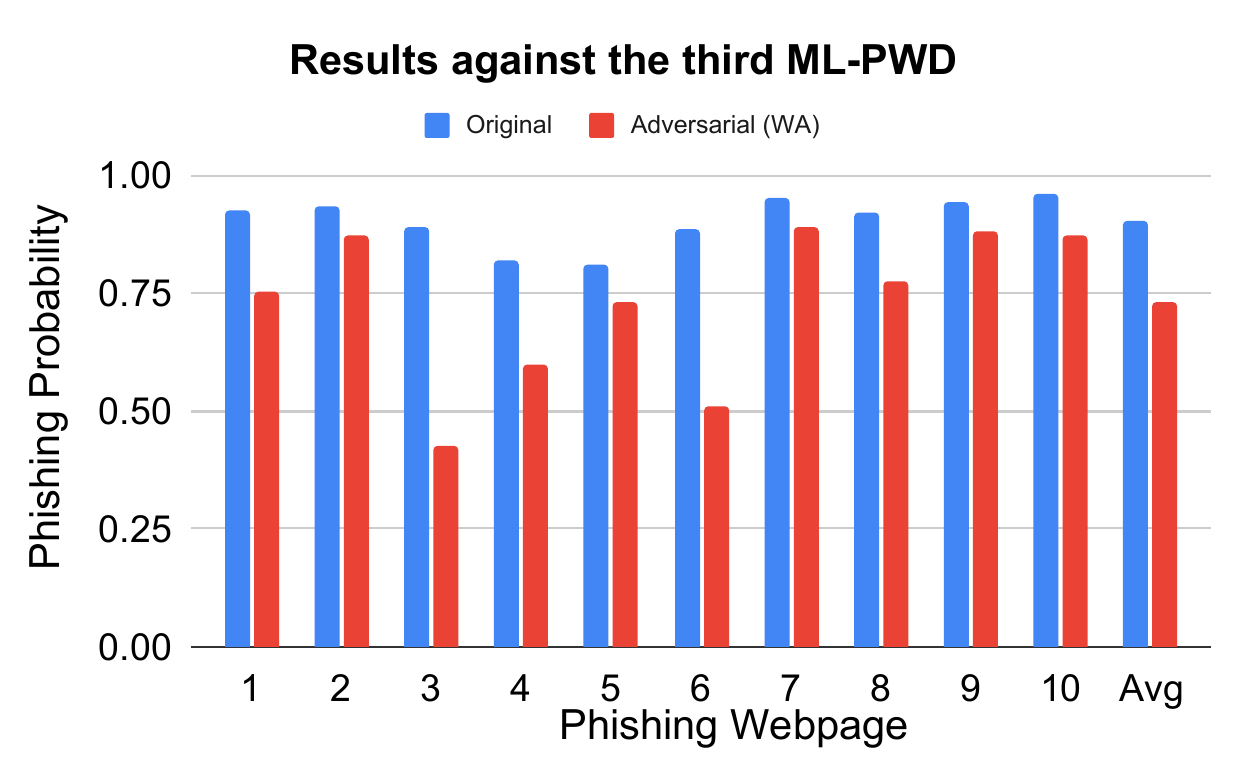}
        \caption{Impact of \ftatks{WA}{r} on the \textit{third} ML-PWD.}
         \label{sfig:pwd3}
    \end{subfigure}
    \caption{Effectiveness of the most likely attacks (\ftatks{WA}{r} on \dataset{$\delta$Phish}) against the ML-PWD provided by the organizers of MLSEC~\cite{mlsec}.}
    \label{fig:mlsec}
\end{figure*}

\subsection{Proof-of-concept: attacks against a competition-grade ML-PWD}
\label{sec:mlsec}

To further prove the impact of our `cheap' attacks (i.e., \atk{WA}), we tested them on a real ML-PWD that is used in a well-known Machine Learning Security Evasion Competition (MLSEC~\cite{mlsec}). Such competition is held yearly, and is organized by leading tech-companies that provide cybersecurity services reliant on ML methods. The 2022 edition of MLSEC envisions a challenge in which participants are asked to \textit{evade} ML-PWD. We took this opportunity to assess whether our attacks had any impact against such `competition-grade' ML-PWD. Short story: they do. A demonstrative video can be found at the \href{https://spacephish.github.io/}{homepage} of our website (which also includes the source-code).

\subsubsection{\textbf{Challenge}}
\label{ssec:context}
Participants of the phishing evasion challenge are given 10 `phishing' webpages, which are provided in their raw HTML form. The purpose of the challenge is to manipulate such webpages so that (i)~they render exactly as the originals, and (ii)~they evade a ML-PWD. Specifically, the organizers provide 8 different ML-PWD, which the participants can use as a black-box: by sending an input (i.e., the HTML of a phishing webpage), they are given an output (i.e., the probability that such webpage is malicious---according to the specific ML-PWD). Such ML-PWDs only analyze the HTML of the webpage (which must render exactly as the original).
Put simply: the objective of the challenge is to tweak the HTML of the 10 webpages with imperceptible modifications that decrease the confidence of the 8 ML-PWD.

\subsubsection{\textbf{Method}}
\label{ssec:method}
Of course, the setting described above perfectly describes the black-box scenarios envisioned in adversarial ML papers: query the detector, and use the response as a guide to craft a more evasive phishing webpage. Our primary attacks (\atk{WA}), however, are query-less. Because we are aware that the target ML-PWD analyzes the HTML (recall that this is an assumption of our threat model), we then craft our `adversarial' phishing webpages by using exactly the same \atks{WA}{r} used in our paper for \dataset{$\delta$Phish}: we add 50 invisible internal links. We apply these WsP to all the 10 webpages \textit{provided by the organizers of the challenge}, and then test whether they had any impact to the \textit{real} ML-PWD involved in the challenge.

\subsubsection{\textbf{Results}}
\label{ssec:results}
By taking into account \textit{all} webpages against \textit{all} ML-PWD, our attacks induced a drop of 3.4\% in the confidence of the ML-PWD, indicating that our WsP had some effect. However, while some ML-PWD were not very affected, others incurred a significant drop. 
Specifically, we focus our attention on the first and third ML-PWD provided by the organizers of MLSEC. The results of our proof-of-concept experiments are shown in Figs.~\ref{fig:mlsec}. These graphs show phishing probability (y-axis) given as output by the corresponding ML-PWD for each of the 10 webpages of the challenge (x-axis). We report two bars: the blue bar are the results of the original webpages, whereas the red bars are the results after applying our WsP.

\subsubsection{\textbf{Analysis}}
\label{ssec:analysis-mlsec}
These two detectors were significantly less certain after our WsP, with an average confidence drop of 17.5\%. We observe that in most cases, the confidences were still above 0.5 (i.e., the webpages would still be classified as `phishing'). A more detailed look, however, reveals that \textbf{these detectors were completely fooled by some webpages} (i.e., their confidence dropped to below 0.5). We report:
\begin{itemize}
    \item Page \#3: from 0.90 down to 0.43 for the 1st and 3rd detectors.
    \item Page \#6: from 0.90 down to 0.49 for the 1st detector.
\end{itemize}
We also attempted the same \atks{WA}{r} by changing the number of fake links, and also by considering a different string\footnote{We also considered the `wrapping' WsP for \dataset{Zenodo}: the effects were negligible---probably because these ML-PWD factored such links into their `count' (i.e., the attacker made a wrong guess). See Appendix~\ref{app:alternative_wspr}}. When applied to, e.g., webpage \#3, adding 280 links dropped the confidence to below 0.2; whereas adding a slightly different string (the first one shown in our pragmatic example in Appendix~B) 280 times, the confidence dropped to 0.2 for the first and third detector, and to 0.49 for the seventh detector. The seventh detector was also fooled by adding such alternative string 50 times to webpage \#4, causing a confidence of 0.46 (down from 0.68). The source-code is available in our repository, and the experiments are entirely reproducible.
Interestingly, these results align with those shown in our primary evaluation: our \textit{query-less} \atk{WA} attacks cannot bypass any ML-PWD, but in some cases \textit{they can induce a miss-classification}.

\section{Additional Experiments: Same-space and Mixed-Space}
\label{sec:new}
We \textit{expand} the evaluation carried out for our ACSAC'22 paper~\cite{apruzzese2022spacephish} with additional experiments. Our goal is twofold: 
\begin{itemize}
    \item assessing other types of perturbations (either WsP, PsP or MsP) \textit{in the same space};
    \item consider a ``stronger'' attacker that applies \textit{multiple} perturbations also in \textit{different} spaces (cf. §\ref{subsec:extensions}).
\end{itemize}
We first describe (§\ref{subsec:same-descr}) and empirically evaluate (§\ref{subsec:same-eva}) the attacks entailing perturbations in the ``same-space''. Then, we describe (§\ref{subsec:mixed-descr}) and evaluate (§\ref{subsec:mixed-eva}) the ``multi-space'' attacks.

\subsection{Same-space Attacks: Description}
\label{subsec:same-descr}
In this section, we elaborate on new attacks in the same evasion space involving our WsP, PsP, and MsP. Building upon the attacks considered in the main evaluation (§\ref{sec:attacks}), we introduce additional perturbations. The motivation behind this extension is to present a more comprehensive range of use cases---all of which are likely to happen, since they are well within the attacker's capabilities (who will never have complete knowledge of the target PWD). Therefore, we explore novel perturbations of the HTML (§\ref{subsubsec:same_html}) and URL (§\ref{subsubsec:same_url}), as well as introduce new variations of MsP, PsP, and WsP. Altogether, the details of the new specific attacks are provided in Table~\ref{tab:atk_types}.

\subsubsection{\textbf{HTML}} 
\label{subsubsec:same_html}
As we know (§\ref{sec:background}), the HTML reflects the visual appearance of a webpage---therefore, changes to the HTML can lead to differences in the way the webpage is presented to its users.\footnote{We recall that our threat model does not assume that the perturbations are ``imperceptible'' to humans. This is because, in a real scenario, phishing is effective because humans are distracted. Hence, even if the webpage changes, the phishing attack can still be successful.} Some of them may be noticed by users (e.g., alterations of the background), while others may not change the appearance at all (e.g., the hidden links considered in our pragmatic use-case §\ref{ssec:pragmatic}). Here, we consider a wide-array of HTML-related perturbations, and scrutinize which are more likely to evade the detection of PWD.
Practically, we propose a total of 37 new HTML-related perturbations---of which, 24 are WsP (i.e., new \atks{WA}{r}), which can be divided into the three following categories:
\begin{enumerate}[label=\arabic*)]
    \item \textit{iWsP} (invisible WsP), which denote perturbations that are inserted into the webpages but remain invisible to users. This means that the webpage appears unchanged before and after the perturbation insertion.
    \item \textit{eWsP} (elusive WsP), which introduce slight changes to the appearance of the webpage. While these changes may require some effort to be noticed by users, they are still discernible upon careful observation.
    \item \textit{rWsP} (recognizable WsP), which result in changes that are clearly visible by users. These modifications have a more pronounced impact on the webpage's appearance, making them readily noticeable.
\end{enumerate} 
The remaining 13 HTML-related perturbations are PsP and MsP (i.e., new \atks{PA}{r} and \atks{MA}{r}). Both of which require write-access to the ML-PWD.
\atks{PA}{r} can bypass some of the checks of ML-PWD. Moreover, in \atks{MA}{r}, attackers may solely focus on evading ML-PWD: as a result, some \atks{MA}{r} might violate the fundamental rules of HTML.

\subsubsection{\textbf{URL}} 
\label{subsubsec:same_url}
Domain and path are two essential components of URL, and most of our URL features in Table~\ref{tab:features} are extracted from them. In this section, we implemented 6 types of perturbations that specifically target the URL. These perturbations, referred to as \atks{WA}{u}, the specific details are provided below.

\begin{itemize}
  \item \textit{replChar}, we replaced the characters in the domain with visually similar characters. 
  \item \textit{sepWrd}, we randomly inserted space within the domain to separate the individual word.
  \item \textit{delChar}, we deleted one character from the domain. 
  \item \textit{swpChar}, we randomly swapped two adjacent characters in domain. 
  \item \textit{addChar}, we randomly inserted an additional character into the domain.
  \item \textit{atkPth},  we also conducted operations of swap, delete, or insert randomly within the path of the URL.
\end{itemize}
We do not consider URL-related perturbations that affect other spaces (i.e., PsP or MsP).

\begin {table}[!htb]\centering
\caption {New Attacks for HTML } \label{tab:title} 
\label{tab:atk_types}
\resizebox{\columnwidth}{!}{
\begin{tabular}{ c|c|c}
\hline
 Category & Perturbation & Description \\
\hline
\multirow{2}{*}{\atks{iWA}{r}} & \textit{addInLnk} & insert internal links  \\ 
& \textit{replOnc} & $\langle a \ href=` link\ ' \rangle$ to $\langle a \ onclick=``this.href=`link \ '"\rangle$  \\ 
& \textit{delHidIt} & delete hidden items from HTML \\ 
& \textit{addHidP} & add hidden large page  \\
& \textit{replJS} & replace  $\langle a \ href=`\#' \rangle$ with $\langle a \ href=`javascript:void(0)'\rangle$ \\
& \textit{replRet} & replace `$\backslash$n' with whitespace  \\
& \textit{htEsc} & escape the whole body content, and write ``document.write(unescape(`\ '))"  to HTML  \\
& \textit{htEncd} & encode HTML with base64 \\
& \textit{replPass} & replace $\langle input \ type=`password' \rangle$ with $\langle input \ type=`text'\ \rangle$  \\
& \textit{replOnfoc} &replace $\langle input \ type=`password'/`email' \rangle$  with  $\langle input \ onfocus=``this.type=`password'/`email'" \rangle$ \\
& \textit{addSusLnk} & add suspicious links $\langle a \rangle$, e.g., $\langle a \ href=`\#skip'\rangle$ \\ 
\hline
\multirow{2}{*}{\atks{eWA}{r}} & \textit{addImgBot} & insert 20 small local images to the webpage bottom  \\ 
& \textit{modFntTyp} & modify the font type italic  \\ 
& \textit{addTps} & randomly insert few typos into HTML text\\
& \textit{modCpy} & modify copyright \\
& \textit{addIcn} & add local icon  \\
& \textit{delSusLnk} & delete suspicious links   \\
& \textit{delSusFrm} & delete suspicious form (i.e., with empty or external `action' links) \\
& \textit{modTtl} & randomly modify the title  \\
& \textit{delCpy} & delete copyright information from HTML \\
\hline
\multirow{2}{*}{\atks{rWA}{r}} & \textit{modBgimg} & change the background image    \\ 
& \textit{modBgClr} & randomly change the background color  \\ 
& \textit{modFntClr} & randomly modify the font color  \\ 
& \textit{modFntSiz} & modify the body font size to 0\\
\hline
\multirow{2}{*}{\atks{PA}{r}} & \textit{delTxt} &delete all text from HTML  \\ 
& \textit{delFrm} & remove forms  \\ 
& \textit{delSpn} & remove all span\\
& \textit{delTtl} & remove title\\
& \textit{addLngTxt} & add long visible text to HTML\\
& \textit{delFtr} & remove footer\\
& \textit{replSusFtrLnk} & replace suspicious links of footer with internal links\\
\hline
\multirow{2}{*}{\atks{MA}{r}} & \textit{brTg} & break the tag $\left\langle html\right\rangle$  \\ 
& \textit{delHt} & remove the whole html\\
& \textit{delHd} & delete the whole $\langle head \rangle$ except style  \\ 
& \textit{delBdy} & delete the whole $\left\langle body\right\rangle$  \\ 
& \textit{brTgs} & break tags  \\
& \textit{hmg} & replace characters with homographic letters \\
\hline
\end{tabular}}
\end{table}

\subsection{Same-space Attacks: Evaluation}
\label{subsec:same-eva}
We now assess the impact of the abovementioned perturbations. For HTML perturbations (§\ref{subsubsec:same-eva-html}), we consider the effects both on the ML-PWD we developed by using the \dataset{$\delta$Phish} and \dataset{Zenodo} datasets, as well as by those provided by MLSEC (we carried out these experiments in December 2022, when the MLSEC API was still open for research purposes). For the URL perturbations (§\ref{subsubsec:same-eva-url}) we consider only the ML-PWD trained on \dataset{$\delta$Phish} and \dataset{Zenodo} because those provided by MLSEC do not consider the URL in their analyses.

\subsubsection{\textbf{Impact of HTML perturbations}} 
\label{subsubsec:same-eva-html}
We begin by considering \dataset{$\delta$Phish}, \dataset{Zenodo}, and then focus on MLSEC. 

\textbf{$\delta$Phish} and \textbf{Zenodo}. In Figs.~\ref{fig:new_wsp}, we present the $tpr$ achieved by ML-PWD trained on \dataset{$\delta$Phish} and \dataset{Zenodo}. We evaluate the performance of these ML-PWD against \atks{iWA}{r}, \atks{eWA}{r} and \atks{rWA}{r} (represented by yellow and red bars)\footnote{Our figures only present the most effective \atk{WsP}, i.e., \atks{iWA}{r} denotes \textit{addHidP}, \atks{eWA}{r} stands for \textit{addImgBot}, and \atks{rWA}{r} represent \textit{modFntClr}.}. To provide a comparison, we also include the $tpr$ achieved by the same ML-PWD on the original set of samples, depicted by the leftmost green bar labeled as ``no-atk". These results aim to address two key questions:

\begin{itemize}
    \item Will different \atk{WsP} have different impacts on ML-PWD and how?
    \item What kind of \atk{WA} is more likely to evade the ML-PWD trained on \dataset{$\delta$Phish} and \dataset{Zenodo}?
\end{itemize}

As shown in Fig.~\ref{sfig:wsp_delta}, the \atks{iWA}{r} perturbation emerges as the most impactful attack, leading to a significant reduction (reduced by $0.68$--$0.95$) in the $tpr$ of $F^r$- and $F^c$-based ML-PWD trained on \dataset{$\delta$Phish}. Specifically, the $tpr$ of RF-PWD trained on $F^c$ drops from $0.945$ to $0.037$, and the $tpr$ of RF-PWD trained on $F^r$ decreases from $0.947$ to $0$. In comparison, the influence of \atks{eWA}{r} and \atks{rWA}{r}is relatively smaller. However, \atks{eWA}{r} still causes a notable drop in the $tpr$ of $F^r$-based LR-PWD, reducing it from $0.78$ to $0.47$. On the other hand, \atk{rWsP} has minimal impact on PWD (only $F^c$-based LR-PWD's $tpr$ decreased by $0.12$). A similar trend is observed in Fig.~\ref{sfig:wsp_zenodo} for the influence on \dataset{Zenodo}, where \atks{iWA}{r} remains the most effective attack. Additionally, \atks{eWA}{r} affects ML-PWD to a greater extend (except for $F^r$-based LR-PWD) compared to \atks{rWA}{r}.
These findings demonstrate that \atks{iWA}{r} poses the greatest challenge to ML-PWD of \dataset{$\delta$Phish} and \dataset{Zenodo}, significantly reducing their detection performance. \atks{eWA}{r} also has a notable impact, while \atks{rWA}{r} has a relatively minor effect on most ML-PWD (except for the ML-PWD using LR to analyze $F^r$).

\begin{figure}[htbp]
    \centering
    \begin{subfigure}[t]{1\columnwidth}
        \centering
        \includegraphics[width=0.8\columnwidth]{figures/new_atks/newatk_wsp_delta.pdf}
        \caption{ Impact of new \atks{WA}{r} (i.e., \atks{iWA}{r}, \atks{eWA}{r} and \atks{rWA}{r}) on ML-PWD trained on \dataset{$\delta$phish}}  
        \label{sfig:wsp_delta}
    \end{subfigure}
    \begin{subfigure}[t]{1\columnwidth}
        \centering
        \includegraphics[width=0.8\columnwidth]{figures/new_atks/newatk_wsp_zenodo.pdf}
        \caption{Impact of new \atks{WA}{r} (i.e., \atks{iWA}{r}, \atks{eWA}{r} and \atks{rWA}{r}) on ML-PWD trained on \dataset{Zenodo}}
        \label{sfig:wsp_zenodo}
    \end{subfigure}
    \caption{Effectiveness of the most likely new attacks \atks{WA}{r}. The three plots in each subfigure represent the algorithm used by a specific ML-PWD. Each plot has bars divided in three groups, each bar denotes a specific $F$ used by the ML-PWD. The green bars show the $tpr$ on the original samples, while the others show the $tpr$ against a specific variant of \atk{WA}.}
    \label{fig:new_wsp}
    \vspace{-1.5em}
\end{figure}

\begin{figure}[htbp]
    \centering
    \begin{subfigure}[t]{1\columnwidth}
        \centering
        \includegraphics[width=0.8\columnwidth]{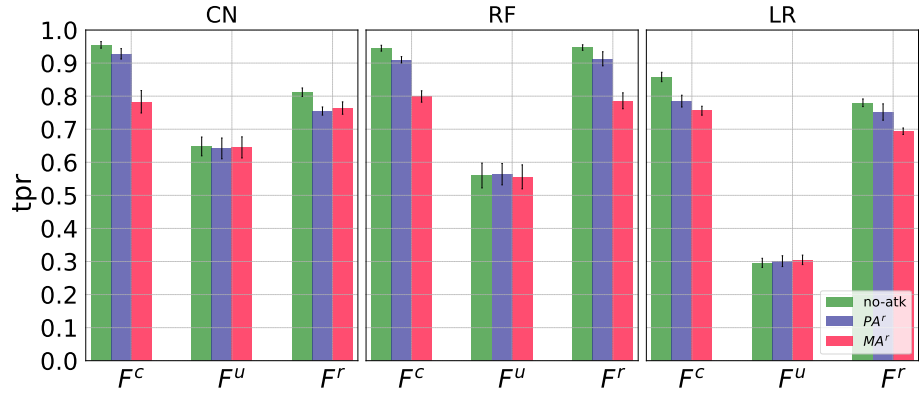}
        \caption{ Impact of new \atks{PA}{r} and \atks{MA}{r} on ML-PWD trained on \dataset{$\delta$phish}}  
        \label{sfig:psp_msp_delta}
    \end{subfigure}
    \begin{subfigure}[t]{1\columnwidth}
        \centering
        \includegraphics[width=0.8\columnwidth]{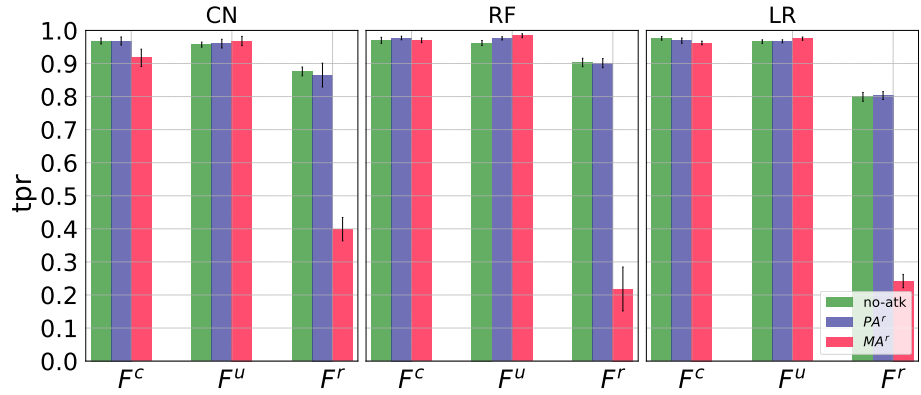}
        \caption{Impact of new \atks{PA}{r} and \atks{MA}{r} on ML-PWD trained on \dataset{Zenodo}}
        \label{sfig:ps_msp_zenodo}
    \end{subfigure}
    \caption{Effectiveness of new attacks \atks{PA}{r} and \atks{MA}{r}. The three plots in each subfigure represent the algorithm used by a specific ML-PWD. Each plot has bars divided in three groups, each denoting a specific $F$ used by the ML-PWD. The green bars show the $tpr$ on the original samples, the blue bars represent $tpr$ against \atks{PA}{r} and the red bars in the rightmost show the $tpr$ against \atks{MA}{r}.}
    \label{fig:new_psp_msp}
    \vspace{-1em}
\end{figure}

Figs.~\ref{fig:new_psp_msp} represents the impact of new \atks{PA}{r} and \atks{MA}{r} on ML-PWD trained on \dataset{$\delta$Phish} and \dataset{Zenodo}. In this context, \atks{PA}{r} refers to \textit{delFrm} (i.e., remove forms from the webpage), while \atks{MA}{r} denotes applying perturbation \textit{hmg} to HTML (i.e., inserting typos to the HTML, both tags and text). Comparing  with the $tpr$ of `no-atk', it is evident that both \atks{PA}{r} and \atks{MA}{r} have negative impact on the $tpr$ of ML-PWD trained on \dataset{$\delta$Phish} and \dataset{$Zenodo$}. Specifally, \atks{PA}{r} reduced the $tpr$ of all $F^c$- and $F^r$-based ML-PWD on \dataset{$\delta$Phish}, with small decreases ranging from $0.01$ to $0.08$. On the other hand, \atks{MA}{r} had a more pronounced effect compared to \atks{PA}{r}, successfully reducing the $tpr$ of $F^r$-based ML-PWD by $0.1$--$0.17$. Nevertheless, \atks{WA}{r} is still the most effective attack compared with them. 

\textbf{MLSEC.}
We have summarized the impact of the new HTML attacks on MLSEC in Table~\ref{tab:new_atk_mlsec}. These attacks are the same HTML attacks used in \dataset{$\delta$Phish} and \dataset{Zenodo}. Our findings reveal several interesting phenomena in the evaluation:

\begin{itemize}
    \item Among the attacks evaluated, \atks{iWA}{r} emerges as the most potent attack, significantly degrading the performance of PWD of MLSEC. The confidence of models $m0$ and $m2$ drop from nearly $0.9$ to $0.02$, indicating a stark decrease in their ability to accurately detect malicious webpages. However, it is worth noting that other attacks also have a substantial impact on degrading the detection capability of PWD. For instance, \atks{PA}{r} reduce the confidence of $m2$ from $0.9$ to $0.61$, while \atks{MA}{r} results in a decrease of $0.76$ in the confidence of $m6$.

    \item Comparing to \atks{eWA}{r} and \atks{rWA}{r}, \atks{iWA}{r} has a greater influence on $m0$--$m3$, leading to a decrease in their confidence by $0.35$--$0.89$. However, for PWD $m4$--$m7$, \atks{iWA}{r} does not decrease their confidence but slightly increase them by $0.01$. On the other hand, \atks{rWA}{r} reduces their confidence by $0.1$ (from nearly $0.8$ to $0.7$), while \atks{eWA}{r} results in a confidence reduction of $0.2$ for PWD $m4$ and $m6$. This phenomenon can be considered reasonable since PWD employed in MLSEC are black-box models which may consist of multiple types of PWD. It implies that the impact of perturbations may vary depending on the specific model characteristics and vulnerabilities. Hence, it is important to note that the goal of this study is not to propose a generalized perturbation set that works for all PWD, but rather to investigate the impact and effectiveness of cheap perturbations on PWD in practice.
    
    \item It is observed that \atks{rWA}{r} has a more widespread impact as it influences all seven PWD on MLSEC, resulting in a reduction of confidence by $0.1$ across the board. 
    
    \item Both \atks{PA}{r} and \atks{MA}{r} are effective attacks that successfully evade the detection of PWD in MLSEC. In particular, \atks{MA}{r} proves to be a potent attack, as it evades five (out of eight) PWD, causing their confidence score to drop below $0.5$. Additionally, the confidence scores of seven PWD decrease to approximately $0.65$ from initial values of around $0.85$. These findings highlight the impact of \atks{PA}{r} and \atks{MA}{r} on the performance of PWD on MLSEC.

\end{itemize}
\begin{table}[!htbp]
    \centering
    \caption{New attack's impact on MLSEC (HTML perturbations)}
    
    \resizebox{0.7\columnwidth}{!}{
        \begin{tabular}{c|c||ccccccccccc}
             \toprule
              \smacal{A} & no-atk &\atks{iWA}{r} & \atks{eWA}{r} & \atks{rWA}{r}& \atks{PA}{r} & \atks{MA}{r} \\
             \toprule
             $m0$ &\res{0.91}{0.052} &\res{0.02}{0.011} &  \res{0.65}{0.185}&\res{0.81}{0.116}&\res{0.91}{0.052}&\res{0.90}{0.062} \\
               $m1$ &\res{0.87}{0.071} &\res{0.52}{0.161}& \res{0.87}{0.085}&\res{0.78}{0.100}&\res{0.67}{0.262}&\res{0.31}{0.051}\\
                $m2$ &\res{0.90}{0.051}&\res{0.02}{0.011}&\res{0.65}{0.185}&\res{0.85}{0.087}&\res{0.61}{0.390}&\res{0.88}{0.096}\\
              $m3$ & \res{0.88}{0.070} &\res{0.51}{0.172}&\res{0.87}{0.079}&\res{0.81}{0.091}&\res{0.66}{0.271}&\res{0.26}{0.080}\\ 
             $m4$ & \res{0.82}{0.106} &\res{0.83}{0.123} & \res{0.64}{0.199}& \res{0.73}{0.112}&\res{0.57}{0.372}&\res{0.80}{0.121}\\
              $m5$ & \res{0.81}{0.120}  &\res{0.82}{0.136}&\res{0.85}{0.107} &\res{0.70}{0.103}&\res{0.64}{0.280}&\res{0.39}{0.166}\\ 
              $m6$ & \res{0.83}{0.108}&\res{0.84}{0.116} &\res{0.64}{0.198} &\res{0.73}{0.111}&\res{0.56}{0.373}&\res{0.07}{0.076}\\
              $m7$ & \res{0.82}{0.121} &\res{0.83}{0.127}&\res{0.85}{0.106}&\res{0.70}{0.097}&\res{0.64}{0.279}&\res{0.36}{0.129}\\
            \bottomrule    
        \end{tabular}
        \label{tab:new_atk_mlsec}
    }
\end{table}

\takeaway{Applying \atk{iWsP} does not change the webpage's appearance but it proves to be highly effective in evading ML-PWD. In contrast, the application of \atk{rWsP} results in obvious changes to the webpage's appearance but it has a relatively minor impact on the performance ML-PWD. \atks{MA}{r} had a more pronounced effect compared to \atks{PA}{r}. Nevertheless, \atks{WA}{r} is still the most effective attack compared with them.}

\subsubsection{\textbf{Impact of URL perturbations}}
\label{subsubsec:same-eva-url}
\begin{figure}[htbp]
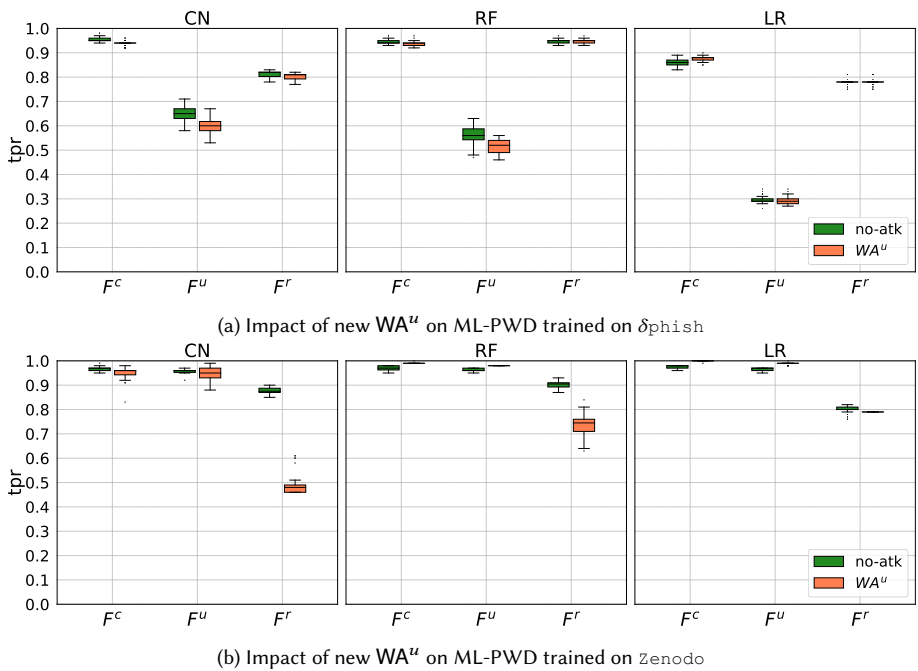

    \centering
    \begin{subfigure}[t]{1\columnwidth}
        \centering
        \includegraphics[width=0.8\columnwidth]{figures/new_atks/delta_urlatk.pdf}
        \caption{ Impact of new \atks{WA}{u} on ML-PWD trained on \dataset{$\delta$phish}}  
        \label{sfig:delta_urlatk}
    \end{subfigure}
    \begin{subfigure}[t]{1\columnwidth}
        \centering
        \includegraphics[width=0.8\columnwidth]{figures/new_atks/zenodo_urlatk.pdf}
        \caption{Impact of new \atks{WA}{u} on ML-PWD trained on \dataset{Zenodo}}
        \label{sfig:zenodo_urlatk}
         
    \end{subfigure}
    \caption{Effectiveness of new attacks \atks{WA}{u}. The three plots in each subfigure represent the algorithm used by a specific ML-PWD. Each plot has box divided into three groups, each denoting a specific $F$ used by the ML-PWD. The green box shows the $tpr$ on the original samples, while the orange box show the $tpr$ against \atks{WA}{u}.}
    \label{fig:new_urlatk}
    \vspace{-1em}
\end{figure}

The impact of \atks{WA}{u} is illustrated in Figs.~\ref{fig:new_urlatk}. Fig.~\ref{sfig:delta_urlatk} reveal the changes when performing \textit{atkPth} on ML-PWDtrained on \dataset{$\delta$Phish}. Green boxes represent the $tpr$ of `no-atk' (i.e., baseline), while the orange boxes indicate the impact of \atks{WA}{u}. Comparing the medians of each box plot, the median line of orange boxes is lower than Green boxes for $F^u$-based ML-PWD, indicating that \atks{WA}{u} can degrade ML-PWD's $tpr$. In contrast, this type of \atks{WA}{u} does not decrease $tpr$ of $F^u$-based CN-PWD trained on \dataset{Zenodo} (as shown in Table~\ref{tab:ze_url} in Appendix~\ref{app:specific_newresults}. However, it is significantly reduces the performance of $F^r$-based ML-PWD. This is because some HTML features require extracting information from both URL and HTML (e.g., HTML\_URLBrand: which checks (in the HTML) if the webpage title includes the brand name that appeared in the URL). Therefore, either URL perturbations or HTML perturbations can possibly affect the $F^r$-based ML-PWD. Furthermore, as shown in Fig.~\ref{sfig:zenodo_urlatk}, another \atks{WA}{u} \textit{sepWrd}, also clearly decreases the $tpr$ of $F^r$-based ML-PWD. Simply put, \atks{WA}{u} will affect ML-PWD's performance.

\subsection{Multi-space Attacks: Description}
\label{subsec:mixed-descr}

Insofar, we have always considered perturbations applied in a single space. However, as mentioned in §\ref{subsec:extensions}, an attacker who can apply PsP or MsP (which require write-access to the ML-PWD) can also also apply WsP (which only require access to the phishing webpage---which the attacker owns). These ``mixed-space'' attacks are worth considering because they are trivial to implement for an attacker---assuming that such an attacker can already apply PsP and/or MsP (we recall that, from a cost viewpoint, WsP$\ll$PsP$<$MsP). Therefore, we introduce 66 types of `mixed-space' attacks (the complete details are in Appendix ~\ref{app:specific_newresults}). These attacks span across all the defined evasion spaces (§\ref{sec:evasion}): Website space, Preprocessing space, and Machine Learning space. In particular, we consider ``accessible'' attacks (which combine \atk{WA} and \atk{PA}), as well as stronger ones (which entail \atk{MA} and \atk{PA}). We also consider ``double'' attacks, entailing multiple perturbations \textit{in the same space} (e.g., WsP+WsP).
We expect that mixed-space attacks, which exploit vulnerabilities and weaknesses present in different stages of the detection process, lead to more evasive samples (at least w.r.t. the corresponding single-space attacks).

\subsection{Multi-space Attacks: Evaluation}
\label{subsec:mixed-eva}

We evaluate the evasion capabilities of our new mixed-space attacks on the ML-PWD trained on the \dataset{$\delta$phish}, \dataset{Zenodo}, as well as those provided by MLSEC.\footnote{Since MLSEC only analyzes the HTML, we do not consider mixed-space attacks entailing perturbations of the URL.} We begin by considering attacks entailing two perturbations \textit{in the same space}, i.e., PsP+PsP (§\ref{subsubsec:psppsp}) and WsP+WsP (§\ref{subsubsec:wspwsp}); then, we consider attacks entailing two perturbations \textit{in different spaces}, i.e., PsP+WsP (§\ref{subsubsec:pspwsp} and PsP+MsP (§\ref{subsubsec:pspmsp}).

\subsubsection{\textbf{Double-PsP}}
\label{subsubsec:psppsp}
\begin{itemize}
    \item \textbf{$\delta$Phish} and \textbf{Zenodo}. Table~\ref{tab:delta_psp_psp} and Table~\ref{tab:ze_psp_psp} demonstrate the impact of 8 kinds of \atks{PA}{r}+\atks{PA}{r} on ML-PWD trained on \dataset{$\delta$Phish} and \dataset{Zenodo}. Even though not all of them significantly impact the PWD of \dataset{$\delta$Phish}. While not all combinations significantly affect the PWD, there are notable influences observed. For instance, when the combination attack occurs(specifically, the perturbation delSpn\_delTtl), the $tpr$ of LR-PWD based on $F^c$ and $F^r$ drops by $0.1$ and $0.16$, respectively. additionally, the $tpr$ of $F^r$-based LR-PWD down from $0.8$ to $0.58$, and CN-PWD's drops from $0.86$ to $0.64$ after being subjected to \atks{PA}{r}+\atks{PA}{r}. In contrast, $F^u$-based PWD is not affected, and most of $F^c$-based PWD remain unchanged. That is because our \atks{PA}{r}+\atks{PA}{r} combinations specifically target HTML, and $F^u$ is the core component when crafting the $F^c$-based ML-PWD.
    \item \textbf{MLSEC}. In the case of MLSEC's PWD, Table~\ref{tab:mlsec_psp_psp} indicates that all cheap \atks{PA}{r}+\atks{PA}{r} combination attacks proposed can decrease the performance of PWD, resulting in the confidence score dropped by $0.01$--$0.32$.
\end{itemize}

\subsubsection{\textbf{Double-WsP}}
\label{subsubsec:wspwsp}

\begin{itemize}
    \item \textbf{$\delta$Phish} and \textbf{Zenodo}. As shown in Table~\ref{tab:delta_wsp_wsp}, the combination attack \atks{WA}{r}+\atks{WA}{r} did not reduce the $tpr$ of ML-PWD trained on \dataset{$\delta$Phish}. In fact, in some cases, the $tpr$ increased to $1.0$, such as the $tpr$ of $F^r$-based CN-PWD increased from $0.79$ to $1$. Similarly, `$replOnfoc\_replRet$' did not affect the ML-PWD of \dataset{Zenodo}, as shown in Table~\ref{tab:ze_wsp_wsp}). However, it is importance to note that under the influence of `$htEsc\_replRet$', the $tpr$ of $F^r$-based LR-PWD reduced to $0.55$ from $0.8$. Moreover, `$htEncd\_replRet$' reduced $tpr$ of $F^r$-based CN-, LR- and RF-PWD to $0$. These findings suggest that while some combinations of \atks{WA}+\atks{WA} attacks may not result in a significant reduction in the $tpr$ of ML-PWD, specific combinations can still have an impact on the detection performance, leading to a decrease in the $tpr$. The effectiveness and impact of these combinations may vary depending on the specific ML-PWD and the nature of the attacks employed.

    \item \textbf{MLSEC}. On the contrary, \atks{WA}{r}+\atks{WA}{r} proves to be a powerful weapon for disrupting PWD of MLSEC. As indicated in Table~\ref{tab:mlsec_wsp_wsp}, the combination attack `$replOnfoc\_replRet$' defeated all detectors, leading to a significant decrease in their confidence scores by $0.12$--$0.58$. Moreover, four PWD have their confidence scores reduced below 0.5, indicating a successful evasion. Furthermore, the attack `$htEsc\_replRet$' evades four detectors, resulting in a substantial reduction in their confidence scores to $0.03$ or near $0.15$. Additionally, the attack `$htEncd\_replRet$' successfully bypasses four detectors and notably decreases the confidence score of model $m0$ from $0.91$ to $0.08$. These findings demonstrate the effectiveness and potency of \atks{WA}{r}+\atks{WA}{r} combination attacks in evading detection and undermining the confidence of PWD in MLSEC. The combination of multiple \atks{WA}{r} proves to be highly disruptive, highlighting the need for robust defense mechanisms against such attacks.
\end{itemize}

\takeaway{The simplest and cheapest attacks can indeed be highly effective in evading PWD, but their effectiveness may vary across different PWD. While these attacks may prove to be successful in bypassing certain PWD, they may not necessarily work equally well on all PWD.}

\subsubsection{\textbf{Mixed: PsP and WsP}} 
\label{subsubsec:pspwsp}

\begin{itemize}
    \item \textbf{$\delta$Phish} and \textbf{Zenodo}. As presented in Tables~\ref{tab:delta_psp_wsp}, we analyze the impact of 52 attacks across the Preprocessing space and Website space of \dataset{$\delta$Phish}. These attacks have a detrimental effect on the detection performance of ML-PWD, particularly those based on $F^r$. Among these attacks, the combination attacks involving `$addHidP$' demonstrate the most significant impact on the ML-PWD. For instance, the attack `$addLngTxt\_addHidP$' mentioned in Table~\ref{subtab:delta_addlongtext_wsp} reduce the $tpr$ of $F^r$-based ML-PWD from $0.79$, $0.95$ and $0.78$ to $0.03$, $0$ and $0$ respectively. This indicates a drastic reduction in the ability of the ML-PWD to detect and classify phishing instances. Similar situation is observed in ML-PWD of \dataset{Zenodo}, as illustrated in Table~\ref{tab:ze_psp_wsp}, the combination attack of \atks{PA}{r}+\atks{WA}{r} demonstrates a decrease in the $tpr$ of ML-PWD trained on \dataset{Zenodo}. Notably, the attack `$delFtr\_addHidP$' leads to a significant reduction in the $tpr$ of $F^r$-based RF-PWD, dropping from $0.9$ to $0.15$. Furthermore, when encountering attack `$delSpn\_addHidP$', the $tpr$ decreases to $0.03$. Other \atks{PA}{r}+\atks{WA}{r} combination attacks also prove effective in bypassing the detection of ML-PWD of \dataset{Zenodo}. For example, the attack `$delFtr\_replPass$' results in a similar drop, and `$delFtr\_addSusLnk$' reduces the $tpr$ by $0.4$--$0.65$. These findings highlight the susceptibility of ML-PWD trained on \dataset{Zenodo} to \atks{PA}{r}+\atks{WA}{r} attacks.  

    \item \textbf{MLSEC}. We executed 53 kinds of \atks{PA}{r}+\atks{WA}{r} on MLSEC's PWD and evaluated their impact, which is reported in Tables~\ref{tab:mlsec_psp_wsp}. All of these combination attacks affected the decision of PWD, with 51 (i.e., except `$delSpn\_modBgClr$' and `$delFtr\_modBgClr$') out of 53 attacks noticeably degrading the confidence of at least one PWD. One particular attack, `$delFrm\_addHidP$' minimizes the confidence of all PWD. Specifically, the confidence of $m0$ and $m2$ dropped from $0.9$ to $0.01$, while the confidence of other PWDs decreased by $0.16$--$0.5$. This substantial reduction caused by this cheap attack is both shocking and expected, as this combination attack simultaneously considers the ``feature space" and ``problem space", i.e., both the high-level definitions of adversarial perturbations~\cite{Pierazzi:Intriguing}.
\end{itemize}

\takeaway{Comparing to other attacks mixing evasion spaces, it is evident that \atks{PA}{r}+\atks{WA}{r} possess greater destructive power and have a substantial impact on the $tpr$ of PWD. These attacks are particularly potent because they traverse both the `feature-space' (e.g., Preprocessing space) and `problem-space' (e.g., Website space).}

\subsubsection{\textbf{Mixed: PsP and MsP}} 
\label{subsubsec:pspmsp}

\begin{itemize}
    \item \textbf{$\delta$Phish} and \textbf{Zenodo}. We showcase 3 combination \atks{PA}{r}+\atks{MA}{r} attacks target ML-PWD trained on \dataset{$\delta$Phish} and \dataset{Zenodo} in Table~\ref{tab:delta_psp_msp} and Table~\ref{tab:ze_psp_msp}, respectively. It is worth noting that these combination attacks are difficult to achieve and require high costs, as attackers must obtain write-access to deeper segments of the ML-PWD. Interestingly, despite the high cost associated with these attacks, they do not consistently and effectively disrupt ML-PWD, except for the attack `$delFtr\_brTgs$' which reduces the $tpr$ of $F^r$-based CN-PWD from $0.86$ to $0.64$. 
    
    \item \textbf{MLSEC}. As depicted in Table~\ref{tab:mlsec_psp_msp}, the combination attack \atks{PA}{r}+\atks{MA}{r} decreases the performance of all considered ML-PWD, but the impact is relatively minor. The largest impact is observed with `$delSpn\_brTgs$' and `$delFtr\_brTgs$'. These attacks lead to a decrease in the confidence of $m0$ by $0.16$ and $0.13$, respectively.
\end{itemize}

\takeaway{Costly attacks (which require both MsP and PsP) do not always possess formidable attacking capabilities. They may slightly affect certain detectors or have no impact on others.}

\section{Related Work}
\label{sec:related}

Countering phishing is a long-standing security problem, which can be considered as a subfield of cyberthreat detection---a research area that is being increasingly investigated also by adversarial ML literature~\cite{Apruzzese:Addressing}.
We focus on the detection of phishing \textit{websites}. Papers that consider phishing in social networks~\cite{binsaeed2020detecting}, darkweb~\cite{yoon2019doppelgangers}, phone calls~\cite{gupta2018towards}, or emails~\cite{duman2016emailprofiler} are complementary to our work---although our findings can also apply to phishing email filters if they analyze the URLs included in the body text (e.g.,~\cite{gupta2021novel}). Our focus is on attacks \textit{against} ML-PWD. For instance, Tian et al.~\cite{tian2018needle} evade PWD that use common blacklists, and their main proposal is to use ML as a detection engine to counter such ``squatting'' phishing websites. Hence, non-ML-PWD (e.g.,~\cite{zhang2021crawlphish}) are outside our scope.

Let us compare our paper with existing works on evasion attacks against ML-PWD. We provide an overview in Table~\ref{tab:related}, highlighting the main differences of our paper with the state-of-the-art. 
Only half of related papers craft their attacks in the problem-space---which requires modifying the raw webpage. Unfortunately, most publicly available datasets do not allow similar procedures. A viable alternative is composing ad-hoc dataset through public feeds as done, e.g., by~\cite{gressel2021feature} and~\cite{sabir2020evasion} (the latter only for URL-based ML-PWD). All these papers, however, do not release the actual dataset, preventing reproducibility and hence introducing experimental bias. The authors of~\cite{song2021advanced} share their dataset, but while the \textit{malicious} websites are provided with complete information (i.e., URL and HTML), the \textit{benign} websites are provided only with their URL---hence preventing complete reproducibility of attacks in the problem-space against ML-PWD inspecting the HTML. 
The latter is a well-known issue in related literature~\cite{panum2020towards}, which does not affect our paper because our entire evaluation is reproducible.
Notably, Aleroud et al.~\cite{aleroud2020bypassing} evaluate attacks both in the problem and feature-space, but on \textit{different} datasets, preventing a fair comparison. Indeed, they evade one ML-PWD trained on \dataset{PhishStorm} (which only includes raw URLs) with attacks in the problem space; and another ML-PWD trained on \dataset{UCI} (which is provided as pre-computed features) through feature space attacks. Hence, it is not possible to compare these two settings. A similar issue affects also~\cite{al2021generating}, which consider 4 datasets, each having a different $F$. Therefore, no prior work \textit{compared the impact} of attacks carried out in distinct evasion-spaces---to the best of our knowledge.
Not many papers consider adversarially robust ML-PWD, and only half consider both SL and DL algorithms---which our evaluation shows to respond differently against adversarial examples (cf. §\ref{ssec:comparison}). It is concerning that few papers overlook the importance of statistically significant comparisons. The most remarkable effort is~\cite{shirazi2019adversarial} which only performs 10 trials (we do 50), which are not enough to compute precise statistical tests.

Most prior work assume stronger attackers than those envisioned in our threat model (cf. §\ref{sec:threat}). Indeed, past threat models portray \textit{black-box} attackers who can freely inspect the output-space and query the ML-PWD (e.g.,~\cite{al2021generating, liang2016cracking, sabir2020evasion}); or \textit{white-box} attackers who perfectly know the target ML model \smacal{M}, such as its configuration, its training data \smacal{D}, or the feature importance (e.g.,~\cite{gressel2021feature, abdelnabi2020visualphishnet, lin2021phishpedia}). 
The only papers considering attackers that are closer to our threat model are~\cite{lee2020building, o2021generative} and~\cite{abdelnabi2020visualphishnet}. However, the ML-PWD considered in~\cite{abdelnabi2020visualphishnet} is specific for \textit{images}, which are tough to implement (cf. §\ref{ssec:limitations}) and also implicitly resembles a ML system for computer vision---a task well-investigated in adversarial ML literature~\cite{Biggio:Wild}. In contrast, the ML-PWD considered in~\cite{lee2020building} and~\cite{o2021generative} is similar to ours, but the adversarial samples are randomly created in the feature space, hence requiring an attacker with write-access to the internal ML-PWD workflow. Such an assumption is not unrealistic, but very unlikely in the context of phishing (cf. §\ref{ssec:considerations}).

\begin{table}[!ht]
    \centering
    \caption{Adversarial attacks against ML-PWD.
    For each paper, we report: the \textit{evasion space} (for simplicity we consider problem and feature-space); which \textit{features} ($F$) are analyzed by the ML-PWD; the ML \textit{algorithms} used by the ML-PWD (SL or DL); if some \textit{defense} is evaluated; how many \textit{datasets} are used (and if they are reproducible); and if the experiments are repeated for \textit{statistical validation}.}
    \resizebox{0.95\columnwidth}{!}{
        \begin{tabular}{c|c?c|c|c|c|c|c}
            \toprule
            
            \begin{tabular}{c} Paper\\(1st Author) \end{tabular} & Year & \begin{tabular}{c} Evasion\\space \end{tabular} &
            \begin{tabular}{c} ML-PWD\\types ($F$) \end{tabular} &
            \begin{tabular}{c} ML\\Algorithms \end{tabular} &            
            Defense &
            \begin{tabular}{c} Datasets\\(reprod.) \end{tabular}  &
            \begin{tabular}{c} Stat.\\Val. \end{tabular} \\ \toprule

            Liang~\cite{liang2016cracking} & 2016 & Problem & $F^c$ & SL & \xmark & 1 (\xmark) & \xmark \\
            Corona~\cite{Corona:Deltaphish} & 2017 & Feature & $F^r$, $F^c$ & SL & \cmark & 1 (\cmark) & \xmark \\
            Bahnsen~\cite{bahnsen2018deepphish} & 2018 & Problem & $F^u$ & DL & \xmark & 1 (\xmark) & \xmark \\
            Shirazi~\cite{shirazi2019adversarial} & 2019 & Feature & $F^c$ & SL & \xmark & 4 (\cmark) & \cmark* \\
            Sabir~\cite{sabir2020evasion} & 2020 & Problem & $F^u$ & SL, DL & \cmark & 1 (\xmark) & \xmark \\
            Lee~\cite{lee2020building} & 2020 & Feature & $F^c$ & SL & \cmark & 1 (\cmark) & \xmark \\
            Abdelnabi~\cite{abdelnabi2020visualphishnet} & 2020 & Problem & $F^r$ & DL & \cmark & 1 (\cmark) & \xmark \\
            Aleroud~\cite{aleroud2020bypassing} & 2020 & Both & $F^u$ & SL & \xmark & 2 (\cmark) & \xmark \\
            Song~\cite{song2021advanced} & 2021 & Problem & $F^c$ & SL & \cmark & 1 (\cmark*) & \xmark \\
            Bac~\cite{bac2021pwdgan} & 2021 & Feature & $F^u$ & SL, DL & \xmark & 1 (\xmark) & \xmark \\
            Lin~\cite{lin2021phishpedia} & 2021 & Feature & $F^c$ & DL & \cmark & 1 (\cmark) & \xmark \\
            
            O'Mara~\cite{o2021generative} & 2021 & Feature & $F^r$ & SL & \xmark & 1 (\cmark)  & \xmark \\

            Al-Qurashi~\cite{al2021generating} & 2021 & Feature & $F^u$, $F^c$ & SL, DL & \xmark & 4 (\cmark)  & \xmark \\
            Gressel~\cite{gressel2021feature} & 2021 & Feature & $F^c$ & SL, DL & \cmark & 1 (\xmark) & \xmark \\
            
            \midrule
            \multicolumn{2}{c?}{Ours} & Both & $F^u$, $F^r$, $F^c$ & DL, SL & \cmark & 2 (\cmark) & \cmark \\
            
            \bottomrule
            
        \end{tabular}
    }
    
    \label{tab:related}
    \vspace{-1.5em}
\end{table}
\section{Conclusions}
\label{sec:conclusions}

We aim to provide a constructive step towards developing ML systems that are secure against adversarial attacks.

Specifically, we focus on the detection of phishing websites, which represent a widespread menace to information systems. Such context entails attackers that actively try to evade `static' detection mechanisms via crafty, but ultimately simple tactics. Machine learning is a reliable tool to catch such phishers, but ML is also prone to evasion. However, realizing the evasion attempts considered by most past work requires a huge resource investment---which contradicts the very nature of phishing. To provide valuable research for ML security, the emphasis should be on attacks that are more likely to occur in the wild. We set this goal as our primary objective.

After dissecting the architecture of ML-PWD, we propose an original interpretation of attacks against ML systems by formalizing the \textsc{evasion-space} of adversarial perturbations. We then carry out a large evaluation of evasion attacks exploiting diverse `spaces', focusing on those requiring less resources to be staged in reality.

\begin{cooltextbox}
\textsc{\textbf{Takeaway:}}
The findings of our paper are useful to both research and practice in the adversarial ML domain.
\begin{itemize}
    \item Our \textit{evasion-space} formalization allows \textbf{researchers} to evaluate adversarial ML attacks without the risk of falling into the ``unrealizable'' perturbation trap (as long as the cost is factored in).
    \item Our \textit{results} raise an alarm for \textbf{practitioners}: some ML-PWD can be evaded with simple tactics that do not rely on gradient computations, days of bruteforcing, or extensive intelligence gathering campaigns.
\end{itemize} 
\end{cooltextbox}




\appendix

\section{Complete Benchmark Tables}
\label{app:benchmark}
We carry out our experiments by developing original software tools, all written in Python3 by leveraging well-known libraries (e.g., scikit-learn, Tensorflow). The ML-PWD using \smamath{RF} and \smamath{LR} are assessed on a system mounting an Intel Xeon W-2223@3.6GHz with 32GB RAM. For the \smamath{CN}, we use an nVidia P100 GPU. (Our results have been reproduced during the ACSAC artifact evaluation.)

\begin{table*}[!htbp]
    \centering
    \caption{Evasion Robustness of the ML-PWD on the \dataset{Zenodo} dataset. The cells report the average (and std. dev.) \smamath{tpr} over the 50 reiterations. Lines correspond to the ML-PWD, while rows correspond to a specific attack.}
    \label{tab:attacks_zenodo}
    \resizebox{0.99\columnwidth}{!}{
        \begin{tabular}{c|c?c||ccc||ccc||ccc||ccc}
             \toprule

              \smacal{A} & $F$ & no-atk & \atks{WA}{u} & \atks{WA}{r} & \atks{WA}{c} & \aatks{u} & \aatks{r} & \aatks{c} & \atks{PA}{u} & \atks{PA}{r} & \atks{PA}{c} & \atks{MA}{u} & \atks{MA}{r} & \atks{MA}{c} \\
             \toprule
             
             \multirow{3}{*}{{\smamath{CN}}}
             
             & $F^u$ & \res{0.96}{0.007} & \res{1.00}{0.000} & \res{0.93}{0.020} & \res{1.00}{0.000} & \res{1.00}{0.000} & \res{0.95}{0.018} & \res{1.00}{0.000} & \res{1.00}{0.017} & \res{0.95}{0.018} & \res{1.00}{0.017} & \res{0.18}{0.222} & \res{0.95}{0.018} & \res{0.18}{0.222} \\
             
             & $F^r$ & \res{0.86}{0.013} & \res{0.88}{0.013} & \res{0.87}{0.056} & \res{0.87}{0.055} & \res{0.88}{0.013} & \res{0.44}{0.153} & \res{0.83}{0.051} & \res{0.54}{0.108} & \res{0.29}{0.120} & \res{0.31}{0.118} & \res{0.88}{0.013} & \res{0.02}{0.095} & \res{0.02}{0.095} \\
             
             & $F^c$ & \res{0.97}{0.009} & \res{0.92}{0.036} & \res{0.93}{0.020} & \res{0.94}{0.063} & \res{0.92}{0.036} & \res{0.92}{0.016} & \res{0.83}{0.115} & \res{1.00}{0.011} & \res{0.90}{0.031} & \res{0.99}{0.017} & \res{0.51}{0.131} & \res{0.92}{0.036} & \res{0.15}{0.211} \\
             
             \midrule 
             
            \multirow{3}{*}{{\smamath{RF}}}
             
             & $F^u$ & \res{0.96}{0.007} & \res{1.00}{0.000} & \res{0.96}{0.008} & \res{1.00}{0.000} & \res{1.00}{0.000} & \res{0.96}{0.008} & \res{1.00}{0.000} & \res{0.54}{0.183} & \res{0.96}{0.007} & \res{0.54}{0.183} & \res{0.04}{0.098} & \res{0.96}{0.007} & \res{0.04}{0.098} \\
             
             & $F^r$ & \res{0.90}{0.013} & \res{0.90}{0.013} & \res{0.88}{0.024} & \res{0.88}{0.025} & \res{0.90}{0.013} & \res{0.71}{0.053} & \res{0.80}{0.025} & \res{0.59}{0.086} & \res{0.47}{0.082} & \res{0.30}{0.088} & \res{0.90}{0.013} & \res{0.04}{0.155} & \res{0.04}{0.155} \\
             
             & $F^c$ & \res{0.97}{0.009} & \res{0.98}{0.064} & \res{0.94}{0.012} & \res{0.94}{0.171} & \res{0.98}{0.063} & \res{0.94}{0.010} & \res{0.94}{0.191} & \res{0.65}{0.101} & \res{0.94}{0.010} & \res{0.21}{0.134} & \res{0.07}{0.115} & \res{0.92}{0.012} & \res{0.03}{0.158} \\
             
             \midrule
             
             \multirow{3}{*}{{\smamath{LR}}}
             
             & $F^u$ & \res{0.97}{0.005} & \res{1.00}{0.000} & \res{0.95}{0.005} & \res{1.00}{0.000} & \res{1.00}{0.000} & \res{0.96}{0.005} & \res{1.00}{0.000} & \res{0.73}{0.071} & \res{0.96}{0.006} & \res{0.73}{0.071} & \res{0.00}{0.000} & \res{0.96}{0.006} & \res{0.00}{0.000} \\
             
             & $F^r$ & \res{0.80}{0.013} & \res{0.80}{0.013} & \res{0.65}{0.043} & \res{0.64}{0.040} & \res{0.80}{0.013} & \res{0.54}{0.027} & \res{0.56}{0.022} & \res{0.61}{0.007} & \res{0.08}{0.013} & \res{0.01}{0.010} & \res{0.80}{0.013} & \res{0.00}{0.000} & \res{0.00}{0.000} \\
             
             & $F^c$ & \res{0.98}{0.005} & \res{0.82}{0.035} & \res{0.95}{0.015} & \res{0.32}{0.079} & \res{0.80}{0.038} & \res{0.93}{0.014} & \res{0.32}{0.132} & \res{0.46}{0.053} & \res{0.91}{0.032} & \res{0.06}{0.025} & \res{0.00}{0.000} & \res{0.76}{0.036} & \res{0.00}{0.000} \\
             
             \bottomrule
             
        \end{tabular}
    }
\end{table*}

\begin{table*}[!htbp]
    \centering
    \caption{Evasion Robustness of the ML-PWD on the \dataset{$\delta$phish} dataset. The cells report the average (and std. dev.) \smamath{tpr} over the 50 reiterations. Lines correspond to the ML-PWD, while rows correspond to a specific attack.}
    \label{tab:attacks_deltaphish}
    \resizebox{0.99\columnwidth}{!}{
        \begin{tabular}{c|c?c||ccc||ccc||ccc||ccc}
             \toprule

              \smacal{A} & $F$ & no-atk & \atks{WA}{u} & \atks{WA}{r} & \atks{WA}{c} & \aatks{u} & \aatks{r} & \aatks{c} & \atks{PA}{u} & \atks{PA}{r} & \atks{PA}{c} & \atks{MA}{u} & \atks{MA}{r} & \atks{MA}{c} \\
             \toprule
             
             \multirow{3}{*}{{\smamath{CN}}}
             
             & $F^u$ & \res{0.65}{0.028} & \res{0.91}{0.276} & \res{0.65}{0.029} & \res{0.91}{0.275} & \res{0.90}{0.299} & \res{0.65}{0.029} & \res{0.90}{0.300} & \res{0.60}{0.165} & \res{0.65}{0.028} & \res{0.60}{0.165} & \res{0.14}{0.346} & \res{0.65}{0.028} & \res{0.14}{0.346} \\
             
             & $F^r$ & \res{0.79}{0.013} & \res{0.80}{0.013} & \res{0.35}{0.018} & \res{0.34}{0.017} & \res{0.80}{0.013} & \res{0.86}{0.033} & \res{0.88}{0.020} & \res{0.46}{0.065} & \res{0.69}{0.038} & \res{0.46}{0.064} & \res{0.81}{0.013} & \res{0.00}{0.000} & \res{0.00}{0.000} \\
             
             & $F^c$ & \res{0.95}{0.010} & \res{0.88}{0.066} & \res{0.93}{0.012} & \res{0.84}{0.113} & \res{0.89}{0.046} & \res{0.89}{0.020} & \res{0.87}{0.058} & \res{0.90}{0.107} & \res{0.58}{0.059} & \res{0.82}{0.163} & \res{0.04}{0.198} & \res{0.01}{0.011} & \res{0.04}{0.196} \\

             \midrule 
             
            \multirow{3}{*}{{\smamath{RF}}}
             
             & $F^u$ & \res{0.56}{0.037} & \res{0.84}{0.330} & \res{0.56}{0.036} & \res{0.84}{0.330} & \res{0.84}{0.330} & \res{0.56}{0.034} & \res{0.84}{0.331} & \res{0.57}{0.238} & \res{0.56}{0.037} & \res{0.57}{0.238} & \res{0.01}{0.053} & \res{0.56}{0.037} & \res{0.01}{0.053} \\
             
             & $F^r$ & \res{0.95}{0.008} & \res{0.95}{0.009} & \res{0.84}{0.003} & \res{0.84}{0.043} & \res{0.95}{0.009} & \res{0.80}{0.038} & \res{0.94}{0.009} & \res{0.84}{0.049} & \res{0.55}{0.090} & \res{0.95}{0.055} & \res{0.95}{0.008} & \res{0.00}{0.000} & \res{0.00}{0.000} \\
             
             & $F^c$ & \res{0.95}{0.009} & \res{0.90}{0.020} & \res{0.92}{0.006} & \res{0.77}{0.047} & \res{0.90}{0.017} & \res{0.86}{0.018} & \res{0.92}{0.015} & \res{0.90}{0.065} & \res{0.68}{0.013} & \res{0.86}{0.097} & \res{0.88}{0.026} & \res{0.00}{0.001} & \res{0.00}{0.000} \\
             
             \midrule
             
             \multirow{3}{*}{{\smamath{LR}}}
             
             & $F^u$ & \res{0.30}{0.014} & \res{0.21}{0.332} & \res{0.30}{0.015} & \res{0.22}{0.341} & \res{0.26}{0.364} & \res{0.30}{0.015} & \res{0.24}{0.359} & \res{0.64}{0.256} & \res{0.30}{0.014} & \res{0.64}{0.256} & \res{0.00}{0.000} & \res{0.30}{0.014} & \res{0.00}{0.000} \\
             
             & $F^r$ & \res{0.78}{0.011} & \res{0.78}{0.011} & \res{0.57}{0.014} & \res{0.56}{0.047} & \res{0.78}{0.011} & \res{0.60}{0.030} & \res{0.63}{0.010} & \res{0.80}{0.029} & \res{0.04}{0.006} & \res{0.45}{0.068} & \res{0.78}{0.011} & \res{0.00}{0.000} & \res{0.00}{0.000} \\
             
             & $F^c$ & \res{0.86}{0.014} & \res{0.47}{0.094} & \res{0.81}{0.011} & \res{0.36}{0.102} & \res{0.73}{0.126} & \res{0.73}{0.018} & \res{0.63}{0.150} & \res{0.65}{0.157} & \res{0.23}{0.014} & \res{0.32}{0.109} & \res{0.00}{0.000} & \res{0.00}{0.000} & \res{0.00}{0.000} \\
             
             \bottomrule
             
        \end{tabular}
    }
\end{table*}

\textbf{Evasion Performance}
We report the complete results of all the 12 considered evasion attacks against all the 18 considered ML-PWD in Table~\ref{tab:attacks_zenodo} (for \dataset{Zenodo}) and Table~\ref{tab:attacks_deltaphish} (for \dataset{$\delta$phish}). These tables also include the performance in non-adversarial settings computed on the 100 phishing samples (drawn from $P_i$ that are used as base for the adversarial samples). We remark that we chose such 100 samples by randomly selecting 100 samples which were correctly detected by the best ML-PWD on each dataset. As such, the \smamath{tpr} reported in the \textit{no-atk} column can slightly differ from the one in Table~\ref{tab:normal} (which is computed on the entire $P_i$).

\textbf{Runtime.}
We report in Table~\ref{tab:times} the runtime for training and testing all our ML-PWD in non-adversarial scenarios. The values denote the average runtime (and standard deviation) across the 50 trials. Training the \smamath{RF} and \smamath{LR} uses all cores/threads of our CPU.

\begin{table}[!htbp]
    \centering
    \caption{Execution Times for training (on \ftcal{D}) and testing (on both $P_i$ and $B_i$) the ML models used by our ML-PWD.}
    \label{tab:times}
    \resizebox{0.99\columnwidth}{!}{
        \begin{tabular}{c|c?cc||cc}
             \toprule
             \multirow{2}{*}{\smacal{A}}& \multirow{2}{*}{$F$} & \multicolumn{2}{c||}{\dataset{Zenodo}} & \multicolumn{2}{c}{\dataset{$\delta$phish}} \\ \cline{3-6}
              & & \small{Train (s)} & \small{Test (ms)} & \small{Train (s)} & \small{Test (ms)} \\
             \toprule

            \multirow{3}{*}{{\smamath{CN}}}
             & $F^u$ & \res{110.88}{{15.318}} & \res{178.13}{9.661} & \res{201.314}{21.753} & \res{301.91}{46.133} \\
             & $F^r$ & \res{76.61}{4.562} & \res{171.95}{10.577} & \res{167.74}{25.197} & \res{273.4}{43.99} \\
             & $F^c$ & \res{152.325}{13.183} & \res{222.696}{86.618} & \res{165.486}{23.367} & \res{274.84}{47.975} \\

            \midrule 
             
            \multirow{3}{*}{{\smamath{RF}}}
             & $F^u$ & \res{0.152}{0.0052} & \res{7.59}{0.208} & \res{0.583}{0.0181} & \res{28.09}{0.402} \\
             & $F^r$ & \res{0.146}{0.0037} & \res{7.85}{0.07} & \res{0.369}{0.0181} & \res{22.39}{0.151} \\
             & $F^c$ & \res{0.179}{0.0035} & \res{9.39}{0.312} & \res{0.44}{0.0062} & \res{23.6}{0.205} \\
             
            \midrule
             
            \multirow{3}{*}{{\smamath{LR}}}
             & $F^u$ & \res{0.045}{0.019} & \res{0.1}{0.005} & \res{0.185}{0.0285} & \res{0.45}{0.895} \\
             & $F^r$ & \res{0.055}{0.0182} & \res{0.09}{0.003} & \res{0.083}{0.0509} & \res{0.74}{1.161} \\
             & $F^c$ & \res{0.063}{0.0179} & \res{0.17}{0.014} & \res{0.301}{0.0678} & \res{0.36}{0.678} \\
             
            \bottomrule
             
        \end{tabular}
    }
\end{table}

\section{Alternative \atks{WA}{r} for \textbf{Zenodo} and \textbf{$\delta$Phish}}
\label{app:alternative_wspr}

As we mentioned in §~\ref{sapp:wsp}, we applied two different \atks{WA}{r} to the ML-PWD of \dataset{$\delta$Phish} and \dataset{Zenodo} (i.e., replOnc: swap <a href=`link'> into <a onclick=``this.href=`link'"> on \dataset{Zenodo}, and addInLnk: insert <a href=`\#' style=`display:none'> can not see</a> to the samples of \dataset{$\delta$Phish}), and report their influence in Figs.~\ref{fig:wa}. In this section, we apply the same \atks{WA}{r}, but with the datasets swapped to see if the influence will change, i.e., applying addInLnk to \dataset{Zenodo} and applying replOnc to \dataset{$\delta$Phish}. The new influence on each dataset is depicted in Table.~\ref{tab:war_alter}. Comparing with the Figs.\ref{fig:wa}, it can be concluded that the \dataset{$\delta$Phish} is more vulnerable to addInLnk, whereas their impact on \dataset{Zenodo} are similar.

\begin{table}[!htbp]
    \centering
    \caption{Impact of Alternative \atks{WA}{r} on ML-PWD generated on \dataset{Zenodo} and \dataset{$\delta$Phish}, reported as the average (and std. dev.) \smamath{tpr} over the 50 trials.}
    \label{tab:normal2}
    \resizebox{0.8\columnwidth}{!}{
        \begin{tabular}{c|c?cc||cc}
             \toprule
             \multirow{2}{*}{\smacal{A}}& \multirow{2}{*}{$F$} & \multicolumn{2}{c||}{\dataset{Zenodo}} & \multicolumn{2}{c}{\dataset{$\delta$phish}} \\ \cline{3-6}
              & & $tpr$ (no-atk) & $tpr$ (addInLnk) & $tpr$ (no-atk) & $tpr$(replOnc) \\
             \toprule

            \multirow{3}{*}{{\smamath{CN}}}
             & $F^u$ & \res{0.96}{0.008} & \res{0.95}{0.018} & \res{0.55}{0.030} & \res{0.65}{0.029} \\
             & $F^r$ & \res{0.88}{0.018} & \res{0.61}{0.034} & \res{0.81}{0.019} & \res{0.89}{0.018} \\
             & $F^c$ & \res{0.97}{0.006} & \res{0.97}{0.021}  & \res{0.93}{0.013} & \res{0.93}{0.012} \\

            \midrule 
             
            \multirow{3}{*}{{\smamath{RF}}}
             & $F^u$ & \res{0.98}{0.004} & \res{0.96}{0.008}& \res{0.45}{0.022} & \res{0.56}{0.036} \\
             & $F^r$ & \res{0.93}{0.013} & \res{0.94}{0.018} & \res{0.94}{0.016} & \res{0.99}{0.003} \\
             & $F^c$ & \res{0.98}{0.006} & \res{0.97}{0.008} & \res{0.97}{0.007} & \res{0.98}{0.006} \\
             
            \midrule
             
            \multirow{3}{*}{{\smamath{LR}}}
             & $F^u$ & \res{0.95}{0.009} & \res{0.96}{0.002} & \res{0.24}{0.017} & \res{0.3}{0.015} \\
             & $F^r$ & \res{0.82}{0.017} & \res{0.95}{0.005} & \res{0.74}{0.025} & \res{0.78}{0.014}\\
             & $F^c$ & \res{0.96}{0.007} & \res{0.98}{0.007} & \res{0.81}{0.020} & \res{0.89}{0.011} \\
             
            \bottomrule
             
        \end{tabular}
    }
    \label{tab:war_alter}
\end{table}

\section{Supplementary Tables for Additional Experiments}
\label{app:specific_newresults}

We now report the \textit{complete} results of \textit{all} our new experiments, which we discussed in §\ref{sec:new}.

\subsection{Perturbation's impact on \dataset{$\delta$Phish}}

We report new \atks{WA}{r}'s impact on the ML-PWD generated on \dataset{$\delta$Phish} in Table ~\ref{tab:iwsp_delta} and Table ~\ref{tab:ewsp_rwsp_delta}. PsP and WsP's influence were depicted in Table~\ref{tab:psp_msp_delta}. And Table~\ref{tab:delta_url} describes the $tpr$ of ML-PWD generated on \dataset{$\delta$Phish} against \atks{WA}{u}. Table~\ref{tab:delta_psp_psp},~\ref{tab:delta_psp_msp},~\ref{tab:delta_psp_wsp} and ~\ref{tab:delta_wsp_wsp} report the influence of hybrid space attacks on \dataset{$\delta$Phish}.

\begin{table}[!htbp]
    \centering
    \caption{Evasion Robustness of the ML-PWD against \atks{iWA}{r} on the \dataset{$\delta$Phish}. The cells report the average (and std. dev.) tpr over the 50 reiterations. Lines correspond to the ML-PWD, while rows correspond to a specific iWsP perturbation.}
    \resizebox{\columnwidth}{!}{
        \begin{tabular}{c|c|c||cccccccccccc}
             \toprule
             
              \smacal{A} & $F$ & no-atk &replOnc & delHidIt & addHidP & replJS & replRet & htEsc & htEncd & replPass & replOnfoc & addSusLnk \\ 
             \toprule
             \multirow{3}{*}{{\smamath{CN}}} 
             & $F^u$ & \res{0.65}{0.028} &\res{0.65}{0.029} &\res{0.65}{0.029} &\res{0.64}{0.031} &\res{0.64}{0.031} &\res{0.64}{0.031} &\res{0.64}{0.031} &\res{0.64}{0.031} &\res{0.64}{0.031} &\res{0.64}{0.035} &\res{0.64}{0.031}\\ 
              & $F^r$ & \res{0.79}{0.013}& \res{0.89}{0.018} &\res{0.81}{0.013} &\res{0.03}{0.006} &\res{0.79}{0.011} &\res{0.81}{0.013} &\res{0.94}{0.03} &\res{1.0}{0.0} &\res{0.81}{0.013} &\res{0.81}{0.013} &\res{0.19}{0.012}\\ 
              & $F^c$ & \res{0.95}{0.010} & \res{0.93}{0.012} &\res{0.95}{0.016} &\res{0.22}{0.059} &\res{0.89}{0.021} &\res{0.96}{0.011} &\res{0.99}{0.01} &\res{0.99}{0.014} &\res{0.95}{0.011} &\res{0.95}{0.013} &\res{0.79}{0.039}\\  
              \midrule  
             \multirow{3}{*}{{\smamath{RF}}}
             & $F^u$ & \res{0.56}{0.037} & \res{0.56}{0.036} &\res{0.56}{0.035} &\res{0.56}{0.033} &\res{0.56}{0.034} &\res{0.57}{0.033} &\res{0.57}{0.031} &\res{0.56}{0.033} &\res{0.57}{0.033} &\res{0.56}{0.037} &\res{0.56}{0.032} \\
             
              & $F^r$ & \res{0.95}{0.008} & \res{0.99}{0.003} &\res{0.88}{0.011} &\res{0.0}{0.0} &\res{0.81}{0.021} &\res{0.95}{0.008} &\res{1.0}{0.003} &\res{1.0}{0.0} &\res{0.95}{0.008} &\res{0.95}{0.008} &\res{0.44}{0.069}\\  
              & $F^c$ &\res{0.95}{0.009} & \res{0.98}{0.006} &\res{0.93}{0.01} &\res{0.04}{0.017} &\res{0.86}{0.015} &\res{0.95}{0.01} &\res{1.0}{0.007} &\res{1.0}{0.0} &\res{0.95}{0.009} &\res{0.94}{0.009} &\res{0.48}{0.043}\\ 
              \midrule 
             \multirow{3}{*}{{\smamath{LR}}}
             & $F^u$ & \res{0.30}{0.014} & \res{0.3}{0.015} &\res{0.29}{0.015} &\res{0.3}{0.015} &\res{0.3}{0.016} &\res{0.3}{0.014} &\res{0.3}{0.014} &\res{0.3}{0.015} &\res{0.3}{0.014} &\res{0.3}{0.021} &\res{0.3}{0.014}\\ 
              & $F^r$ & \res{0.78}{0.011} & \res{0.78}{0.014} &\res{0.68}{0.017} &\res{0.0}{0.0} &\res{0.68}{0.005} &\res{0.78}{0.011} &\res{0.84}{0.006} &\res{1.0}{0.0} &\res{0.78}{0.011} &\res{0.78}{0.011} &\res{0.3}{0.009}\\  
              & $F^c$ & \res{0.86}{0.014} & \res{0.89}{0.011} &\res{0.82}{0.016} &\res{0.17}{0.015} &\res{0.78}{0.01} &\res{0.86}{0.014} &\res{0.92}{0.015} &\res{1.0}{0.005} &\res{0.87}{0.014} &\res{0.74}{0.042} &\res{0.62}{0.025} \\
            \bottomrule 
        \end{tabular}
        \label{tab:iwsp_delta}
    }
\end{table}

\begin{table}[!htbp]
    \centering
    \caption{Evasion Robustness of the ML-PWD against \atks{eWA}{r} and \atks{rWA}{r} on the \dataset{$\delta$Phish}. The cells report the average (and std. dev.) tpr over the 50 reiterations. Lines correspond to the ML-PWD, while rows correspond to a specific eWsP or rWsP attack.}
    
    \resizebox{\columnwidth}{!}{
        \begin{tabular}{c|c|c||cccccccc||cccc}
             \toprule 
              \multirow{2}{*}{\smacal{A}}& \multirow{2}{*}{$F$} & \multirow{2}{*}{no-atk}& \multicolumn{8}{c||}{eWsP} & \multicolumn{4}{c}{rWsP} \\ \cline{4-15}
              
              & & &addImgBot & modFntTyp&modCpy&addIcn&delSusLnk&delSusFrm&modTtl&delCpy&modBgimg&modBgClr&modFntClr&modFntSiz \\
             \toprule
             \multirow{3}{*}{{\smamath{CN}}} 
             & $F^u$ & \res{0.65}{0.028} & \res{0.64}{0.031} &\res{0.64}{0.031} &\res{0.64}{0.031} &\res{0.64}{0.031} &\res{0.64}{0.031} &\res{0.64}{0.031} &\res{0.64}{0.031} &\res{0.64}{0.031} &\res{0.64}{0.031} &\res{0.64}{0.031} &\res{0.63}{0.036} &\res{0.64}{0.031} \\ 
              & $F^r$ & \res{0.79}{0.013}& \res{0.63}{0.063} &\res{0.81}{0.013} &\res{0.77}{0.016} &\res{0.71}{0.024} &\res{0.84}{0.021} &\res{0.75}{0.012} &\res{0.81}{0.013} &\res{0.77}{0.016} &\res{0.81}{0.013} &\res{0.81}{0.013} &\res{0.81}{0.013} &\res{0.81}{0.013}  \\ 
              & $F^c$ & \res{0.95}{0.010} & \res{0.92}{0.032} &\res{0.95}{0.011} &\res{0.94}{0.014} &\res{0.92}{0.021} &\res{0.93}{0.012} &\res{0.93}{0.016} &\res{0.95}{0.011} &\res{0.94}{0.014} &\res{0.95}{0.011} &\res{0.95}{0.011} &\res{0.94}{0.017} &\res{0.95}{0.011} \\  
              \midrule 
             \multirow{3}{*}{{\smamath{RF}}}
             & $F^u$ & \res{0.56}{0.037} &  \res{0.57}{0.034} &\res{0.56}{0.033} &\res{0.56}{0.033} &\res{0.56}{0.033} &\res{0.56}{0.033} &\res{0.56}{0.032} &\res{0.56}{0.033} &\res{0.56}{0.034} &\res{0.56}{0.034} &\res{0.57}{0.034} &\res{0.56}{0.036} &\res{0.56}{0.033} \\ 
              & $F^r$ & \res{0.95}{0.008} &\res{0.88}{0.026} &\res{0.95}{0.008} &\res{0.95}{0.007} &\res{0.89}{0.019} &\res{0.92}{0.011} &\res{0.91}{0.021} &\res{0.95}{0.008} &\res{0.95}{0.007} &\res{0.95}{0.008} &\res{0.95}{0.008} &\res{0.95}{0.008} &\res{0.95}{0.008}  \\ 
              & $F^c$ &\res{0.95}{0.009} &  \res{0.88}{0.015} &\res{0.95}{0.009} &\res{0.94}{0.009} &\res{0.89}{0.015} &\res{0.92}{0.007} &\res{0.91}{0.009} &\res{0.95}{0.009} &\res{0.94}{0.009} &\res{0.95}{0.009} &\res{0.95}{0.009} &\res{0.94}{0.009} &\res{0.95}{0.009}\\ 
              \midrule  
             \multirow{3}{*}{{\smamath{LR}}}
             & $F^u$ & \res{0.30}{0.014} & \res{0.3}{0.014} &\res{0.3}{0.014} &\res{0.3}{0.014} &\res{0.3}{0.015} &\res{0.3}{0.015} &\res{0.3}{0.016} &\res{0.3}{0.015} &\res{0.3}{0.016} &\res{0.3}{0.016} &\res{0.3}{0.014} &\res{0.3}{0.024} &\res{0.3}{0.014} \\ 
              & $F^r$ & \res{0.78}{0.011} & \res{0.47}{0.026} &\res{0.78}{0.011} &\res{0.77}{0.011} &\res{0.61}{0.015} &\res{0.83}{0.007} &\res{0.75}{0.025} &\res{0.79}{0.011} &\res{0.77}{0.011} &\res{0.78}{0.011} &\res{0.78}{0.011} &\res{0.78}{0.011} &\res{0.78}{0.011} \\  
              & $F^c$ & \res{0.86}{0.014} & \res{0.66}{0.028} &\res{0.87}{0.014} &\res{0.89}{0.013} &\res{0.82}{0.013} &\res{0.91}{0.009} &\res{0.78}{0.018} &\res{0.87}{0.014} &\res{0.89}{0.013} &\res{0.87}{0.013} &\res{0.87}{0.014} &\res{0.74}{0.044} &\res{0.87}{0.014} \\
            \bottomrule 
        \end{tabular}
        \label{tab:ewsp_rwsp_delta}
    }
\end{table}

\begin{table}[!htbp]
    \centering
    \caption{Impact of \atks{PA}{r} and \atks{MA}{r} on ML-PWD generated on \dataset{$\delta$phish}. The cells report the average (and std. dev.) tpr over the 50 reiterations. Lines correspond to the ML-PWD, while rows correspond to a specific PsP or MsP attack.}
    
    \resizebox{\columnwidth}{!}{
        \begin{tabular}{c|c|c||ccccccc||cccccc}
             \toprule
             
             
             \multirow{2}{*}{\smacal{A}}& \multirow{2}{*}{$F$} & \multirow{2}{*}{no-atk}& \multicolumn{7}{c||}{PsP} & \multicolumn{6}{c}{MsP} \\ \cline{4-16}

             & & & delTxt&delFrm&delSpn&delTtl&addLngTxt&delFtr&replSusFtrLnk&brTg&delHt&delHd&delBdy&brTgs&hmg \\
             
             \toprule
             \multirow{3}{*}{{\smamath{CN}}} 
             & $F^u$ & \res{0.65}{0.028} &\res{0.64}{0.031} &\res{0.64}{0.031} &\res{0.64}{0.031} &\res{0.64}{0.031} &\res{0.64}{0.031} &\res{0.64}{0.031} &\res{0.64}{0.031} &\res{0.64}{0.031} &\res{0.64}{0.031} &\res{0.64}{0.031} &\res{0.64}{0.031} &\res{0.64}{0.031} &\res{0.65}{0.032}  \\ 
              & $F^r$ & \res{0.79}{0.013}& \res{0.78}{0.014} &\res{0.75}{0.012} &\res{0.8}{0.013} &\res{0.81}{0.013} &\res{0.81}{0.013} &\res{0.76}{0.015} &\res{0.79}{0.011} &\res{0.81}{0.013} &\res{1.0}{0.0} &\res{0.79}{0.009} &\res{0.87}{0.018} &\res{0.81}{0.012} &\res{0.76}{0.019}  \\ 
              & $F^c$ & \res{0.95}{0.010} &\res{0.89}{0.034} &\res{0.93}{0.016} &\res{0.95}{0.012} &\res{0.91}{0.027} &\res{0.95}{0.011} &\res{0.93}{0.013} &\res{0.95}{0.011} &\res{0.95}{0.011} &\res{0.99}{0.014} &\res{0.82}{0.045} &\res{0.98}{0.015} &\res{0.95}{0.011} &\res{0.78}{0.034}   \\ 
              \midrule  
             \multirow{3}{*}{{\smamath{RF}}}
             & $F^u$ & \res{0.56}{0.037} & \res{0.56}{0.033} &\res{0.56}{0.032} &\res{0.57}{0.032} &\res{0.57}{0.032} &\res{0.57}{0.033} &\res{0.56}{0.033} &\res{0.56}{0.035} &\res{0.56}{0.035} &\res{0.56}{0.035} &\res{0.56}{0.033} &\res{0.57}{0.033} &\res{0.56}{0.034} &\res{0.56}{0.036}    \\ 
              & $F^r$ & \res{0.95}{0.008} & \res{0.94}{0.012} &\res{0.91}{0.021} &\res{0.95}{0.007} &\res{0.94}{0.012} &\res{0.95}{0.008} &\res{0.91}{0.01} &\res{0.94}{0.011} &\res{0.95}{0.008} &\res{1.0}{0.0} &\res{0.83}{0.019} &\res{1.0}{0.003} &\res{0.95}{0.008} &\res{0.79}{0.024}  \\  
              & $F^c$ &\res{0.95}{0.009} & \res{0.92}{0.012} &\res{0.91}{0.009} &\res{0.94}{0.009} &\res{0.93}{0.011} &\res{0.95}{0.009} &\res{0.94}{0.01} &\res{0.94}{0.009} &\res{0.95}{0.009} &\res{1.0}{0.0} &\res{0.86}{0.015} &\res{1.0}{0.007} &\res{0.94}{0.009} &\res{0.8}{0.017}  \\ 
              \midrule  
             \multirow{3}{*}{{\smamath{LR}}}
             & $F^u$ & \res{0.30}{0.014} & \res{0.3}{0.015} &\res{0.3}{0.016} &\res{0.3}{0.015} &\res{0.3}{0.016} &\res{0.3}{0.016} &\res{0.3}{0.015} &\res{0.3}{0.015} &\res{0.3}{0.015} &\res{0.3}{0.014} &\res{0.3}{0.014} &\res{0.3}{0.016} &\res{0.3}{0.016} &\res{0.3}{0.014} \\ 
              & $F^r$ & \res{0.78}{0.011} & \res{0.64}{0.024} &\res{0.75}{0.025} &\res{0.75}{0.016} &\res{0.64}{0.025} &\res{0.78}{0.011} &\res{0.79}{0.016} &\res{0.76}{0.011} &\res{0.78}{0.011} &\res{1.0}{0.0} &\res{0.65}{0.01} &\res{0.84}{0.006} &\res{0.78}{0.011} &\res{0.69}{0.01}  \\ 
              & $F^c$ & \res{0.86}{0.014} &\res{0.78}{0.018} &\res{0.78}{0.018} &\res{0.87}{0.014} &\res{0.76}{0.02} &\res{0.87}{0.014} &\res{0.89}{0.014} &\res{0.85}{0.012} &\res{0.87}{0.013} &\res{1.0}{0.004} &\res{0.76}{0.03} &\res{0.95}{0.008} &\res{0.87}{0.013} &\res{0.76}{0.013}  \\
            \bottomrule
             
        \end{tabular}
        \label{tab:psp_msp_delta}
    }
\end{table}
\begin{table}[!htbp]
    \centering
    \caption{Impact of \atks{WA}{u} on ML-PWD of \dataset{$\delta$Phish}.}
    
    \resizebox{\columnwidth}{!}{
        \begin{tabular}{c|c|c||cccccccc}
             \toprule 
              \smacal{A} & $F$ & no-atk &replChar&sepWrd&delChar&swpChar&addChar&atkPth\\
              
             \toprule
             \multirow{3}{*}{{\smamath{CN}}}
             
             & $F^u$ & \res{0.65}{0.028} & \res{0.64}{0.043} &\res{0.64}{0.038} &\res{0.63}{0.033} &\res{0.63}{0.037} &\res{0.64}{0.044} &\res{0.6}{0.029}  \\ 
              & $F^r$ & \res{0.79}{0.013}& \res{0.81}{0.013} &\res{0.79}{0.016} &\res{0.8}{0.014} &\res{0.81}{0.014} &\res{0.81}{0.014} &\res{0.8}{0.013} \\ 
              & $F^c$ & \res{0.95}{0.010} & \res{0.95}{0.009} &\res{0.95}{0.01} &\res{0.94}{0.01} &\res{0.95}{0.011} &\res{0.94}{0.012} &\res{0.94}{0.009} \\ 
              \midrule 
             \multirow{3}{*}{{\smamath{RF}}}
             & $F^u$ & \res{0.56}{0.037} &  \res{0.56}{0.03} &\res{0.59}{0.024} &\res{0.56}{0.029} &\res{0.56}{0.032} &\res{0.56}{0.031} &\res{0.52}{0.027} \\
              & $F^r$ & \res{0.95}{0.008} &\res{0.95}{0.009} &\res{0.95}{0.008} &\res{0.95}{0.009} &\res{0.95}{0.009} &\res{0.95}{0.009} &\res{0.95}{0.008}  \\
              & $F^c$ &\res{0.95}{0.009} &  \res{0.94}{0.009} &\res{0.94}{0.009} &\res{0.92}{0.011} &\res{0.94}{0.009} &\res{0.94}{0.009} &\res{0.94}{0.01}\\
              \midrule 
             \multirow{3}{*}{{\smamath{LR}}}
             & $F^u$ & \res{0.30}{0.014} & \res{0.3}{0.02} &\res{0.31}{0.024} &\res{0.28}{0.019} &\res{0.28}{0.02} &\res{0.29}{0.019} &\res{0.29}{0.015}\\ 
              & $F^r$ & \res{0.78}{0.011} & \res{0.78}{0.011} &\res{0.79}{0.012} &\res{0.77}{0.012} &\res{0.78}{0.012} &\res{0.78}{0.011} &\res{0.78}{0.011} \\  
              & $F^c$ & \res{0.86}{0.014} & \res{0.83}{0.018} &\res{0.85}{0.028} &\res{0.84}{0.016} &\res{0.83}{0.019} &\res{0.83}{0.021} &\res{0.88}{0.01} \\
            \bottomrule 
        \end{tabular}
        \label{tab:delta_url}
    }
\end{table}

\begin{table}[!htbp]
    \centering
    \caption{Impact of \atks{PA}{r}+\atks{WA}{r} on ML-PWD generated on \dataset{$\delta$phish}. The cells report the average (and std. dev.) tpr over the 50 reiterations. Lines correspond to the ML-PWD, while rows correspond to a specific \atk{PsP}+\atk{WsP} perturbation.}
    \begin{subtable}{1\textwidth}
        \centering
        \caption{\atks{PA}{r}+\atks{WA}{r}}
        \resizebox{\columnwidth}{!}{
\begin{tabular}{c|c|c||ccccccccccccc}
                 \toprule 

                  \multirow{3}{*}{\smacal{A}}& \multirow{3}{*}{$F$} & \multirow{3}{*}{no-atk}& \multirow{3}{*}{addLngTxt\_}&\multirow{3}{*}{addLngTxt\_}&\multirow{3}{*}{addLngTxt\_}&\multirow{3}{*}{addLngTxt\_}&\multirow{3}{*}{addLngTxt\_}&\multirow{3}{*}{addLngTxt\_}&\multirow{3}{*}{addLngTxt\_}&\multirow{3}{*}{addLngTxt\_}&\multirow{3}{*}{addLngTxt\_}&\multirow{3}{*}{addLngTxt\_}&\multirow{3}{*}{addLngTxt\_}&\multirow{3}{*}{addLngTxt\_}&\multirow{3}{*}{addLngTxt\_} \\  
              & & & &&& &&&& &&&&&\\
            & & &addInLnk&delHidIt&replOnfoc&replPass&addHidP&replJS&delSusLnk&modCpy&modTtl&addIcn&addSusLnk&addImgBot&modBgClr \\ 
                 \toprule
                 \multirow{3}{*}{{\smamath{CN}}}
                 & $F^u$ & \res{0.65}{0.028} & \res{0.64}{0.031} &\res{0.64}{0.031} &\res{0.64}{0.031} &\res{0.64}{0.031} &\res{0.64}{0.032} &\res{0.64}{0.031} &\res{0.63}{0.036} &\res{0.64}{0.031} &\res{0.64}{0.031} &\res{0.64}{0.031} &\res{0.64}{0.031} &\res{0.64}{0.031} &\res{0.64}{0.031}   \\
                 
                  & $F^r$ & \res{0.79}{0.013}&\res{0.35}{0.017} &\res{0.81}{0.013} &\res{0.81}{0.013} &\res{0.81}{0.013} &\res{0.03}{0.006} &\res{0.79}{0.011} &\res{0.84}{0.021} &\res{0.77}{0.016} &\res{0.81}{0.013} &\res{0.71}{0.024} &\res{0.2}{0.013} &\res{0.63}{0.063} &\res{0.81}{0.013} \\
\bottomrule
            \end{tabular}
            \label{subtab:delta_addlongtext_wsp}
            }
            \end{subtable}
            \vspace{0.5cm}
         
    
        \begin{subtable}{1\textwidth}
        \centering
        \caption{\atks{PA}{r}+\atks{WA}{r}}
        
        \resizebox{\columnwidth}{!}{
            \begin{tabular}{c|c|c||ccccccccccccccc}
                 \toprule
                 
                 \multirow{3}{*}{\smacal{A}}& \multirow{3}{*}{$F$} & \multirow{3}{*}{no-atk}& \multirow{3}{*}{delFrm\_}&\multirow{3}{*}{delFrm\_}&\multirow{3}{*}{delFrm\_}&\multirow{3}{*}{delFrm\_}&\multirow{3}{*}{delFrm\_}&\multirow{3}{*}{delFrm\_}&\multirow{3}{*}{delFrm\_}&\multirow{3}{*}{delFrm\_}&\multirow{3}{*}{delFrm\_}&\multirow{3}{*}{delFrm\_}&\multirow{3}{*}{delFrm\_}&\multirow{3}{*}{delFrm\_}&\multirow{3}{*}{delFrm\_}&\multirow{3}{*}{delFrm\_}&\multirow{3}{*}{delFrm\_} \\  
              & & & &&& &&&& &&&&& &&\\
            & & &addInLnk&onclick&delHidIt&addHidP&replJS&delSusLnk& addImgBot&modFntSiz &modBgimg&modBgClr&delCpy& modTtl&modCpy& addIcn&replRet \\

                 \toprule
                 \multirow{3}{*}{{\smamath{CN}}}
                 &$F^u$ & \res{0.65}{0.028} &  \res{0.64}{0.031} &\res{0.64}{0.031} &\res{0.64}{0.031} &\res{0.64}{0.031} &\res{0.64}{0.031} &\res{0.64}{0.031} &\res{0.64}{0.031} &\res{0.64}{0.031} &\res{0.64}{0.031} &\res{0.64}{0.031} &\res{0.64}{0.031} &\res{0.64}{0.031} &\res{0.64}{0.031} &\res{0.64}{0.031} &\res{0.64}{0.031}  \\
\bottomrule
                 
            \end{tabular}
            \label{subtab:delta_delform_wsp} 
        }
        \end{subtable}
        \vspace{0.5cm} 

        \begin{subtable}{1\textwidth}
        \centering
        \caption{ \atks{PA}{r}+\atks{WA}{r}}
        
        \resizebox{\columnwidth}{!}{
            \begin{tabular}{c|c|c||cccccccccccc}
                 \toprule 

                  \multirow{3}{*}{\smacal{A}}& \multirow{3}{*}{$F$} & \multirow{3}{*}{no-atk}& \multirow{3}{*}{delFtr\_}&\multirow{3}{*}{delFtr\_}&\multirow{3}{*}{delFtr\_}&\multirow{3}{*}{delFtr\_}&\multirow{3}{*}{delFtr\_}&\multirow{3}{*}{delFtr\_}&\multirow{3}{*}{delFtr\_}&\multirow{3}{*}{delFtr\_}&\multirow{3}{*}{delFtr\_}&\multirow{3}{*}{delFtr\_}&\multirow{3}{*}{delFtr\_}&\multirow{3}{*}{delFtr\_}\\  
              & & & &&& &&&& &&&&\\
            & & &addInLnk&delHidIt&replOnfoc&replPass& addHidP&replJS&delSusLnk& modCpy&modTtl& addIcn&addSusLnk &modBgClr \\ 
                 \toprule
                 \multirow{3}{*}{{\smamath{CN}}}
                 
                 & $F^u$ & \res{0.65}{0.028} &  \res{0.64}{0.031} &\res{0.64}{0.031} &\res{0.64}{0.031} &\res{0.64}{0.031} &\res{0.64}{0.031} &\res{0.64}{0.031} &\res{0.64}{0.031} &\res{0.64}{0.031} &\res{0.64}{0.031} &\res{0.64}{0.031} &\res{0.64}{0.031} &\res{0.64}{0.031}  \\
                 
                  & $F^r$ & \res{0.79}{0.013}&  \res{0.26}{0.012} &\res{0.76}{0.014} &\res{0.76}{0.015} &\res{0.76}{0.015} &\res{0.01}{0.01} &\res{0.76}{0.014} &\res{0.83}{0.021} &\res{0.75}{0.017} &\res{0.76}{0.015} &\res{0.64}{0.022} &\res{0.14}{0.015} &\res{0.76}{0.015} \\
                 
                  & $F^c$ & \res{0.95}{0.010} &  \res{0.9}{0.031} &\res{0.92}{0.019} &\res{0.93}{0.013} &\res{0.93}{0.013} &\res{0.22}{0.058} &\res{0.87}{0.019} &\res{0.92}{0.011} &\res{0.92}{0.013} &\res{0.93}{0.013} &\res{0.88}{0.019} &\res{0.77}{0.032} &\res{0.93}{0.013}\\
\bottomrule
                 
            \end{tabular}
            \label{subtab:delta_delfoot_wsp}
            }
        \end{subtable}
        \vspace{0.5cm}
    
        \begin{subtable}{1\textwidth}
        \centering
        \caption{\atks{PA}{r}+\atks{WA}{r}} 
        \resizebox{\columnwidth}{!}{
            \begin{tabular}{c|c|c||cccccccccccc}
                 \toprule 

            \multirow{3}{*}{\smacal{A}}& \multirow{3}{*}{$F$} & \multirow{3}{*}{no-atk}& \multirow{3}{*}{delSpn\_}&\multirow{3}{*}{delSpn\_}&\multirow{3}{*}{delSpn\_}&\multirow{3}{*}{delSpn\_}&\multirow{3}{*}{delSpn\_}&\multirow{3}{*}{delSpn\_}&\multirow{3}{*}{delSpn\_}&\multirow{3}{*}{delSpn\_}&\multirow{3}{*}{delSpn\_}&\multirow{3}{*}{delSpn\_}&\multirow{3}{*}{delSpn\_}&\multirow{3}{*}{delSpn\_}\\  
              & & & &&& &&&& &&&&\\
            & & &addInLnk&replOnfoc&replPass& addHidP&replJS&delSusLnk& modCpy&modTtl& addIcn&addSusLnk &addImgBot&modBgClr \\
            
                 \toprule
                 \multirow{3}{*}{{\smamath{CN}}}
                 
                 & $F^u$ & \res{0.65}{0.028} & \res{0.64}{0.031} &\res{0.64}{0.031} &\res{0.64}{0.031} &\res{0.64}{0.031} &\res{0.64}{0.031} &\res{0.64}{0.031} &\res{0.64}{0.031} &\res{0.64}{0.031} &\res{0.64}{0.031} &\res{0.64}{0.031} &\res{0.64}{0.031} &\res{0.64}{0.031}  \\
                 
                  & $F^r$ & \res{0.79}{0.013}&\res{0.35}{0.019} &\res{0.8}{0.013} &\res{0.8}{0.013} &\res{0.06}{0.025} &\res{0.78}{0.011} &\res{0.82}{0.019} &\res{0.78}{0.014} &\res{0.8}{0.013} &\res{0.69}{0.022} &\res{0.22}{0.023} &\res{0.61}{0.067} &\res{0.8}{0.013} \\
                 
                  & $F^c$ & \res{0.95}{0.010} & \res{0.92}{0.033} &\res{0.95}{0.012} &\res{0.95}{0.012} &\res{0.22}{0.059} &\res{0.89}{0.021} &\res{0.93}{0.013} &\res{0.94}{0.014} &\res{0.95}{0.012} &\res{0.92}{0.022} &\res{0.8}{0.035} &\res{0.89}{0.042} &\res{0.95}{0.012}\\
\bottomrule
                 
            \end{tabular}
            \label{subtab:delta_delspan_wsp}
            }
        \end{subtable}
        \vspace{0.5cm}
        \label{tab:delta_psp_wsp}
        
\end{table}

\begin{table}[!htbp]
    \centering
    \caption{Impact of \atks{PA}{r}+\atks{PA}{r} on ML-PWD generated on \dataset{$\delta$Phish}.The cells report the average (and std. dev.) tpr over the 50 reiterations. Lines correspond to the ML-PWD, while rows correspond to a specific PsP+PsP perturbation.}
    \resizebox{\columnwidth}{!}{
        \begin{tabular}{c|c|c||cccccccc}
             \toprule
             \smacal{A} & $F$ & no-atk & addLngTxt\_delTtl & delFtr\_delTtl & delFtr\_addLngTxt & delSpn\_delTtl & delSpn\_delFtrer & delSpn\_addLngTxt & delFrm\_delFtr & delFrm\_delSpn\\ 
             \toprule
             \multirow{3}{*}{{\smamath{CN}}}
             & $F^u$ & \res{0.65}{0.028} & \res{0.64}{0.031} &\res{0.64}{0.031} &\res{0.64}{0.031} &\res{0.64}{0.031} &\res{0.64}{0.031} &\res{0.64}{0.031} &\res{0.64}{0.031} &\res{0.64}{0.031}\\
              & $F^r$ & \res{0.79}{0.013}&\res{0.81}{0.013} &\res{0.76}{0.015} &\res{0.76}{0.015} &\res{0.8}{0.013} &\res{0.75}{0.015} &\res{0.8}{0.013} &\res{1.0}{0.0} &\res{0.74}{0.01} \\
               & $F^c$ & \res{0.95}{0.010} &\res{0.91}{0.027} &\res{0.89}{0.026} &\res{0.93}{0.013} &\res{0.9}{0.03} &\res{0.92}{0.014} &\res{0.95}{0.012} &\res{0.99}{0.014} &\res{0.9}{0.018} \\
               \midrule  
             \multirow{3}{*}{{\smamath{RF}}}
             & $F^u$ & \res{0.56}{0.037}&\res{0.56}{0.033} &\res{0.57}{0.033} &\res{0.56}{0.032} &\res{0.56}{0.033} &\res{0.57}{0.033} &\res{0.56}{0.034} &\res{0.56}{0.034} &\res{0.56}{0.032} \\
             & $F^r$ & \res{0.95}{0.008} &\res{0.94}{0.012} &\res{0.89}{0.014} &\res{0.91}{0.01} &\res{0.94}{0.011} &\res{0.91}{0.011} &\res{0.95}{0.007} &\res{1.0}{0.0} &\res{0.91}{0.01}\\
             & $F^c$ &\res{0.95}{0.009} & \res{0.93}{0.011} &\res{0.91}{0.016} &\res{0.94}{0.01} &\res{0.92}{0.012} &\res{0.94}{0.01} &\res{0.94}{0.009} &\res{1.0}{0.0} &\res{0.92}{0.008}\\
              \midrule  
             \multirow{3}{*}{{\smamath{LR}}}
             & $F^u$ & \res{0.30}{0.014}&\res{0.3}{0.015} &\res{0.3}{0.015} &\res{0.3}{0.014} &\res{0.3}{0.015} &\res{0.3}{0.015} &\res{0.3}{0.014} &\res{0.3}{0.015} &\res{0.3}{0.014} \\
\bottomrule
        \end{tabular}
        \label{tab:delta_psp_psp}
    }
\end{table}  
\begin{table}[!htbp]
    \centering
    \caption{Impact of \atks{PA}{r}+\atks{MA}{r} attacks on \dataset{$\delta$Phish}.The cells report the average (and std. dev.) tpr over the 50 reiterations. Lines correspond to the ML-PWD, while rows correspond to a specific PsP+MsP.}
    \resizebox{\columnwidth}{!}{
        \begin{tabular}{c|c|c||ccc}
             \toprule
              \smacal{A} & $F$ & no-atk & addLngTxt\_delBdy& delfoot\_delBdy & delSpn\_delBdy\\
             \toprule
             \multirow{3}{*}{{\smamath{CN}}}
             & $F^u$ & \res{0.65}{0.028} & \res{0.64}{0.031} &\res{0.64}{0.031} &\res{0.64}{0.031}\\
              & $F^r$ & \res{0.79}{0.013}& \res{0.81}{0.012} &\res{0.76}{0.015} &\res{0.8}{0.013}\\
               & $F^c$ & \res{0.95}{0.010} &\res{0.95}{0.011} &\res{0.93}{0.013} &\res{0.95}{0.012} \\
               \midrule 
             
             \multirow{3}{*}{{\smamath{RF}}}
             & $F^u$ & \res{0.56}{0.037}& \res{0.56}{0.031} &\res{0.56}{0.034} &\res{0.56}{0.033}\\
             & $F^r$ & \res{0.95}{0.008} &\res{0.95}{0.008} &\res{0.91}{0.01} &\res{0.95}{0.006}\\
             & $F^c$ &\res{0.95}{0.009} &\res{0.94}{0.009} &\res{0.94}{0.01} &\res{0.94}{0.009} \\
              \midrule 
             
             \multirow{3}{*}{{\smamath{LR}}}
             & $F^u$ & \res{0.30}{0.014}& \res{0.3}{0.014} &\res{0.3}{0.014} &\res{0.3}{0.015} \\
             & $F^r$ & \res{0.78}{0.011}& \res{0.78}{0.011} &\res{0.79}{0.016} &\res{0.75}{0.016} \\
             & $F^c$ &\res{0.86}{0.014}  & \res{0.87}{0.014} &\res{0.89}{0.014} &\res{0.87}{0.014}\\

           \bottomrule
        \end{tabular}
        \label{tab:delta_psp_msp}
    }
\end{table} 
\begin{table}[!htbp]
    \centering
    \caption{Impact of \atks{WA}{r}+\atks{WA}{r} on \dataset{$\delta$Phish}.The cells report the average (and std. dev.) tpr over the 50 reiterations. Lines correspond to the ML-PWD, while rows correspond to a specific WsP+WsP.}
    \resizebox{\columnwidth}{!}{
        \begin{tabular}{c|c|c||ccc}
             \toprule
              \smacal{A} & $F$ & no-atk & replOnfoc\_replRet & htEsc\_replRet & htEncd\_replRet\\
              
             \toprule
             \multirow{3}{*}{{\smamath{CN}}}
             & $F^u$ & \res{0.65}{0.028} & \res{0.64}{0.031} &\res{0.63}{0.035} &\res{0.64}{0.031}\\
              & $F^r$ & \res{0.79}{0.013}& \res{0.81}{0.013} &\res{1.0}{0.0} &\res{1.0}{0.0}\\
               & $F^c$ & \res{0.95}{0.010} & \res{0.96}{0.011} &\res{0.97}{0.033} &\res{0.99}{0.014}\\
               \midrule 
             
             \multirow{3}{*}{{\smamath{RF}}}
             & $F^u$ & \res{0.56}{0.037}&\res{0.56}{0.034} &\res{0.56}{0.036} &\res{0.56}{0.032} \\
             & $F^r$ & \res{0.95}{0.008} &\res{0.95}{0.008} &\res{1.0}{0.0} &\res{1.0}{0.0} \\
             & $F^c$ &\res{0.95}{0.009} & \res{0.95}{0.01} &\res{1.0}{0.0} &\res{1.0}{0.0}\\
              \midrule 
             
             \multirow{3}{*}{{\smamath{LR}}}
             & $F^u$ & \res{0.30}{0.014}& \res{0.3}{0.016} &\res{0.3}{0.023} &\res{0.3}{0.015}\\
             & $F^r$ & \res{0.78}{0.011}&\res{0.78}{0.011} &\res{1.0}{0.0} &\res{1.0}{0.0} \\
             & $F^c$ &\res{0.86}{0.014} &\res{0.86}{0.014} &\res{0.91}{0.069} &\res{1.0}{0.004}\\

           \bottomrule
        \end{tabular}
        \label{tab:delta_wsp_wsp}
    }
\end{table}  

\subsection{Perturbation's impact on MLSEC}
We executed 37 kinds of single attacks and report the influence of MLSEC's PWD in Table~\ref{tab:iwsp_mlsec}, ~\ref{tab:ewsp_rwsp_mlsec} and ~\ref{tab:psp_msp_mlsec}, and the influence of hybrid space attacks in Table ~\ref{tab:mlsec_psp_msp}, ~\ref{tab:mlsec_psp_psp}, ~\ref{tab:mlsec_psp_wsp} and ~\ref{tab:mlsec_wsp_wsp}.

\begin{table}[!htbp]
    \centering
    \caption{Impact of \atks{iWA}{r} on the PWD of MLSEC. The cells report the average (and std. dev.) tpr over the 50 reiterations. Lines correspond to the ML-PWD, while rows correspond to a specific iWsP perturbation.}
    \resizebox{\columnwidth}{!}{
\begin{tabular}{c|c||ccccccccccc}
             \toprule 
              \smacal{A} & no-atk &addInLnk&replOnc & delHidIt & addHidP & replJS & replRet & htEsc & htEncd & replPass & replOnfoc & addSusLnk\\
             \toprule
             $m0$ &\res{0.91}{0.052} &\res{0.73}{0.159} &\res{0.91}{0.052} &\res{0.88}{0.049} &\res{0.02}{0.011} &\res{0.9}{0.059} &\res{0.79}{0.123} &\res{0.06}{0.046} &\res{0.06}{0.036} &\res{0.48}{0.259} &\res{0.5}{0.254} &\res{0.5}{0.203} \\
               $m1$ &\res{0.87}{0.071}& \res{0.86}{0.083} &\res{0.88}{0.07} &\res{0.85}{0.093} &\res{0.52}{0.161} &\res{0.85}{0.088} &\res{0.74}{0.115} &\res{0.37}{0.113} &\res{0.41}{0.126} &\res{0.85}{0.068} &\res{0.84}{0.09} &\res{0.85}{0.1}  \\
                $m2$ &\res{0.9}{0.051}& \res{0.73}{0.158} &\res{0.9}{0.052} &\res{0.88}{0.052} &\res{0.02}{0.011} &\res{0.9}{0.058} &\res{0.83}{0.105} &\res{0.85}{0.127} &\res{0.9}{0.051} &\res{0.47}{0.266} &\res{0.51}{0.263} &\res{0.5}{0.202} \\
              $m3$ & \res{0.88}{0.07}& \res{0.86}{0.08} &\res{0.87}{0.07} &\res{0.85}{0.096} &\res{0.51}{0.172} &\res{0.85}{0.087} &\res{0.79}{0.108} &\res{0.86}{0.099} &\res{0.88}{0.07} &\res{0.85}{0.066} &\res{0.85}{0.093} &\res{0.85}{0.1}\\ 
             $m4$ & \res{0.82}{0.106} & \res{0.83}{0.108} &\res{0.82}{0.111} &\res{0.8}{0.136} &\res{0.83}{0.123} &\res{0.82}{0.126} &\res{0.69}{0.122} &\res{0.09}{0.067} &\res{0.07}{0.045} &\res{0.46}{0.256} &\res{0.46}{0.242} &\res{0.47}{0.171}\\
              $m5$ & \res{0.81}{0.12} & \res{0.82}{0.12} &\res{0.81}{0.124} &\res{0.8}{0.141} &\res{0.82}{0.136} &\res{0.81}{0.136} &\res{0.67}{0.114} &\res{0.43}{0.098} &\res{0.46}{0.139} &\res{0.79}{0.125} &\res{0.79}{0.119} &\res{0.8}{0.157} \\ 
              $m6$ & \res{0.83}{0.108}& \res{0.83}{0.11} &\res{0.83}{0.112} &\res{0.81}{0.131} &\res{0.84}{0.116} &\res{0.82}{0.127} &\res{0.73}{0.148} &\res{0.84}{0.126} &\res{0.83}{0.108} &\res{0.47}{0.256} &\res{0.49}{0.236} &\res{0.47}{0.174}\\
\bottomrule
             
        \end{tabular}
        \label{tab:iwsp_mlsec}
    }
\end{table}

\begin{table}[!htbp]
    \centering
    \caption{Evasion Robustness of the MLSEC's PWD against \atks{eWA}{r} and \atks{rWA}{r}. The cells report the average (and std. dev.) tpr over the 50 reiterations. Lines correspond to the PWD, while rows correspond to a specific eWsP or rWsP attack.}
    \resizebox{\columnwidth}{!}{
        \begin{tabular}{c|c||cccccccc||cccc}
             \toprule
              
              \multirow{2}{*}{\smacal{A}}& \multirow{2}{*}{no-atk}& \multicolumn{8}{c||}{eWsP} & \multicolumn{4}{c}{rWsP} \\ \cline{3-14}
              
              & &addImgBot & modFntTyp&modCpy&addIcn&delSusLnk&delSusFrm&modTtl&delCpy&modBgimg&modBgClr&modFntClr&modFntSiz\\ 
             \toprule
             
             $m0$ &\res{0.91}{0.052} &  \res{0.65}{0.185} &\res{0.92}{0.051} &\res{0.89}{0.061} &\res{0.76}{0.126} &\res{0.86}{0.095} &\res{0.8}{0.253} &\res{0.87}{0.102} &\res{0.89}{0.061}&\res{0.92}{0.053} &\res{0.91}{0.056} &\res{0.81}{0.116} &\res{0.92}{0.051} \\
               $m1$ &\res{0.87}{0.071}&  \res{0.87}{0.085} &\res{0.89}{0.063} &\res{0.85}{0.091} &\res{0.77}{0.089} &\res{0.82}{0.146} &\res{0.81}{0.106} &\res{0.84}{0.12} &\res{0.85}{0.089}&\res{0.89}{0.064} &\res{0.88}{0.077} &\res{0.78}{0.1} &\res{0.89}{0.063} \\
                $m2$ &\res{0.9}{0.051}&  \res{0.65}{0.185} &\res{0.91}{0.05} &\res{0.89}{0.06} &\res{0.76}{0.122} &\res{0.86}{0.095} &\res{0.79}{0.262} &\res{0.87}{0.101} &\res{0.89}{0.06} &\res{0.91}{0.052} &\res{0.91}{0.055} &\res{0.85}{0.087} &\res{0.91}{0.05}\\
              $m3$ & \res{0.88}{0.07}& \res{0.87}{0.079} &\res{0.89}{0.064} &\res{0.85}{0.09} &\res{0.77}{0.081} &\res{0.82}{0.146} &\res{0.8}{0.124} &\res{0.84}{0.119} &\res{0.85}{0.088}&\res{0.89}{0.066} &\res{0.88}{0.076} &\res{0.81}{0.091} &\res{0.89}{0.064}  \\ 
             $m4$ & \res{0.82}{0.106} & \res{0.64}{0.199} &\res{0.87}{0.065} &\res{0.81}{0.124} &\res{0.83}{0.109} &\res{0.8}{0.156} &\res{0.73}{0.257} &\res{0.8}{0.156} &\res{0.81}{0.127}&\res{0.87}{0.066} &\res{0.86}{0.079} &\res{0.73}{0.112} &\res{0.87}{0.065} \\
\bottomrule
        \end{tabular}
        \label{tab:ewsp_rwsp_mlsec}
        }
\end{table}

\begin{table}[!htbp]
    \centering
    \caption{Impact of \atks{PA}{r} and \atks{MA}{r} on PWD of MLSEC. The cells report the average (and std. dev.) tpr over the 50 reiterations. Lines correspond to the ML-PWD, while rows correspond to a specific PsP or MsP attack.}
    \resizebox{\columnwidth}{!}{
        \begin{tabular}{c|c|ccccccc||cccccc} 
             \toprule

              \multirow{2}{*}{\smacal{A}}& \multirow{2}{*}{no-atk}& \multicolumn{7}{c||}{PsP} & \multicolumn{6}{c}{MsP} \\ \cline{3-15}

             & & delTxt&delFrm&delSpn&delTtl&addLngTxt&delFtr&replSusFtrLnk&brTg&delHt&delHd&delBdy&brTgs&hmg \\

             \toprule
             $m0$ &\res{0.91}{0.052} & \res{0.64}{0.272} &\res{0.91}{0.052} &\res{0.85}{0.089} &\res{0.88}{0.062} &\res{0.86}{0.095} &\res{0.87}{0.085} &\res{0.9}{0.059}&\res{0.9}{0.062} &\res{0.9}{0.062} &\res{0.9}{0.062} &\res{0.9}{0.062} &\res{0.9}{0.062} &\res{0.9}{0.062} \\
               $m1$ &\res{0.87}{0.071}& \res{0.83}{0.133} &\res{0.67}{0.262} &\res{0.67}{0.262} &\res{0.82}{0.117} &\res{0.83}{0.133} &\res{0.82}{0.117} &\res{0.67}{0.262}&\res{0.31}{0.051} &\res{0.31}{0.051} &\res{0.31}{0.051} &\res{0.31}{0.051} &\res{0.31}{0.051} &\res{0.31}{0.051}  \\
                $m2$ &\res{0.9}{0.051}&  \res{0.84}{0.148} &\res{0.61}{0.39} &\res{0.61}{0.39} &\res{0.85}{0.087} &\res{0.84}{0.148} &\res{0.85}{0.087} &\res{0.61}{0.39}&\res{0.88}{0.096} &\res{0.88}{0.096} &\res{0.88}{0.096} &\res{0.88}{0.096} &\res{0.88}{0.096} &\res{0.88}{0.096} \\
              $m3$ & \res{0.88}{0.07}&\res{0.83}{0.131} &\res{0.66}{0.271} &\res{0.66}{0.271} &\res{0.82}{0.115} &\res{0.83}{0.131} &\res{0.82}{0.115} &\res{0.66}{0.271}&\res{0.26}{0.08} &\res{0.26}{0.08} &\res{0.26}{0.08} &\res{0.26}{0.08} &\res{0.26}{0.08} &\res{0.26}{0.08}  \\ 
             $m4$ & \res{0.82}{0.106} &\res{0.8}{0.169} &\res{0.57}{0.372} &\res{0.57}{0.372} &\res{0.79}{0.149} &\res{0.8}{0.169} &\res{0.79}{0.149} &\res{0.57}{0.372} &\res{0.8}{0.121} &\res{0.8}{0.121} &\res{0.8}{0.121} &\res{0.8}{0.121} &\res{0.8}{0.121} &\res{0.8}{0.121} 
             \\
\bottomrule
        \end{tabular}
        \label{tab:psp_msp_mlsec}
    }
\end{table}

\begin{table}[!htbp]
    \centering
    \caption{ Impact of \atks{PA}{r}+\atks{PA}{r} on MLSEC.The cells report the average (and std. dev.) tpr over the 50 reiterations. Lines correspond to the PWD, while rows correspond to a specific PsP+PsP perturbation.}
    \resizebox{\columnwidth}{!}{
        \begin{tabular}{c|c||cccccccccc}
             \toprule
             \smacal{A} & no-atk &  addLngTxt\_delTtl & delFtr\_delTtl & delFtr\_addLngTxt & delSpn\_delTtl & delSpn\_delFtrer & delSpn\_addLngTxt & delFrm\_delFtr & delFrm\_delSpn \\ 
             
             \toprule
              $m0$ &\res{0.91}{0.052}& \res{0.82}{0.16} &\res{0.85}{0.092} &\res{0.8}{0.167} &\res{0.84}{0.093} &\res{0.84}{0.096} &\res{0.79}{0.173} &\res{0.59}{0.375} &\res{0.58}{0.375} \\
             $m1$ & \res{0.87}{0.071}& \res{0.81}{0.145} &\res{0.84}{0.089} &\res{0.81}{0.13} &\res{0.81}{0.117} &\res{0.83}{0.105} &\res{0.79}{0.157} &\res{0.66}{0.269} &\res{0.62}{0.28} \\
              $m2$ & \res{0.9}{0.051}  &\res{0.81}{0.159} &\res{0.85}{0.094} &\res{0.8}{0.166} &\res{0.83}{0.095} &\res{0.84}{0.096} &\res{0.79}{0.172} &\res{0.59}{0.376} &\res{0.58}{0.376} \\
              $m3$ & \res{0.88}{0.07}& \res{0.81}{0.143} &\res{0.83}{0.093} &\res{0.81}{0.128} &\res{0.81}{0.118} &\res{0.83}{0.106} &\res{0.79}{0.154} &\res{0.65}{0.276} &\res{0.62}{0.288} \\
              $m4$ &  \res{0.82}{0.106}&\res{0.77}{0.178} &\res{0.78}{0.146} &\res{0.75}{0.2} &\res{0.78}{0.145} &\res{0.77}{0.156} &\res{0.76}{0.204} &\res{0.57}{0.358} &\res{0.56}{0.367} \\
              $m5$ & \res{0.81}{0.12}& \res{0.77}{0.172} &\res{0.8}{0.111} &\res{0.77}{0.157} &\res{0.78}{0.141} &\res{0.8}{0.124} &\res{0.77}{0.191} &\res{0.66}{0.259} &\res{0.63}{0.275} \\
             $m6$ &  \res{0.83}{0.108}&\res{0.77}{0.177} &\res{0.77}{0.155} &\res{0.74}{0.198} &\res{0.77}{0.152} &\res{0.77}{0.154} &\res{0.76}{0.201} &\res{0.56}{0.358} &\res{0.56}{0.367} \\
              $m7$ & \res{0.82}{0.121}&\res{0.77}{0.17} &\res{0.8}{0.12} &\res{0.77}{0.155} &\res{0.78}{0.146} &\res{0.8}{0.122} &\res{0.77}{0.186} &\res{0.67}{0.255} &\res{0.63}{0.273}\\
           \bottomrule
        \end{tabular}
        \label{tab:mlsec_psp_psp}
    }
\end{table} 
\begin{table}[!htbp]
    \centering
    \caption{Impact of \atks{PA}{r}+\atks{MA}{r} on MLSEC. The cells report the average (and std. dev.) tpr over the 50 reiterations. Lines correspond to the PWD, while rows correspond to a specific PsP+MsP perturbation.}

    \resizebox{\columnwidth}{!}{
        \begin{tabular}{c|c||cccc}
             \toprule
              \smacal{A} & no-atk & addLngTxt\_delBdy & delfoot\_delBdy & delSpn\_delBdy \\
              
             \toprule  
              $m0$ & \res{0.91}{0.052}&\res{0.78}{0.167} &\res{0.79}{0.137} &\res{0.75}{0.156}\\
              $m1$ & \res{0.87}{0.071}&\res{0.8}{0.146} &\res{0.81}{0.115} &\res{0.78}{0.14}\\
              $m2$ & \res{0.9}{0.051}&\res{0.8}{0.161} &\res{0.8}{0.13} &\res{0.77}{0.149} \\  
              $m3$ & \res{0.88}{0.07}& \res{0.83}{0.125} &\res{0.82}{0.108} &\res{0.79}{0.133}\\
              $m4$ & \res{0.82}{0.106}&\res{0.79}{0.136} &\res{0.76}{0.154} &\res{0.72}{0.184} \\
              $m5$ & \res{0.81}{0.12}&\res{0.8}{0.139} &\res{0.8}{0.122} &\res{0.77}{0.148} \\ 
              $m6$ & \res{0.83}{0.108}&\res{0.79}{0.137} &\res{0.76}{0.154} &\res{0.72}{0.183}  \\
              $m7$ & \res{0.82}{0.121}&\res{0.8}{0.139} &\res{0.8}{0.123} &\res{0.77}{0.145}\\ 
           \bottomrule
        \end{tabular}
        \label{tab:mlsec_psp_msp}
    }
\end{table} 

\begin{table}[!htbp]
    \centering
    \caption{Impact of \atks{PA}{r}+\atks{WA}{r} on PWD of MLSEC. The cells report the average (and std. dev.) tpr over the 50 reiterations. Lines correspond to the PWD, while rows correspond to a specific \atk{PsP}+\atk{WsP} perturbation.}
    \begin{subtable}{1\textwidth}
        \centering
        \caption{\atks{PA}{r}+\atks{WA}{r}} 
        \resizebox{\columnwidth}{!}{
            \begin{tabular}{c|c||cccccccccccccccc}
                 \toprule

                  \multirow{3}{*}{\smacal{A}} & \multirow{3}{*}{no-atk}& \multirow{3}{*}{delFrm\_}&\multirow{3}{*}{delFrm\_}&\multirow{3}{*}{delFrm\_}&\multirow{3}{*}{delFrm\_}&\multirow{3}{*}{delFrm\_}&\multirow{3}{*}{delFrm\_}&\multirow{3}{*}{delFrm\_}&\multirow{3}{*}{delFrm\_}&\multirow{3}{*}{delFrm\_}&\multirow{3}{*}{delFrm\_}&\multirow{3}{*}{delFrm\_}&\multirow{3}{*}{delFrm\_}&\multirow{3}{*}{delFrm\_}&\multirow{3}{*}{delFrm\_}&\multirow{3}{*}{delFrm\_}&\multirow{3}{*}{delFrm\_} \\  
              & & & &&& &&&& &&&&&&&\\
            & & addInLnk&onclick&delHidIt&addHidP&replJS&delSusLnk&addImgBot&modFntSiz& modBgimg&modBgClr&delCpy&modTtl&modCpy &addIcn&replRet&combine \\ 
            
                 \toprule
                 $m0$ &\res{0.91}{0.052} & \res{0.49}{0.335} &\res{0.61}{0.388} &\res{0.6}{0.384} &\res{0.01}{0.008} &\res{0.61}{0.388} &\res{0.59}{0.379} &\res{0.43}{0.325} &\res{0.63}{0.388} &\res{0.63}{0.388} &\res{0.63}{0.387} &\res{0.61}{0.387} &\res{0.59}{0.383} &\res{0.61}{0.386} &\res{0.49}{0.329} &\res{0.51}{0.341} &\res{0.01}{0.008}\\
                   $m1$ &\res{0.87}{0.071}& \res{0.66}{0.268} &\res{0.67}{0.262} &\res{0.63}{0.284} &\res{0.41}{0.181} &\res{0.67}{0.262} &\res{0.62}{0.291} &\res{0.69}{0.243} &\res{0.69}{0.265} &\res{0.69}{0.265} &\res{0.69}{0.265} &\res{0.67}{0.261} &\res{0.65}{0.27} &\res{0.66}{0.261} &\res{0.55}{0.254} &\res{0.53}{0.257} &\res{0.37}{0.196}  \\
\bottomrule
            \end{tabular}
            \label{subtab:delform_wsp_mlsec}
            }
        \end{subtable}
        \vspace{0.5cm} 

    
        \begin{subtable}{1\textwidth}
        \centering
        \caption{ \atks{PA}{r}+\atks{WA}{r}}
        
        \resizebox{\columnwidth}{!}{
            \begin{tabular}{c|c||ccccccccccccc}
                 \toprule

             \multirow{3}{*}{\smacal{A}} & \multirow{3}{*}{no-atk}& \multirow{3}{*}{delSpn\_}&\multirow{3}{*}{delSpn\_}&\multirow{3}{*}{delSpn\_}&\multirow{3}{*}{delSpn\_}&\multirow{3}{*}{delSpn\_}&\multirow{3}{*}{delSpn\_}&\multirow{3}{*}{delSpn\_}&\multirow{3}{*}{delSpn\_}&\multirow{3}{*}{delSpn\_}&\multirow{3}{*}{delSpn\_}&\multirow{3}{*}{delSpn\_}&\multirow{3}{*}{delSpn\_}&\multirow{3}{*}{delSpn\_}\\  
              & & & &&& &&&& &&&&\\
            & & addInLnk &replOnfoc & replPass & addHidP & replJS & delSusLnk & modCpy & modTtl & addIcn &addSusLnk & addImgBot & modBgClr & combine \\

                 \toprule
                 $m0$ &\res{0.91}{0.052} & \res{0.73}{0.137} &\res{0.41}{0.28} &\res{0.39}{0.277} &\res{0.02}{0.007} &\res{0.85}{0.087} &\res{0.84}{0.117} &\res{0.85}{0.091} &\res{0.83}{0.12} &\res{0.7}{0.127} &\res{0.68}{0.195} &\res{0.6}{0.199} &\res{0.88}{0.075} &\res{0.0}{0.004} \\
                   $m1$ &\res{0.87}{0.071}& \res{0.83}{0.112} &\res{0.78}{0.129} &\res{0.8}{0.116} &\res{0.52}{0.149} &\res{0.82}{0.115} &\res{0.79}{0.165} &\res{0.82}{0.119} &\res{0.8}{0.14} &\res{0.7}{0.1} &\res{0.83}{0.117} &\res{0.85}{0.097} &\res{0.86}{0.096} &\res{0.55}{0.106}   \\
                    $m2$ &\res{0.9}{0.051}& \res{0.73}{0.131} &\res{0.42}{0.287} &\res{0.38}{0.282} &\res{0.02}{0.008} &\res{0.85}{0.085} &\res{0.84}{0.117} &\res{0.85}{0.089} &\res{0.83}{0.118} &\res{0.7}{0.125} &\res{0.68}{0.194} &\res{0.59}{0.199} &\res{0.88}{0.073} &\res{0.0}{0.003}  \\
\bottomrule
            \end{tabular}
            \label{subtab:delspan_wsp_mlsec}
             }
        \end{subtable}
        \vspace{0.5cm}
    
    \begin{subtable}{1\textwidth}
        \centering
        \caption{\atks{PA}{r}+\atks{WA}{r}} 
        \resizebox{\columnwidth}{!}{
            \begin{tabular}{c|c||cccccccccccccc}
                 \toprule

                  \multirow{3}{*}{\smacal{A}} & \multirow{3}{*}{no-atk}& \multirow{3}{*}{delFtr\_}&\multirow{3}{*}{delFtr\_}&\multirow{3}{*}{delFtr\_}&\multirow{3}{*}{delFtr\_}&\multirow{3}{*}{delFtr\_}&\multirow{3}{*}{delFtr\_}&\multirow{3}{*}{delFtr\_}&\multirow{3}{*}{delFtr\_}&\multirow{3}{*}{delFtr\_}&\multirow{3}{*}{delFtr\_}&\multirow{3}{*}{delFtr\_}&\multirow{3}{*}{delFtr\_}&\multirow{3}{*}{delFtr\_}&\multirow{3}{*}{delFtr\_}\\  
              & & & &&& &&&& &&&&&\\
            & & addInLnk & delHidIt & replOnfoc & replPass & addHidP & replJS & delSusLnk & modCpy & modTtl & addicon &addSusLnk&addImgBot&modBgClr&combine \\ 
                 \toprule
                 $m0$ &\res{0.91}{0.052} & \res{0.77}{0.132} &\res{0.77}{0.241} &\res{0.49}{0.197} &\res{0.51}{0.196} &\res{0.02}{0.009} &\res{0.87}{0.085} &\res{0.84}{0.118} &\res{0.87}{0.085} &\res{0.85}{0.099} &\res{0.72}{0.148} &\res{0.71}{0.194} &\res{0.67}{0.187} &\res{0.89}{0.071} &\res{0.0}{0.002} \\
                   $m1$ &\res{0.87}{0.071}&  \res{0.85}{0.092} &\res{0.78}{0.165} &\res{0.81}{0.086} &\res{0.83}{0.089} &\res{0.53}{0.148} &\res{0.85}{0.086} &\res{0.81}{0.156} &\res{0.85}{0.086} &\res{0.84}{0.092} &\res{0.73}{0.093} &\res{0.85}{0.099} &\res{0.87}{0.077} &\res{0.87}{0.076} &\res{0.36}{0.096} \\
                    $m2$ &\res{0.9}{0.051}& \res{0.77}{0.131} &\res{0.77}{0.253} &\res{0.48}{0.214} &\res{0.49}{0.215} &\res{0.02}{0.011} &\res{0.87}{0.084} &\res{0.84}{0.117} &\res{0.87}{0.084} &\res{0.85}{0.098} &\res{0.72}{0.147} &\res{0.71}{0.192} &\res{0.67}{0.187} &\res{0.89}{0.07} &\res{0.0}{0.001}  \\
\bottomrule
            \end{tabular}
            \label{subtab:delfoot_wsp_mlsec}
            }
        \end{subtable}
        \vspace{0.5cm}
    \begin{subtable}{1\textwidth}
        \centering
        \caption{\atks{PA}{r}+\atks{WA}{r}} 
        \resizebox{\columnwidth}{!}{
            \begin{tabular}{c|c||cccccccccccccc}
                 \toprule

                  \multirow{3}{*}{\smacal{A}} & \multirow{3}{*}{no-atk}& \multirow{3}{*}{addLngTxt\_}&\multirow{3}{*}{addLngTxt\_}&\multirow{3}{*}{addLngTxt\_}&\multirow{3}{*}{addLngTxt\_}&\multirow{3}{*}{addLngTxt\_}&\multirow{3}{*}{addLngTxt\_}&\multirow{3}{*}{addLngTxt\_}&\multirow{3}{*}{addLngTxt\_}&\multirow{3}{*}{addLngTxt\_}&\multirow{3}{*}{addLngTxt\_}&\multirow{3}{*}{addLngTxt\_}&\multirow{3}{*}{addLngTxt\_}&\multirow{3}{*}{addLngTxt\_}&\multirow{3}{*}{addLngTxt\_}\\  
              & & & &&& &&&& &&&&&\\
            & & addInLnk&delHidIt&replOnfoc&replPass&addHidP&replJS&delSusLnk&modCpy&modTtl&addIcn&addSusLnk&addImgBot&modBgClr&combine \\ 
                 \toprule
                 $m0$ &\res{0.91}{0.052} & \res{0.69}{0.214} &\res{0.73}{0.277} &\res{0.41}{0.245} &\res{0.42}{0.272} &\res{0.02}{0.011} &\res{0.83}{0.158} &\res{0.68}{0.218} &\res{0.83}{0.158} &\res{0.83}{0.158} &\res{0.67}{0.187} &\res{0.68}{0.218} &\res{0.59}{0.227} &\res{0.84}{0.157} &\res{0.0}{0.004}  \\
                   $m1$ &\res{0.87}{0.071}& \res{0.85}{0.105} &\res{0.73}{0.216} &\res{0.8}{0.144} &\res{0.82}{0.116} &\res{0.5}{0.189} &\res{0.82}{0.141} &\res{0.86}{0.103} &\res{0.82}{0.141} &\res{0.82}{0.141} &\res{0.69}{0.136} &\res{0.86}{0.103} &\res{0.84}{0.126} &\res{0.84}{0.131} &\res{0.48}{0.139}  \\
                    $m2$ &\res{0.9}{0.051}& \res{0.69}{0.213} &\res{0.73}{0.281} &\res{0.42}{0.25} &\res{0.41}{0.275} &\res{0.02}{0.011} &\res{0.83}{0.157} &\res{0.68}{0.215} &\res{0.83}{0.157} &\res{0.83}{0.157} &\res{0.66}{0.189} &\res{0.68}{0.215} &\res{0.58}{0.227} &\res{0.84}{0.156} &\res{0.0}{0.004}  \\
\bottomrule
            \end{tabular}
            \label{subtab:addlongtext_wsp_mlsec}
        }
            \end{subtable}
        \vspace{0.5cm}
        \label{tab:mlsec_psp_wsp}
\end{table}

\begin{table}[!htbp]
    \centering
    \caption{Impact of \atks{WA}{r}+\atks{WA}{r} on MLSEC. The cells report the average (and std. dev.) tpr over the 50 reiterations. Lines correspond to the PWD, while rows correspond to a specific WsP+WsP.}
    \resizebox{\columnwidth}{!}{
        \begin{tabular}{c|c||ccccc}
             \toprule
              \smacal{A} & no-atk & replOnfoc\_replRet & htEsc\_replRet & htEncd\_replRet \\
              
             \toprule 
              $m0$ &\res{0.91}{0.052}& \res{0.33}{0.165} &\res{0.03}{0.021} &\res{0.08}{0.04} \\
              $m1$ & \res{0.87}{0.071}& \res{0.63}{0.136} &\res{0.16}{0.065} &\res{0.22}{0.082}\\
              $m2$ & \res{0.9}{0.051}& \res{0.46}{0.229} &\res{0.86}{0.137} &\res{0.9}{0.051}\\ 
              $m3$ & \res{0.88}{0.07}&\res{0.76}{0.13} &\res{0.86}{0.105} &\res{0.88}{0.07}  \\
              $m4$ & \res{0.82}{0.106}&\res{0.31}{0.176} &\res{0.03}{0.017} &\res{0.11}{0.05} \\
              $m5$ & \res{0.81}{0.12} & \res{0.55}{0.124} &\res{0.15}{0.053} &\res{0.28}{0.093}\\ 
              $m6$ & \res{0.83}{0.108}& \res{0.36}{0.221} &\res{0.85}{0.124} &\res{0.83}{0.108}\\
              $m7$ & \res{0.82}{0.121}& \res{0.62}{0.182} &\res{0.85}{0.109} &\res{0.82}{0.121} \\ 

           \bottomrule
        \end{tabular}
        \label{tab:mlsec_wsp_wsp}
    }
\end{table}  

\subsection{Perturbation's impact on \dataset{Zenodo}}

In this section, we present new perturbation's influence on \dataset{Zenodo}, single attacks' influence is shown in Table~\ref{tab:iwsp_zenodo},~\ref{tab:ewsp_rwsp_zenodo},~\ref{tab:psp_msp_zenodo}, and hybrid attacks' impact is shown in Table~\ref{tab:ze_psp_msp},~\ref{tab:ze_psp_psp},~\ref{tab:ze_psp_wsp} ~\ref{tab:ze_wsp_wsp}.

\begin{table}[!htbp]
    \centering
    \caption{Evasion Robustness of the ML-PWD against \atks{iWA}{r} on the \dataset{Zenodo}. The cells report the average (and std. dev.) tpr over the 50 reiterations. Lines correspond to the ML-PWD, while rows correspond to a specific iWsP perturbation.}
    
    \resizebox{\columnwidth}{!}{
\begin{tabular}{c|c|c||cccccccccc}
             \toprule 
              \smacal{A} & $F$ & no-atk & addInLnk & delHidIt & addHidP & replJS & replRet & htEsc & htEncd & replPass & replOnfoc & addSusLnk \\ 
             \toprule
             \multirow{3}{*}{{\smamath{CN}}}
             
             & $F^u$ & \res{0.96}{0.007}&\res{0.95}{0.018} &\res{0.95}{0.018} &\res{0.93}{0.017} &\res{0.96}{0.013} &\res{0.96}{0.013} &\res{0.92}{0.023} &\res{0.96}{0.013} &\res{0.96}{0.013} &\res{0.95}{0.018} &\res{0.96}{0.013} \\
             
              & $F^r$ & \res{0.86}{0.013} & \res{0.61}{0.034} &\res{0.88}{0.012} &\res{0.28}{0.008} &\res{0.74}{0.05} &\res{0.88}{0.013} &\res{0.87}{0.025} &\res{0.0}{0.0} &\res{0.88}{0.013} &\res{0.88}{0.013} &\res{0.48}{0.022}\\
             
              & $F^c$ & \res{0.97}{0.009}& \res{0.97}{0.021} &\res{0.96}{0.016} &\res{0.86}{0.027} &\res{0.95}{0.013} &\res{0.97}{0.012} &\res{0.92}{0.019} &\res{0.9}{0.033} &\res{0.97}{0.012} &\res{0.97}{0.012} &\res{0.97}{0.021}\\  
              \midrule 
             
             \multirow{3}{*}{{\smamath{RF}}}
             & $F^u$ &\res{0.96}{0.007} & \res{0.96}{0.008} &\res{0.96}{0.008} &\res{0.98}{0.005} &\res{0.98}{0.005} &\res{0.98}{0.005} &\res{0.98}{0.005} &\res{0.98}{0.005} &\res{0.98}{0.005} &\res{0.96}{0.007} &\res{0.98}{0.005} \\
             
              & $F^r$ & \res{0.90}{0.013} &\res{0.94}{0.018} &\res{0.84}{0.01} &\res{0.03}{0.064} &\res{0.71}{0.016} &\res{0.9}{0.013} &\res{0.84}{0.027} &\res{0.0}{0.0} &\res{0.9}{0.013} &\res{0.9}{0.013} &\res{0.64}{0.062}\\ 
              & $F^c$ & \res{0.97}{0.009}& \res{0.97}{0.008} &\res{0.97}{0.01} &\res{0.96}{0.006} &\res{0.96}{0.006} &\res{0.98}{0.004} &\res{0.96}{0.007} &\res{0.96}{0.007} &\res{0.98}{0.005} &\res{0.97}{0.01} &\res{0.97}{0.008}\\
\bottomrule
             
        \end{tabular}
        \label{tab:iwsp_zenodo}
    }
\end{table}

\begin{table}[!htbp]
    \centering
    \caption{Evasion Robustness of the ML-PWD against \atks{eWA}{r} and \atks{rWA}{r} on the \dataset{Zenodo}. The cells report the average (and std. dev.) tpr over the 50 reiterations. Lines correspond to the ML-PWD, while rows correspond to a specific eWsP or rWsP perturbation.}
    \resizebox{\columnwidth}{!}{
        \begin{tabular}{c|c|c||cccccccc||cccc}
             \toprule

              \multirow{2}{*}{\smacal{A}}& \multirow{2}{*}{$F$} & \multirow{2}{*}{no-atk}& \multicolumn{8}{c||}{eWsP} & \multicolumn{4}{c}{rWsP} \\ \cline{4-15}
              
              & & &addImgBot & modFntTyp&modCpy&addIcn&delSusLnk&delSusFrm&modTtl&delCpy&modBgimg&modBgClr&modFntClr&modFntSiz \\
              
             \toprule
             \multirow{3}{*}{{\smamath{CN}}}
             
             & $F^u$ & \res{0.96}{0.007}& \res{0.96}{0.013} &\res{0.96}{0.013} &\res{0.96}{0.013} &\res{0.96}{0.013} &\res{0.96}{0.013} &\res{0.96}{0.013} &\res{0.96}{0.013} &\res{0.96}{0.013} &\res{0.96}{0.013} &\res{0.96}{0.013} &\res{0.96}{0.013} &\res{0.96}{0.013} \\
             
              & $F^r$ & \res{0.86}{0.013} &\res{0.74}{0.06} &\res{0.88}{0.013} &\res{0.76}{0.045} &\res{0.82}{0.019} &\res{0.91}{0.016} &\res{0.86}{0.036} &\res{0.88}{0.013} &\res{0.76}{0.045}&\res{0.88}{0.013} &\res{0.88}{0.013} &\res{0.88}{0.013} &\res{0.88}{0.013}  \\
             
              & $F^c$ & \res{0.97}{0.009}&\res{0.97}{0.01} &\res{0.97}{0.012} &\res{0.97}{0.013} &\res{0.96}{0.012} &\res{0.96}{0.012} &\res{0.97}{0.012} &\res{0.97}{0.013} &\res{0.97}{0.013}&\res{0.97}{0.012} &\res{0.97}{0.012} &\res{0.97}{0.012} &\res{0.97}{0.012}  \\  
              \midrule  
             \multirow{3}{*}{{\smamath{RF}}}
             & $F^u$ &\res{0.96}{0.007} & \res{0.98}{0.005} &\res{0.98}{0.005} &\res{0.98}{0.005} &\res{0.98}{0.005} &\res{0.98}{0.005} &\res{0.98}{0.005} &\res{0.98}{0.005} &\res{0.98}{0.005} &\res{0.98}{0.005} &\res{0.98}{0.005} &\res{0.98}{0.005} &\res{0.98}{0.005}  \\
\bottomrule 
        \end{tabular}
        \label{tab:ewsp_rwsp_zenodo}
    }
\end{table}

\begin{table}[!htbp]
    \centering
    \caption{Impact of \atks{WA}{u} on ML-PWD of \dataset{Zenodo}. The cells report the average (and std. dev.) tpr over the 50 reiterations. Lines correspond to the ML-PWD, while rows correspond to a specific \atk{iWsP} perturbation.}
    
    \resizebox{\columnwidth}{!}{
        \begin{tabular}{c|c|c||ccccccc}
             \toprule
        
              \smacal{A} & $F$ & no-atk & replChar & sepWrd & delChar & swpChar & addChar & atkPth \\ 
             \toprule
             \multirow{3}{*}{{\smamath{CN}}} 
             & $F^u$ & \res{0.96}{0.007}& \res{0.98}{0.012} &\res{0.95}{0.024} &\res{0.99}{0.009} &\res{0.99}{0.013} &\res{0.99}{0.007} &\res{0.97}{0.017} \\ 
              & $F^r$ & \res{0.86}{0.013} &\res{0.5}{0.043} &\res{0.49}{0.04} &\res{0.5}{0.043} &\res{0.49}{0.04} &\res{0.49}{0.038} &\res{0.5}{0.043}  \\ 
              & $F^c$ & \res{0.97}{0.009}& \res{0.98}{0.017} &\res{0.95}{0.025} &\res{0.99}{0.024} &\res{0.99}{0.019} &\res{0.99}{0.017} &\res{0.97}{0.021} \\  
              \midrule  
             \multirow{3}{*}{{\smamath{RF}}}
             & $F^u$ &\res{0.96}{0.007} & \res{1.0}{0.004} &\res{0.98}{0.0} &\res{1.0}{0.005} &\res{1.0}{0.004} &\res{1.0}{0.006} &\res{0.99}{0.002}  \\ 
              & $F^r$ & \res{0.90}{0.013} & \res{0.73}{0.043} &\res{0.73}{0.043} &\res{0.74}{0.043} &\res{0.74}{0.041} &\res{0.75}{0.041} &\res{0.73}{0.043}   \\ 
              & $F^c$ & \res{0.97}{0.009}& \res{1.0}{0.005} &\res{0.99}{0.001} &\res{1.0}{0.0} &\res{1.0}{0.002} &\res{1.0}{0.003} &\res{0.98}{0.006} \\ 
              \midrule  
             \multirow{3}{*}{{\smamath{LR}}}
             & $F^u$ & \res{0.97}{0.005} &  \res{0.99}{0.002} &\res{0.99}{0.003} &\res{1.0}{0.003} &\res{1.0}{0.0} &\res{0.99}{0.001} &\res{0.97}{0.007} \\ 
              & $F^r$ & \res{0.80}{0.013} &\res{0.78}{0.0} &\res{0.79}{0.0} &\res{0.79}{0.0} &\res{0.79}{0.0} &\res{0.8}{0.0} &\res{0.78}{0.0}    \\ 
              & $F^c$ & \res{0.98}{0.005} & \res{0.99}{0.0} &\res{1.0}{0.001} &\res{1.0}{0.0} &\res{1.0}{0.0} &\res{1.0}{0.0} &\res{0.97}{0.004} \\
            \bottomrule
             
        \end{tabular}
        \label{tab:ze_url}
    }
\end{table}

\begin{table}[!htbp]
    \centering
    \caption{Impact of \atks{PA}{r} and \atks{MA}{r} on ML-PWD generated on \dataset{Zenodo}. The cells report the average (and std. dev.) tpr over the 50 reiterations. Lines correspond to the ML-PWD, while rows correspond to a specific PsP or MsP attack.}
    \resizebox{\columnwidth}{!}{
        \begin{tabular}{c|c|c||ccccccc||cccccc}
             \toprule 

            \multirow{2}{*}{\smacal{A}}& \multirow{2}{*}{$F$} & \multirow{2}{*}{no-atk}& \multicolumn{7}{c||}{PsP} & \multicolumn{6}{c}{MsP} \\ \cline{4-16}

             & & & delTxt & delFrm & delSpn & delTtl & addLngTxt & delFtr & replSusFtrLnk & brTg & delHt & delHd & delBdy & brTgs & hmg\\

             \toprule
             \multirow{3}{*}{{\smamath{CN}}} 
             & $F^u$ & \res{0.96}{0.007}& \res{0.96}{0.013} &\res{0.96}{0.013} &\res{0.96}{0.013} &\res{0.96}{0.013} &\res{0.96}{0.013} &\res{0.96}{0.013} &\res{0.96}{0.013} &\res{0.96}{0.013} &\res{0.96}{0.013} &\res{0.96}{0.013} &\res{0.96}{0.013} &\res{0.96}{0.013} &\res{0.97}{0.014}  \\ 
              & $F^r$ & \res{0.86}{0.013} & \res{0.78}{0.046} &\res{0.86}{0.036} &\res{0.88}{0.015} &\res{0.88}{0.013} &\res{0.88}{0.013} &\res{0.64}{0.052} &\res{0.74}{0.05} &\res{0.88}{0.013} &\res{0.0}{0.0} &\res{0.82}{0.011} &\res{0.43}{0.049} &\res{0.88}{0.013} &\res{0.4}{0.035} \\ 
              & $F^c$ & \res{0.97}{0.009}&  \res{0.97}{0.012} &\res{0.97}{0.012} &\res{0.97}{0.012} &\res{0.98}{0.011} &\res{0.97}{0.012} &\res{0.96}{0.017} &\res{0.96}{0.012} &\res{0.97}{0.012} &\res{0.9}{0.033} &\res{0.95}{0.02} &\res{0.93}{0.019} &\res{0.97}{0.012} &\res{0.92}{0.026} \\  
              \midrule  
             \multirow{3}{*}{{\smamath{RF}}}
             & $F^u$ &\res{0.96}{0.007} &  \res{0.98}{0.005} &\res{0.98}{0.005} &\res{0.98}{0.005} &\res{0.98}{0.005} &\res{0.98}{0.005} &\res{0.98}{0.005} &\res{0.98}{0.005} &\res{0.97}{0.007} &\res{0.98}{0.005} &\res{0.98}{0.005} &\res{0.98}{0.005} &\res{0.98}{0.005} &\res{0.98}{0.006}  \\
\bottomrule   
        \end{tabular}
        \label{tab:psp_msp_zenodo}
    }
\end{table}
\begin{table}[!htbp]
    \centering
    \caption{Impact of \atks{PA}{r}+\atks{PA}{r} on Zenodo}
    \resizebox{\columnwidth}{!}{
        \begin{tabular}{c|c|c|cccccccccccccc}
             \toprule 
              \smacal{A} & $F$ & no-atk & addLngTxt\_delTtl & delFtr\_delTtl & delFtr\_addLngTxt & delSpn\_delTtl & delSpn\_delFtrer & delSpn\_addLngTxt & delFrm\_delFtr & delFrm\_delSpn \\
             \toprule
             \multirow{3}{*}{{\smamath{CN}}} 
             & $F^u$ & \res{0.96}{0.007}&\res{0.96}{0.013} &\res{0.96}{0.013} &\res{0.96}{0.013} &\res{0.96}{0.013} &\res{0.96}{0.013} &\res{0.96}{0.013} &\res{0.96}{0.013} &\res{0.96}{0.013}\\
             & $F^r$ & \res{0.86}{0.013}&\res{0.88}{0.013} &\res{0.64}{0.052} &\res{0.64}{0.052} &\res{0.88}{0.015} &\res{0.64}{0.052} &\res{0.88}{0.015} &\res{0.78}{0.064} &\res{0.83}{0.045}\\
             & $F^c$ & \res{0.97}{0.009}&\res{0.98}{0.011} &\res{0.96}{0.015} &\res{0.96}{0.017} &\res{0.98}{0.011} &\res{0.96}{0.016} &\res{0.98}{0.012} &\res{0.96}{0.015} &\res{0.97}{0.015} \\
             \midrule 
             \multirow{3}{*}{{\smamath{RF}}}
             & $F^u$ &\res{0.96}{0.007}&\res{0.98}{0.005} &\res{0.98}{0.005} &\res{0.98}{0.005} &\res{0.98}{0.005} &\res{0.98}{0.005} &\res{0.98}{0.005} &\res{0.98}{0.005} &\res{0.98}{0.005}\\
             & $F^r$ & \res{0.90}{0.013}&\res{0.86}{0.032} &\res{0.8}{0.042} &\res{0.86}{0.018} &\res{0.87}{0.025} &\res{0.86}{0.018} &\res{0.9}{0.012} &\res{0.7}{0.052} &\res{0.79}{0.034}\\
             & $F^c$ & \res{0.97}{0.009}&\res{0.98}{0.004} &\res{0.97}{0.007} &\res{0.97}{0.006} &\res{0.98}{0.005} &\res{0.97}{0.007} &\res{0.98}{0.005} &\res{0.97}{0.007} &\res{0.98}{0.005}\\
             \midrule 
             \multirow{3}{*}{{\smamath{LR}}}
             & $F^u$ & \res{0.97}{0.005}&\res{0.97}{0.004} &\res{0.97}{0.004} &\res{0.97}{0.004} &\res{0.97}{0.004} &\res{0.97}{0.004} &\res{0.97}{0.004} &\res{0.97}{0.004} &\res{0.97}{0.004}\\
\bottomrule
        \end{tabular}
        \label{tab:ze_psp_psp}
    }
\end{table} 

\begin{table}[!htbp]
    \centering
    \caption{Impact of \atks{PA}{r}+\atks{WA}{r} on ML-PWD generated on \dataset{Zenodo}. The cells report the average (and std. dev.) tpr over the 50 reiterations. Lines correspond to the ML-PWD, while rows correspond to a specific \atk{PsP}+\atk{WsP} perturbation.}
    \begin{subtable}{1\textwidth}
    \centering
    \caption{\atks{PA}{r}+\atks{WA}{r}} 
        \resizebox{\columnwidth}{!}{
            \begin{tabular}{c|c|c||ccccccccccccc}
                 \toprule

                \multirow{3}{*}{\smacal{A}}& \multirow{3}{*}{$F$} & \multirow{3}{*}{no-atk}& \multirow{3}{*}{addLngTxt\_}&\multirow{3}{*}{addLngTxt\_}&\multirow{3}{*}{addLngTxt\_}&\multirow{3}{*}{addLngTxt\_}&\multirow{3}{*}{addLngTxt\_}&\multirow{3}{*}{addLngTxt\_}&\multirow{3}{*}{addLngTxt\_}&\multirow{3}{*}{addLngTxt\_}&\multirow{3}{*}{addLngTxt\_}&\multirow{3}{*}{addLngTxt\_}&\multirow{3}{*}{addLngTxt\_}&\multirow{3}{*}{addLngTxt\_}&\multirow{3}{*}{addLngTxt\_} \\  
              & & & &&& &&&& &&&&&\\
            & & &addInLnk&delHidIt&replOnfoc&replPass&addHidP&replJS&delSusLnk&modCpy&modTtl&addIcn&addSusLnk&addImgBot&modBgClr \\

                 \toprule
                 \multirow{3}{*}{{\smamath{CN}}} 
                 & $F^u$ & \res{0.96}{0.007}&  \res{0.96}{0.013} &\res{0.96}{0.013} &\res{0.96}{0.013} &\res{0.96}{0.013} &\res{0.96}{0.013} &\res{0.96}{0.013} &\res{0.96}{0.013} &\res{0.96}{0.013} &\res{0.96}{0.013} &\res{0.96}{0.013} &\res{0.96}{0.013} &\res{0.96}{0.013} &\res{0.96}{0.013} \\ 
                  & $F^r$ & \res{0.86}{0.013} & \res{0.61}{0.034} &\res{0.88}{0.012} &\res{0.88}{0.013} &\res{0.88}{0.013} &\res{0.28}{0.008} &\res{0.74}{0.05} &\res{0.91}{0.016} &\res{0.76}{0.045} &\res{0.88}{0.013} &\res{0.82}{0.019} &\res{0.5}{0.043} &\res{0.72}{0.058} &\res{0.88}{0.013} \\ 
                  & $F^c$ & \res{0.97}{0.009}& \res{0.97}{0.021} &\res{0.97}{0.016} &\res{0.98}{0.013} &\res{0.98}{0.013} &\res{0.92}{0.024} &\res{0.95}{0.015} &\res{0.96}{0.012} &\res{0.98}{0.013} &\res{0.98}{0.013} &\res{0.96}{0.013} &\res{0.97}{0.02} &\res{0.96}{0.013} &\res{0.98}{0.013} \\
\bottomrule
                 
            \end{tabular}
            \label{subtab:ze_addlongtext_wsp}
        }
            \end{subtable}
            \vspace{0.5cm}
            
        \begin{subtable}{1\textwidth}
        \centering
        \caption{\atks{PA}{r}+\atks{WA}{r}}
    
        \resizebox{\columnwidth}{!}{
            \begin{tabular}{c|c|c||cccccccccccc}
                 \toprule

                \multirow{3}{*}{\smacal{A}}& \multirow{3}{*}{$F$} & \multirow{3}{*}{no-atk}& \multirow{3}{*}{delFtr\_}&\multirow{3}{*}{delFtr\_}&\multirow{3}{*}{delFtr\_}&\multirow{3}{*}{delFtr\_}&\multirow{3}{*}{delFtr\_}&\multirow{3}{*}{delFtr\_}&\multirow{3}{*}{delFtr\_}&\multirow{3}{*}{delFtr\_}&\multirow{3}{*}{delFtr\_}&\multirow{3}{*}{delFtr\_}&\multirow{3}{*}{delFtr\_}&\multirow{3}{*}{delFtr\_}\\  
              & & & &&& &&&& &&&&\\
            & & &addInLnk & delHidIt & replOnfoc & replPass & addHidP & replJS & delSusLnk & modCpy & modTtl & addIcn & addSusLnk & modBgClr \\  
                 \toprule
                 \multirow{3}{*}{{\smamath{CN}}} 
                 & $F^u$ & \res{0.96}{0.007}& \res{0.96}{0.013} &\res{0.96}{0.013} &\res{0.96}{0.013} &\res{0.94}{0.03} &\res{0.96}{0.013} &\res{0.96}{0.013} &\res{0.96}{0.013} &\res{0.96}{0.013} &\res{0.96}{0.013} &\res{0.96}{0.013} &\res{0.96}{0.013} &\res{0.96}{0.013} \\ 
                  & $F^r$ & \res{0.86}{0.013} & \res{0.33}{0.053} &\res{0.64}{0.052} &\res{0.64}{0.052} &\res{0.54}{0.049} &\res{0.54}{0.049} &\res{0.64}{0.052} &\res{0.85}{0.057} &\res{0.63}{0.05} &\res{0.64}{0.052} &\res{0.55}{0.055} &\res{0.28}{0.02} &\res{0.64}{0.052} \\ 
                  & $F^c$ & \res{0.97}{0.009}&  \res{0.96}{0.028} &\res{0.96}{0.016} &\res{0.96}{0.017} &\res{0.83}{0.035} &\res{0.91}{0.025} &\res{0.94}{0.012} &\res{0.96}{0.011} &\res{0.96}{0.017} &\res{0.96}{0.017} &\res{0.95}{0.015} &\res{0.96}{0.027} &\res{0.96}{0.017} \\
\bottomrule
                 
            \end{tabular}
            \label{subtab:ze_delfoot_wsp}
       }
        \end{subtable}
        \vspace{0.5cm} 
        
    \begin{subtable}{1\textwidth}
    \centering
    \caption{\atks{PA}{r}+\atks{WA}{r}} 
        
        \resizebox{\columnwidth}{!}{
            \begin{tabular}{c|c|c||ccccccccccccccc}
                 \toprule
                  \multirow{3}{*}{\smacal{A}}& \multirow{3}{*}{$F$} & \multirow{3}{*}{no-atk}& \multirow{3}{*}{delFrm\_}&\multirow{3}{*}{delFrm\_}&\multirow{3}{*}{delFrm\_}&\multirow{3}{*}{delFrm\_}&\multirow{3}{*}{delFrm\_}&\multirow{3}{*}{delFrm\_}&\multirow{3}{*}{delFrm\_}&\multirow{3}{*}{delFrm\_}&\multirow{3}{*}{delFrm\_}&\multirow{3}{*}{delFrm\_}&\multirow{3}{*}{delFrm\_}&\multirow{3}{*}{delFrm\_}&\multirow{3}{*}{delFrm\_}&\multirow{3}{*}{delFrm\_}&\multirow{3}{*}{delFrm\_} \\  
                  
              & & & &&& &&&& &&&&& &&\\
            & & &addInLnk & onclick & delHidIt & addHidP &
                  replJS\ & delSusLnk & addImgBot & modFntSiz &
                  modBgimg & modBgClr & delCpy & modTtl  & 
                  modCpy & addIcn & replRet \\ 
                 \toprule
                 \multirow{3}{*}{{\smamath{CN}}}
                 
                 & $F^u$ & \res{0.96}{0.007}& \res{0.96}{0.013} &\res{0.96}{0.013} &\res{0.96}{0.013} &\res{0.96}{0.013} &\res{0.94}{0.018} &\res{0.96}{0.013} &\res{0.96}{0.013} &\res{0.96}{0.013} &\res{0.96}{0.013} &\res{0.96}{0.013} &\res{0.96}{0.013} &\res{0.96}{0.013} &\res{0.96}{0.013} &\res{0.96}{0.013} &\res{0.96}{0.013}  \\
                 
                  & $F^r$ & \res{0.86}{0.013} & \res{0.56}{0.043} &\res{0.87}{0.024} &\res{0.84}{0.038} &\res{0.27}{0.008} &\res{0.76}{0.054} &\res{0.87}{0.028} &\res{0.75}{0.052} &\res{0.84}{0.038} &\res{0.84}{0.038} &\res{0.84}{0.038} &\res{0.78}{0.064} &\res{0.84}{0.038} &\res{0.78}{0.064} &\res{0.82}{0.042} &\res{0.84}{0.038}  \\
\bottomrule
                 
            \end{tabular}
            \label{subtab:ze_delform_wsp}
        } 
        \end{subtable}
        \vspace{0.5cm} 
        
    \begin{subtable}{1\textwidth}
        \centering
        \caption{ \atks{PA}{r}+\atks{WA}{r}} 
        \resizebox{\columnwidth}{!}{
            \begin{tabular}{c|c|c|cccccccccccc}
                 \toprule 

                \multirow{3}{*}{\smacal{A}}& \multirow{3}{*}{$F$} & \multirow{3}{*}{no-atk}& \multirow{3}{*}{delSpn\_}&\multirow{3}{*}{delSpn\_}&\multirow{3}{*}{delSpn\_}&\multirow{3}{*}{delSpn\_}&\multirow{3}{*}{delSpn\_}&\multirow{3}{*}{delSpn\_}&\multirow{3}{*}{delSpn\_}&\multirow{3}{*}{delSpn\_}&\multirow{3}{*}{delSpn\_}&\multirow{3}{*}{delSpn\_}&\multirow{3}{*}{delSpn\_}&\multirow{3}{*}{delSpn\_}\\  
                  & & & &&& &&&& &&&&\\
                & & &addInLnk & replOnfoc & replPass & addHidP & replJS & delSusLnk & modCpy & modTtl & addIcn & addSusLnk & addImgBot & modBgClr \\
                 \toprule
                 \multirow{3}{*}{{\smamath{CN}}} 
                 & $F^u$ & \res{0.96}{0.007}& \res{0.94}{0.021} &\res{0.96}{0.013} &\res{0.96}{0.013} &\res{0.96}{0.013} &\res{0.96}{0.013} &\res{0.96}{0.013} &\res{0.96}{0.013} &\res{0.96}{0.013} &\res{0.96}{0.013} &\res{0.96}{0.013} &\res{0.96}{0.013} &\res{0.96}{0.013} \\ 
                  & $F^r$ & \res{0.86}{0.013} & \res{0.63}{0.025} &\res{0.88}{0.015} &\res{0.88}{0.014} &\res{0.27}{0.008} &\res{0.74}{0.052} &\res{0.89}{0.019} &\res{0.76}{0.044} &\res{0.88}{0.015} &\res{0.83}{0.02} &\res{0.4}{0.02} &\res{0.75}{0.062} &\res{0.88}{0.015}   \\ 
                  & $F^c$ & \res{0.97}{0.009}&  \res{0.95}{0.02} &\res{0.98}{0.012} &\res{0.98}{0.011} &\res{0.92}{0.024} &\res{0.95}{0.015} &\res{0.96}{0.012} &\res{0.98}{0.012} &\res{0.98}{0.012} &\res{0.96}{0.013} &\res{0.95}{0.019} &\res{0.97}{0.011} &\res{0.98}{0.012} \\
\bottomrule   
            \end{tabular}
            \label{subtab:ze_delspan_com}
       }
        \end{subtable}
        \vspace{0.5cm}
        \label{tab:ze_psp_wsp}
        
\end{table}

\begin{table}[!htbp]
    \centering
    \caption{Impact of \atks{PA}{r}+\atks{MA}{r} in \dataset{Zenodo}}
    \resizebox{\columnwidth}{!}{
        \begin{tabular}{c|c|c|cccccccccccccc}
             \toprule 
              \smacal{A} & $F$ & no-atk& addLngTxt\_delBdy& delFtr\_delBdy & delSpn\_delBdy \\ 
             \toprule
             \multirow{3}{*}{{\smamath{CN}}} 
             & $F^u$ & \res{0.96}{0.007}&\res{0.96}{0.013} &\res{0.96}{0.013} &\res{0.96}{0.013}  \\
             & $F^r$ & \res{0.86}{0.013}&\res{0.88}{0.013} &\res{0.64}{0.052} &\res{0.88}{0.015}  \\
             & $F^c$ & \res{0.97}{0.009}&\res{0.98}{0.013} &\res{0.96}{0.017} &\res{0.98}{0.012}  \\
             \midrule 
             \multirow{3}{*}{{\smamath{RF}}}
             & $F^u$ &\res{0.96}{0.007}&\res{0.98}{0.005} &\res{0.98}{0.005} &\res{0.98}{0.005}  \\
             & $F^r$ & \res{0.90}{0.013}&\res{0.9}{0.013} &\res{0.86}{0.018} &\res{0.9}{0.012}  \\
             & $F^c$ & \res{0.97}{0.009}&\res{0.98}{0.005} &\res{0.97}{0.006} &\res{0.98}{0.005} \\
             \midrule 
             \multirow{3}{*}{{\smamath{LR}}}
             & $F^u$ & \res{0.97}{0.005}&\res{0.97}{0.004} &\res{0.97}{0.004} &\res{0.97}{0.004}  \\
             & $F^r$ & \res{0.80}{0.013}&\res{0.8}{0.013} &\res{0.76}{0.008} &\res{0.8}{0.013}  \\
             & $F^c$ & \res{0.98}{0.005}&\res{0.98}{0.002} &\res{0.98}{0.004} &\res{0.98}{0.001}  \\
            \bottomrule        
        \end{tabular}
        \label{tab:ze_psp_msp}
    }
\end{table} 

\begin{table}[!htbp]
    \centering
    \caption{Impact of \atks{WA}{r}+\atks{WA}{r} on \dataset{Zenodo}}
    \resizebox{\columnwidth}{!}{
        \begin{tabular}{c|c|c|cccccccccccccc}
             \toprule
              \smacal{A} & $F$ & no-atk& replOnfoc\_replRet & htEsc\_replRet & htEncd\_replRet \\
              
             \toprule
            \multirow{3}{*}{{\smamath{CN}}} 
             & $F^u$ & \res{0.96}{0.007}&\res{0.96}{0.013} &\res{0.96}{0.013} &\res{0.96}{0.013}\\
             & $F^r$ & \res{0.86}{0.013}&\res{0.88}{0.013} &\res{0.87}{0.025} &\res{0.0}{0.0}\\
             & $F^c$ & \res{0.97}{0.009}&\res{0.98}{0.013} &\res{0.96}{0.007} &\res{0.91}{0.032} \\
             \midrule 
             \multirow{3}{*}{{\smamath{RF}}}
             & $F^u$ &\res{0.96}{0.007}&\res{0.98}{0.005} &\res{0.98}{0.005} &\res{0.98}{0.005}\\
             & $F^r$ & \res{0.90}{0.013}&\res{0.9}{0.013} &\res{0.84}{0.027} &\res{0.0}{0.0}\\
             & $F^c$ & \res{0.97}{0.009}&\res{0.98}{0.005} &\res{0.96}{0.007} &\res{0.96}{0.007}\\
             \midrule 
             \multirow{3}{*}{{\smamath{LR}}}
             & $F^u$ & \res{0.97}{0.005}&\res{0.97}{0.004} &\res{0.97}{0.004} &\res{0.97}{0.004}\\
             & $F^r$ & \res{0.80}{0.013}&\res{0.8}{0.013} &\res{0.55}{0.009} &\res{0.0}{0.0}\\
             & $F^c$ & \res{0.98}{0.005}&\res{0.97}{0.005} &\res{0.97}{0.001} &\res{0.97}{0.004}\\
            \bottomrule
        \end{tabular}
        \label{tab:ze_wsp_wsp}
    }
\end{table}

\end{document}